\documentclass[aps,prb,twocolumn,nofootinbib,superscriptaddress]{revtex4-2}
\usepackage[a4paper,top=1.5cm,bottom=1.5cm,left=1.5cm,right=1.5cm]{geometry}
\usepackage{amsfonts}
\usepackage{amscd}
\usepackage{bm}
\usepackage{amsmath}
\usepackage{amssymb}
\usepackage{mathptmx}
\renewcommand{\onlinecite}[1]{\hspace{-1 ex} \nocite{#1}\citenum{#1}}
\setcitestyle{numbers,super}
\usepackage{ifpdf}
\usepackage[pdftex]{graphicx}
\usepackage{epstopdf}
\DeclareGraphicsExtensions{.eps, .pdf, .jpg, .tif}
\usepackage[pdftex]{hyperref}
\usepackage{color}
\newcommand{\onefourth}{\frac{\mbox{\small 1}}{\mbox{\small 4}}}
\newcommand{\upt}{UPt$_{3}$}
\newcommand{\sro}{Sr$_{2}$RuO$_{4}$}
\newcommand{\He}{$^3$He}
\newcommand{\Hea}{$^3$He-A}
\newcommand{\Heb}{$^3$He-B}
\newcommand{\Hefour}{$^4$He}
\newcommand{\be}{\begin{equation}}
\newcommand{\ee}{\end{equation}}
\newcommand{\ber}{\begin{eqnarray}}
\newcommand{\eer}{\end{eqnarray}}
\newcommand{\sgn}{\mbox{sgn}}
\newcommand{\eps}{\varepsilon}
\newcommand{\grad}{\mbox{\boldmath$\nabla$}}
\newcommand{\dive}{\mbox{\boldmath$\nabla$}\cdot}
\def\pder#1#2{\mbox{$\displaystyle\frac{\partial #1}{\partial #2}$}}
\def\ket#1{\mbox{$\displaystyle\vert\,#1\,\rangle$}}

\def\com#1#2{\left[#1\,,\,#2\right]} 
\def\ul#1{\underline{#1}}
\def\cH{{\mathcal H}}
\def\cN{{\mathcal N}}
\def\cY{{\mathcal Y}}
\def\vd{{\bf d}}
\def\ve{{\bf e}}
\def\vp{{\bf p}}
\def\vv{{\bf v}}
\def\vj{{\bf j}}
\def\vm{{\bf m}}
\def\vn{{\bf n}}
\def\vl{{\bf l}}
\def\vr{{\bf r}}
\def\vx{{\bf x}}
\def\vy{{\bf y}}
\def\vz{{\bf z}}
\def\vJ{{\bf J}}
\def\vL{{\bf L}}
\def\vQ{{\bf Q}}

\def\nicefrac#1#2{\genfrac{}{}{}{1}{#1}{#2}}

\def\ns{\negthickspace}

\def\hDelta{\hat{\Delta}}
\def\hp{\hat{p}}
\newcommand{\hvp}{\hat\vp}
\newcommand{\hvx}{\hat\vx}

\newcommand{\hvz}{\hat\vz}
\newcommand{\hvm}{\hat\vm}
\newcommand{\hvn}{\hat\vn}
\newcommand{\hvl}{\hat\vl}
\newcommand{\hvd}{\hat\vd}
\def\g{\mathfrak{G}}
\def\whg{\widehat\g}
\def\f{\mathfrak{F}}
\def\tone{\widehat{1}}

\def\tz{{\widehat{\tau}_3}}
\def\whDelta{\widehat{\Delta}}
\newcommand{\taux}{{\tau}_1}
\newcommand{\tauy}{{\tau}_2}
\newcommand{\tauz}{{\tau}_3}
\begin{document}
\title{
Weyl Fermions \& Broken Symmetry Phases of Laterally Confined $^3$He Films
}
\author{Hao Wu} 
\affiliation{Department of Physics and Astronomy, Northwestern University, Evanston, IL 60208}
\author{J.A. Sauls} 
\email{sauls@lsu.edu} 
\affiliation{Hearne Institute of Theoretical Physics, Louisiana State University, Baton Rouge, LA 70803}
\affiliation{Department of Physics and Astronomy, Northwestern University, Evanston, IL 60208}
\date{\today}
\begin{abstract}
Broken symmetries in topological condensed matter systems have implications for the spectrum of Fermionic excitations confined on surfaces or topological defects. The Fermionic spectrum of confined (quasi-2D) \Hea\ consists of branches of chiral edge states. The negative energy states are related to the ground-state angular momentum, $L_z = (N/2) \hbar$, for $N/2$ Cooper pairs. The power law suppression of the angular momentum, $L_z(T) \simeq (N/2)\,\hbar\,[1 - \nicefrac{2}{3}(\pi T/\Delta)^2 ]$ for $0 \le T \ll T_c$, in the fully gapped 2D chiral A-phase reflects the thermal excitation of the chiral edge Fermions. We discuss the effects of wave function overlap, and hybridization between edge states confined near opposing edge boundaries on the edge currents, ground-state angular momentum and ground-state order parameter of superfluid \He\ thin films.
Under strong lateral confinement, the chiral A phase undergoes a sequence of phase transitions, first to a pair density wave (PDW) phase with broken translational symmetry at $D_{c2} \sim 16 \xi_0$. The PDW phase is described by a periodic array of chiral domains with alternating chirality, separated by domain walls. The period of PDW phase diverges as the confinement length $D\rightarrow D_{c_2}$. 
The PDW phase breaks time-reversal symmetry, translation invariance, but is invariant under the combination of time-reversal and translation by a one-half period of the PDW. The mass current distribution of the PDW phase reflects this combined symmetry, and originates from the spectra of edge Fermions and the chiral branches bound to the domain walls.
Under sufficiently strong confinement a second-order transition occurs to the non-chiral ``polar phase'' at $D_{c1} \sim 9\xi_0$, in which a single p-wave orbital state of Cooper pairs is aligned along the channel.
\end{abstract}
\maketitle
\vspace*{-7mm}
\section{Introduction}
\vspace*{-5mm}

The superfluid phases of liquid \He\ are BCS condensates with topologically non-trivial ground states. The topology of the ground state of \He\ has sparked research into the properties of liquid \He\ in confined geometries.\cite{chu09,vol09,vol09a,sau11,miz12,miz12a,wu13,vor18,miz18d} In parallel with new theoretical investigations, the physical properties of superfluid \He\ confined in one- or two dimensions have become accessible through a combination of nano-scale device fabrication, precision NMR and transport measurements at low temperatures.\cite{gon11,lev13,lev13b}

For thin films, or \He\  confined in thin rectangular cavities, the ground state of \He\ is the \emph{chiral} ABM state,\cite{fre88,lev13,ben10} defined by the spin-triplet, p-wave order parameter, $\vd(\vp) = \Delta\,\hat{\vz}\,(p_x \pm i p_y)$, for a condensate of Cooper pairs with orbital angular momentum $L_z = \pm\,\hbar$ quantized along the normal to the film surface, and spin projection $S_z = 0$.\cite{and61,bri74}$^{,}$\footnote{The spin-quantization axis is aligned with the orbital angular momentum in order to minimize the nuclear dipolar energy.} This phase breaks time-reversal symmetry as well as parity, and is realized at all pressures below melting pressure in films with thickness $w \lesssim 300\,\mbox{nm}$. Experimental confirmation of broken time-reversal and mirror symmetry of \Hea\ was recently made by the observation of an anomalous Hall effect for electron bubbles moving in \Hea\ films.\cite{ike13,she16} Thus, thin films and confined geometries are ideal for investigating the topological excitations of the chiral phase of superfluid \He.

The ABM state, and its generalization to higher angular momenta, is also a model for topological superconductors with a chiral ground state, including the model for the superconducting state of \sro\ originally proposed as an electronic analog of superfluid \Hea,\cite{ric95}$^{,}$\footnote{The identification of the pairing symmetry and ground state of \sro, including whether or not it is chiral, as well as whether it is spin-triplet or spin-singlet, has been a focus of several research groups.~\cite{kal09,rag10,sca14,pus19,nga20,kiv20,leg21}} as well as the multi-component, heavy fermion superconductor, \upt, which exhibits broken time-reversal symmetry in the low-temperature B-phase; a spin-triplet, chiral order parameter of the form,\cite{sau94} $\vd(\vp)=\Delta\,\hat{\vz}\,p_z(p_x + i p_y)^2$, is consistent with Josephson interferometry\cite{str09} and polar Kerr effect measurements,\cite{sch14} and is predicted to exhibit chiral edge states, an intrinsic thermal Hall effect,\cite{gos15} and other novel transport properties.\cite{nga20}
There is also broad interest in inhomogeneous superconducting ground states, often associated with competing orders or pairing breaking mechanisms,~\cite{agt08,vor18,hol18a,jia23} as well as novel pair density wave phases in chiral superconductors~\cite{bar23} and paired fractional quantum Hall fluids.~\cite{san19} 
Here we predict novel inhomogeneous ground states to emerge, via a sequence of phase transitions, when the chiral phase of superfluid \He\ is confined along one direction of its orbital motion. Confinment may also provide a technique for creating novel ground states in unconventional and topological superconductors.
 
Much is now understood about topological superfluids and superconductors, particularly superfluid \He.\cite{vol16,miz16} 
Many new phases of confined \He\ with novel broken symmetries have been predicted based on Ginzburg-Landau theory that incorporates strong-coupling corrections to weak-coupling BCS theory for spin-triplet, p-wave pairing and boundary scattering and confinement.~\cite{wim13,wim15,wim16,wim18}
For recent reviews of broken symmetry phases of superfluid \He\ in confined geometries and emergent topology in these phases see Refs.~\onlinecite{vor18,miz18d}, and Refs.~\onlinecite{zhe17,lev19,hei21} for recent experimental studies of new phases of confined superfluid \He.
In the 2D limit the chiral ABM state is fully gapped and belongs to a topological class related to that of integer quantum Hall systems, but with a topological index defined in terms of the Nambu-Bogoliubov Hamiltonian.\cite{rea00,vol88,vol92}
The topology of the 2D chiral ABM state requires the presence of gapless Weyl fermions confined on the edge of a thin film of superfluid \Hea, or a domain wall separating degenerate topologically distinct phases with opposite chirality.\cite{volovik03} For an isolated boundary a single branch of Weyl fermioms disperses linearly with momentum $p_{||}$ along the boundary, i.e. $\varepsilon(p_{||}) = c\,p_{||}$, where the velocity of the Weyl Fermions, $c=v_f\,(\Delta/2E_f)$, is much smaller than the group velocity of Fermions in the normal phase of the Fermi liquid. 

Broken time-reversal symmetry is reflected in the asymmetry of the Weyl branch under time-reversal, $\varepsilon(-p_{||})=-\varepsilon(p_{||})$, which implies the existence of nonvanishing ground-state edge current derived from the occupation of the negative energy states. For superfluid \Hea\ confined in a thin cylindrically symmetric cavity, the \emph{edge sheet current}, $J=\nicefrac{1}{4}\,n\,\hbar$, is the origin of the ground-state angular momentum, $L_z = (N/2)\,\hbar$, predicted by McClure and Takagi based on symmetry properties of an N-particle BCS ground-state of chiral p-wave pairs confined by a cylindrically symmetric potential.\cite{mcc79}
Many unique features of \Hea, particularly anomalous transport properties, are directly related to Weyl Fermions confined on edges, domain walls, vortices or bound to electron bubbles.\cite{she16, tsu14, ike13, ike13b, ike15}

\begin{figure}[t]
\includegraphics[width=0.49\textwidth]{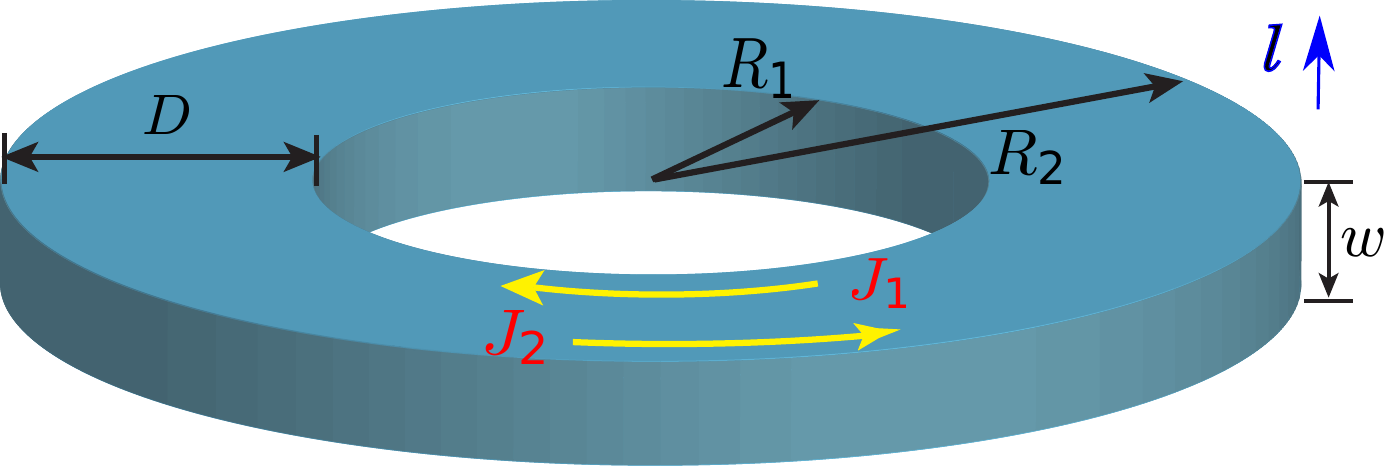}
\caption{Counter-propagating edge currents for \Hea\ confined in a thin toroidal cavity. For $w\ll\xi_0$, $R_{1,2}\gg\xi_0$ and $D\gg\xi_0$ the edge currents are: $J_2=-J_1=\nicefrac{1}{4}\,n\,\hbar$.} \label{fig-Toroidal_Film}
\end{figure}

We consider the effects of \emph{hybridization} between distinct branches of Weyl Fermions confined on neighboring boundaries and domain walls. We focus on thin (thickness $w\ll\xi_0$) films of \Hea\ that are \emph{laterally} confined, as shown for example in Fig. \ref{fig-Toroidal_Film}; $D$ is the lateral confinement dimension. In Sec. \ref{sec-hybridization-Andreev} we relate bound-state formation to the multiple scattering problem of Nambu-Bogoliubov Fermions in a laterally confined chiral superfluid film. We develop a quasiclassical analysis for Bogoliubov quasiparticles in a laterally confined film, map the multiple-reflection of wave packets onto ballistic porpagation of wavepackets through a periodic array of domain walls separating degenerate, time-reversed domains of chiral order parameter. The sub-gap and continuum spectra exhibit a band-structure resulting from hybridization of counter-propagating Weyl Fermions on opposite edges. 
In Sec. \ref{sec-hybridization-Eilenberger} we employ a quasiclassical Green's function formalism to calculate the local density of states (DOS), resolved in both energy and momentum, as well as the edge currents and order parameter. Self-consistent solutions for the order parameter and spectral function (Sec. \ref{sec-Self-consistent_QC}) are essential for obtaining and interpreting the ground-state of \He\ films under strong confinement. In section \ref{sec-Edge_Current_vs-TD}, we show that hybridization of the Weyl branches leads to suppression of the chiral ABM state, and a reduction of the edge current. Even weak lateral confinement ($D\gg\xi_0$ where $\xi_0=\hbar v_f/2\pi\,k_{\text{B}}T_c\approx 200-800\,\mbox{\AA}$ is the superfluid correlation length) suppresses the chiral ABM state and stabilizes a non-chiral \emph{polar} state, $\vd(\vp) = \Delta\,\hat\vz\,p_y$ for $T\lesssim T_c$. In Sec. \ref{sec-Phase_Diagram} we develop linear instability analysis to identify phase transitions and phase boundaries as a function of temperature $T$ and lateral confinement $D$.

The phase diagram for \He\ films as a function of temperature and lateral confinement is obtained based on numerical calculations of the thermodynamic potential (Sec. \ref{sec-Free_Energy}) from self-consistent solutions for the order parameter and spectral function. A central result of this report is the prediction of a \emph{pair density wave} (PDW) phase of \He\ that spontaneously breaks translational symmetry, in addition to parity and time-reversal symmetries, and is stable over a wide range of temperature and confinement (Sec. \ref{sec-PDW_Phase}).
As a function of confinement, the PDW phase evolves from the polar phase as a single-Q mode chiral instability at $D_{c_1} \simeq 9\xi_0$ into to a periodic array of domain walls separating pairs of time-reversed chiral domains. The onset of the PDW phase from the translationally invariant chiral ABM state defines the upper critical confinement length, $D_{c_2}\approx 16\xi_0$. The structure, evolution and signatures of the PDW phase are discussed in Secs. \ref{sec-PDW_Phase}-\ref{sec-Chiral_DWs}.
We begin with a brief introduction to the Fermionic spectrum and boundary current on an isolated edge of a thin film of \Hea.  

\vspace*{-5mm}
\subsubsection*{Symmetry and Topology of \Hea}
\vspace*{-5mm}

The spectrum and wave functions for the Fermionic excitations of the chiral ABM state are obtained from solutions of the Bogoliubov equations,\cite{vol88,kur90}
\ber\label{eq-Bogoliubov_Equations}
+\xi(\vp)\,u(\vr) +\hDelta(\vp)\,v(\vr) = \varepsilon\,u(\vr)
\,,
\\
-\xi(\vp)\,v(\vr) +\hDelta^{\dag}(\vp)\,u(\vr) = \varepsilon\,v(\vr)
\,,
\eer
for the two-component particle ($u(\vr)$) and hole ($v(\vr)$) spinor wavefunctions. 
The kinetic energy, $\xi(\vp)=|\vp|^2/2m^{*}-\mu$, is measured relative to the chemical potential, $\mu$, while $\hDelta(\vp)$ is the $2\times 2$ spin-matrix order parameter,
\be\label{ESP_A-phase}
\hDelta(\vp)=\vd\cdot(i\vec\sigma\sigma_y)\,\Delta^{(\pm)}(\vp)
\,,
\ee
 and $\vp\rightarrow\nicefrac{\hbar}{i}\grad$ in the coordinate basis. 
For the chiral ABM state the orbital order parameter is $\Delta^{(\pm)}(\vp) \equiv \Delta\,(\hvm \pm i \hvn)\cdot\vp/p_f$, where $\{\hvm\,,\,\hvn\,,\,\hvl\}$ is an orthogonal triad defining the orbital orientation of the Cooper pairs with chirality $\pm 1$. The real vector $\hvd$ is the direction in spin space along which Cooper pairs have zero spin projection.
For a dipole-locked \Hea\ film $\hvd \parallel \hvl=\hvz$. Thus, the spin state of the pairs, $i\vec\sigma\sigma_y\cdot\hvd=\sigma_x$, is the ESP state with equal amplitudes for spins polarized along $+\hvx$ and $-\hvx$.

The 3D bulk A-phase has gapless Fermions for momenta along the nodal directions, $\vp\, || \pm\,\hvl$.\cite{volovik92} These zero-energy Fermions are protected by the integer phase winding of $\Delta^{\pm}\propto e^{\pm i \varphi_{\hp}}$ about the chiral axis $\hvl$. For thin films of \Hea, dimensional confinement discretizes the 3D Fermi sphere into the set of 2D Fermi disks; the Fermi surface is the set of boundaries of the Fermi disks. The resulting chiral phase is fully gapped on each disk. In the 2D limit the topology of \Hea\ can be expressed via the Bogoliubov Hamiltonian written in the momentum and iso-spin representation using the Nambu-Pauli matrices, $\vec{\tau}$, 
\be
\cH \equiv \vec{\vm}(\vp)\cdot\vec{\tau} = \xi(\vp)\tauz + c\,p_x\taux \mp c\,p_y\tauy
\,,
\label{eq-Bogoliubov_Hamiltonian_ABM}
\ee
for both ESP components, with $c = \Delta/p_f$, which emerge as the velocity of Weyl Fermions confined on a boundary of \Hea. The dispersion relation for excitations in the bulk of the chiral condensate is obtained from $\cH^2=|\vec{\vm}(\vp)|^2=\xi(\vp)^2 + c^2 |\vp|^2$. Note that $|\vec{\vm}(\vp)| \ne 0$ for all $\vp$ provided $\mu\ne 0$. The point $\mu=0$ defines a quantum critical point separating a chiral BEC ($\mu<0$) and a chiral BCS ($\mu>0$) condensate. 
For \Hea\ the chemical potential is large and positive, $\mu = E_f > 0$, thus the low-energy excitations are confined near the Fermi surface with $|\vp|\approx p_f$, with bulk excitation energies $\varepsilon^{(\pm)}(\vp) = \pm\sqrt{\xi(\vp)^2 + \Delta^2}$.

For $\mu > 0$ the iso-spin vector,
\be 
\vec{\vm}(\vp) =
c\,p_x\,\hat{\ve}_1\mp\,c\,p_y\,\hat{\ve}_2+(p^2/2m^{*} - \mu)\,\hat{\ve}_3
\,,
\ee 
defines a skyrmion field in 2D momentum space with Chern number\cite{vol88}
\be
N_{\text{2D}} = \pi\,\int\,
	        \frac{d^2p}{(2\pi)^2}\,\hvm(\vp)\cdot
	        \left(\pder{\hvm}{p_x}\times\pder{\hvm}{p_y}\right) = \pm 1
\,,
\ee
where $\hat{\vm}(\vp) = \vec{\vm}(\vp)/|\vec{\vm}(\vp)|$. However, the chiral BEC with $\mu < 0$ is topologically trivial with $N_{\text{2D}} = 0$; thus, the point $\mu=0$ is a topological phase transition.

\vspace*{-5mm}
\subsubsection*{Andreev's Equation}
\vspace*{-5mm}

At a domain wall or an interface the Chern number changes abruptly. A discontinous change in the topological index forces the excitation gap to close and a spectrum of gapless Fermions to be confined on the boundary. In \Hea\ such a change in topology is also accompanied by the breaking of chiral Cooper pairs, with one or more components of the order parameter suppressed at the boundary. Thus, we consider a more general parametrization, $\Delta(\vr,\vp) = \left(\Delta_{1}(\vr)\,p_x + i\,\Delta_{2}(\vr)\,p_y\right)/p_f$, where the two order parameter components heal to their bulk values far from the boundary, $\Delta_{1,2}(\vr)\rightarrow\Delta$. The Hamiltonian generalizes to
\be
\label{eq-Bogoliubov-Hamiltonian-component}
\cH = \xi(\vp)\tauz + \Delta_{1}(\vr,\vp)\taux - \Delta_{2}(\vr,\vp)\tauy
\,,
\ee
for both ESP components defined by $\hvd=\hvz$, with the pair potentials interpreted as symmetrized differential operators, $\Delta_{1,2}(\vr,\vp)=\nicefrac{\hbar}{2i}\left(\Delta_{1,2}(\vr)\,\partial_{x,y}+\partial_{x,y}\,\Delta_{1,2}(\vr)\right)$.

A quasiclassical approximation to the Bogoliubov equations is obtained by separating the Bogoliubov spinor, $\ket{\varphi}\equiv \left(u\,,\, v\right)^{\text{T}}$, into fast- and slow spatial variations: $\ket{\varphi} = e^{ip_f\hvp\cdot\vr/\hbar}\,\ket{\psi}$, where the quasi-classical spinor amplitudes, $\ket{\psi}\equiv (u_{\vp}, v_{\vp})^{\text{T}}$, vary in space on a length scale $L\gtrsim\xi$. By retaining the leading order terms in $\hbar/p_f L\ll 1$ we obtain Andreev's equation,\cite{and64,sau18}
\ber
i\hbar\,\vv_{\vp}\cdot\grad\ket{\psi}
+
\begin{pmatrix}
\varepsilon			&	-\Delta(\hvp,\vr)	\cr
\Delta^{\dag}(\hvp,\vr)		&	-\varepsilon	
\end{pmatrix}
\ket{\psi}
 = 0
\,,
\label{eq-Andreev_Equation}
\eer
where 
$\Delta(\hvp,\vr) = \Delta_{1}(\vr)\,\hp_x+i\,\Delta_{2}(\vr)\,\hp_y$, $\hp_{x,y}$ are the direction cosines of the Fermi momentum, $\vp=p_f\hat{\vp}$, and Fermi velocity, $\vv_{\vp}=v_f\hvp$.
The latter define straight-line trajectories in classical phase space for the propagation of wavepackets of Bogoliubov Fermions, i.e. coherent superpositions of particles and holes with amplitudes given by $\ket{\psi}$.
 
\vspace*{-5mm}
\subsubsection*{Weyl Fermions on the edge}
\vspace*{-5mm}

At an edge boundary, or a domain wall, the topology of \Hea\ changes abruptly leading to a spectrum of gapless Weyl Fermions propagating along the edge with dispersion relation
\be
\varepsilon(\vp) = \mp c\,p_{||}\,,\quad - p_f \le p_{||} \le p_f
\,,
\ee
where the $\mp$ sign corresponds to the chirality, $\pm\hvl$, of the ground state. A $T=0$ the negative energy states are occupied, and generate a ground-state edge current, $\vj(x_{\perp})$, parallel to the edge, where $x_{\perp}$ is the distance into \Hea\ normal to the surface. The total sheet current, including the back action from the negative energy continuum to the formation of the gapless edge states, is given by\cite{sau11}
\be
\vJ = \int_{0}^{\infty}\ns dx_{\perp}\,\vj(x_{\perp}) = \pm\onefourth\,n\,\hbar\,\hat{\ve}_{||}
\,,
\ee
where $n$ is the density of \He\ atoms, and $\hat{\ve}_{\parallel,\perp}$ are unit vectors tangent ($\parallel$) and outward normal ($\perp$) to the boundary. The sign of the current is defined by the chirality of the ground state. For a thin film ($w\ll\xi_0$) of \Hea\ with chirality $\pm\hvz$ confined in a cylindrical cavity with a macroscopic radius $R\ggg\xi_0$, a sheet current circulates on the boundary given by $\vJ = \pm\onefourth\,n\,\hbar\,\hat{\ve}_{\varphi}$. This current gives rise to the ground-state angular momentum, 
\be
\vL_{\pm} = \int_{V}dV\vr\times\vj(\vr) = \pm\frac{N}{2}\,\hbar\,\hat{\vz}
\,,
\ee
where $N$ is the total number of \He\ atoms.\cite{sto04,sau11}

\vspace*{-5mm}
\subsubsection*{Symmetry Protection of the Edge Current} 
\vspace*{-5mm}

McClure and Takagi (MT) obtained the above prediction for the ground-state angular momentum of \Hea\ by considering an N-particle BCS ground state (condensate) of chiral p-wave pairs.\cite{mcc79} For a cylindrically symmetric confining potential they showed that a condensate wave function of $N/2$ chiral pairs is an eigenfunction of the angular momentum operator with eigenvalue $L_z=(N/2)\hbar$, and that this result is independent of the radial size of the pairs, i.e. the result is valid in both the BEC and BCS limits. 
Since a BEC of chiral p-wave molecules is topologically trivial ($N_{\text{2D}}=0$), while the chiral BCS condensate is characterized by a non-trivial topological winding number $N_{\text{2D}}=\pm 1$, the MT result, $L_z=(N/2)\hbar$, is not simply a consequence of the topology of the bulk Hamiltonian in Eq.  \ref{eq-Bogoliubov_Hamiltonian_ABM}, nor is the edge current in the BCS condensate protected by topology. Rather, the MT result implies that the ground-state angular momentum of $(N/2) \hbar$ is protected by symmetry. Indeed, for a BCS condensate of chiral pairs confined by a non-specular boundary that violates axial symmetry on any length scale 
$L \ll \xi_0$, 
the ground-state edge current, and thus the angular momentum, is reduced compared to the MT result.\cite{sau11}
By contrast, for a chiral BEC with a non-specular boundary the ground-state angular momentum is still $(N/2)\hbar$ because the angular momentum is confined within each molecule. Thus, the total angular momentum of the BEC is then $\hbar$ times the volume integral of the condensate density which is
insensitive to the boundary conditions.\cite{vol81b}

Nevertheless, the topology of the chiral BCS condensate plays a key role in determining the origin of the ground-state angular momentum. The point $\mu=0$ separating BEC and BCS condensation is a phase transition in which the topology of the Bogoliubov Hamiltonian changes discontinuously. The change in topology is accompanied by ``spectral flow'' in which an odd number of branches of Fermionic excitations appear as $\mu$ is ``adiabatically tuned'' through zero from $\mu<0$ to $\mu>0$.\cite{bal86,bal87}
In the 2D limit the spectral flow yields a single branch of chiral Fermions, which disperses through the Fermi level. The zero energy mode is protected by the non-trivial topology of the bulk Hamiltonian, while the occupation of the negative energy states leads to a ground-state edge current. However, the magnitude of that current depends on the surface boundary conditions, and thus the symmetry of the confining potential, and is equal to the MT result only for specular reflection by a cylindrically symmetric potential.

\vspace*{-5mm}
\section{Laterally confined \He\ films}
\label{sec-hybridization-Andreev} 
\vspace*{-5mm}

\begin{figure}[t]
\includegraphics[width=0.8\columnwidth]{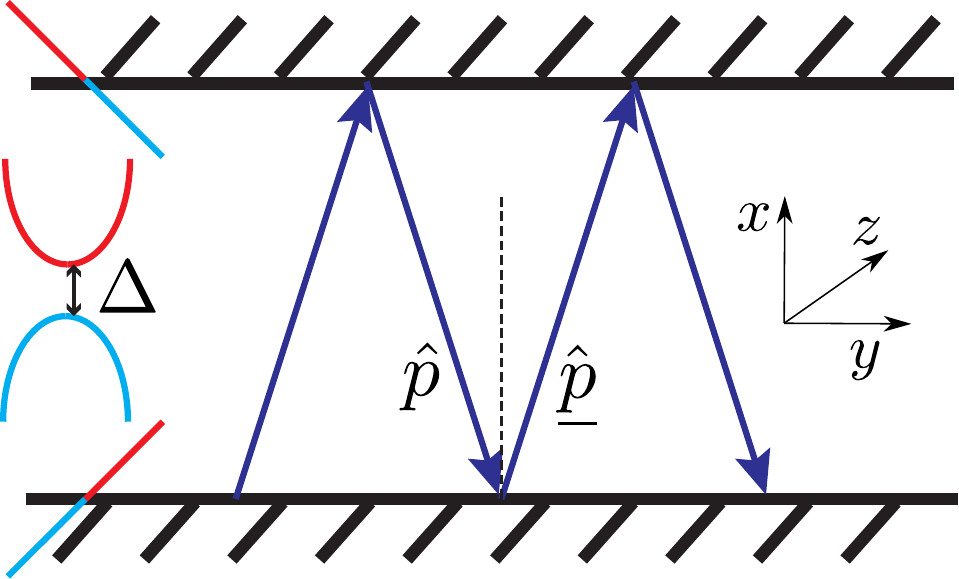}
\caption{Thin film of \He\ with parallel edges and lateral separation, $D$. A quasiclassical trajectory with multiple reflections couples Fermionic states on opposite edges.}
\label{fig-Chiral_Ribbon-Trajectories}
\end{figure}

A key feature of the edge current is that the \emph{direction} of the edge current is defined by the chirality and the outward normal to the surface, i.e. $\vJ\sim \hat\vl\times\hat\ve_{\perp}$. Thus, for a toroidal geometry the currents counter propagate on the inner and outer edges as shown in Fig. \ref{fig-Toroidal_Film}. If the inner and outer edges are both specularly reflecting, then $\vJ_2 = -\vJ_1 = \nicefrac{1}{4}\,n\,\hbar\,\hat\ve_{\varphi}$, and the ground-state angular momentum is then $\vL = \ve_{r}\times\left(2\pi\,R_1^2\vJ_1 + 2\pi\,R_2^2\vJ_2\right)\,w = (N/2)\hbar\,\hat\vz$. Thus, we recover the MT value as a result of compensation of the angular momentum from the outer edge current by that from counter-propagating inner edge current. However, this result can only be valid when there is no overlap between the states on the inner and outer edges. 
So, ``what is the effect of lateral confinement ($D=R_2-R_1$) on the ground state of \He\ films?'', and at ``what length scales $D$ are the ground state properties of the film strongly modified?'' In what follows we neglect the effects of boundary curvature and consider a thin film, laterally confined in a narrow channel with parallel edges separated by distance $D$ as shown Fig. \ref{fig-Chiral_Ribbon-Trajectories}.
Our analysis also assumes zero magnetic field. However, all of the ground states of laterally confined, thin \He\ films that we report are equal spin pairing states. Thus, the magnetic field does not play a role in the relative stability of the different orbital ground states of thin \He\ films under lateral confinement.
The nuclear Zeeman and dipole energies are however important for determining the NMR signatures of these phases, but this is outside the scope of this report.

We first investigate the effects of lateral confinement by solving Andreev's equation (Eq. \ref{eq-Andreev_Equation}) subject to boundary conditions connecting the Nambu spinor amplitudes defined on classical trajectories. Wave packets of normal-state particles and holes propagate at the Fermi velocity along classical, straight-line trajectories defined by the direction of the group velocity $\hvp$. The wave packets undergo specular reflection at a translationally invariant boundary.\cite{sto04,kur87,mil88} Thus, for specular reflection, $\vp\rightarrow \underline{\vp} = \vp -2\vn\,(\vn\cdot\vp)$, where $\vn$ is the normal to the boundary directed into \He, the Nambu spinors obey the conditions,
\ber
\hspace*{-6mm}
\ket{\psi(\hvp;0,y)} = \ket{\psi(\ul\hvp;0,y)}
\,,
\nonumber
\\
\ket{\psi(\hvp;D,y)} = \ket{\psi(\ul\hvp;D,y)}
\,,
\label{eq-specular_bc}
\eer
for parallel boundaries as shown in Fig. \ref{fig-Chiral_Ribbon-Trajectories} with $x$ being the coordinate normal to the boundaries, and $y$ is the coordinate along the channel, in which case $\ul\hp_x = -\hp_x$ and $\ul\hp_y = \hp_y$. Note that we have assumed that translational symmetry is preserved along the laterally confined film by restricting the order parameter to depend only on $x$. Andreev's equations reduce to coupled scalar equations,
\ber
i\partial{u} + \varepsilon u - \Delta(\hvp,x)\,\tilde{v} &=& 0	
\,,
\nonumber
\\
i\partial{\tilde{v}} - \varepsilon\tilde{v} + \Delta^{*}(\hvp,x)\,u &=& 0 
\,,
\label{eq-Andreev_Equation2}
\eer
where $\partial\equiv \hbar v_f\hvp\cdot\grad$ and $\tilde{v} \equiv \sigma_x v$. 

\begin{figure}[t]
\includegraphics[width=0.65\columnwidth]{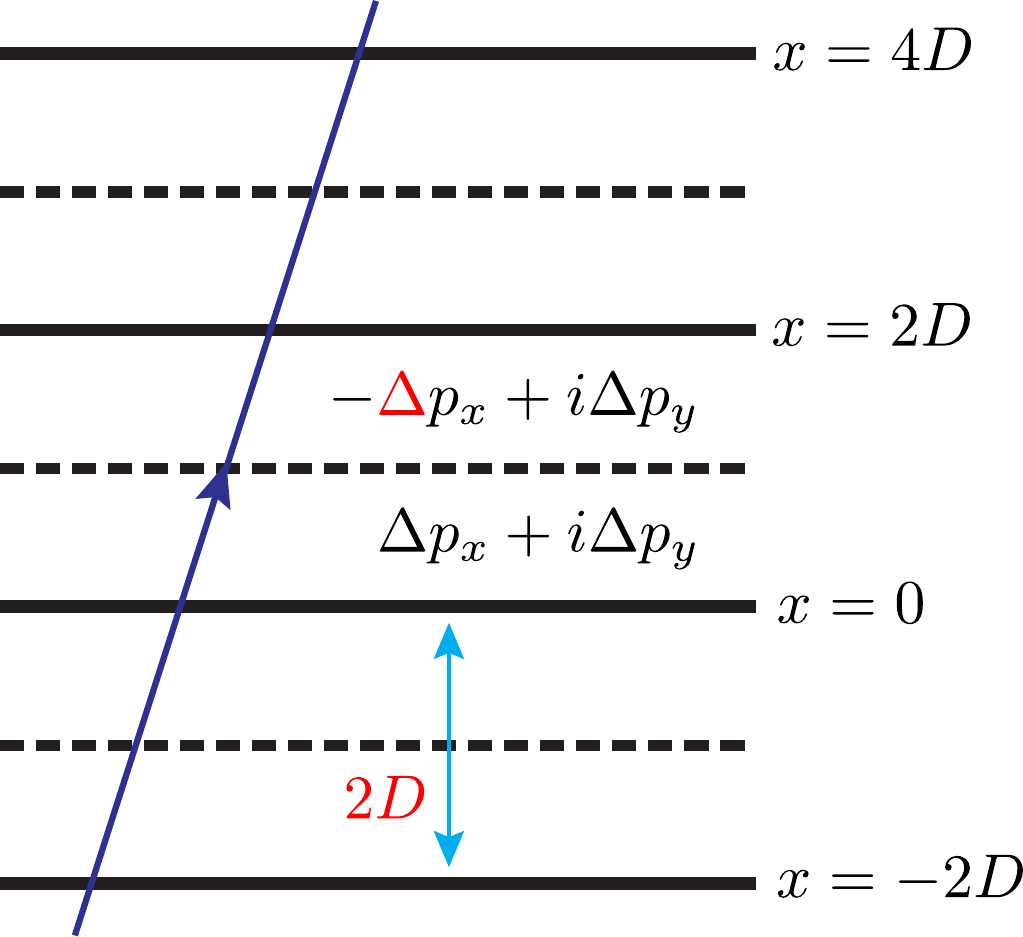}
\caption{Mapping of multiply-reflected trajectories to a single trajectory traversing a periodic array of oppositely aligned chiral domains with period $2D$.}
\label{fig-Chiral_Ribbon-Mapping}
\end{figure}

Equations \ref{eq-Andreev_Equation2} and \ref{eq-specular_bc} defined on the domain, $x\in[0,D]$ are equivalent to Eq. \ref{eq-Andreev_Equation2} extended to $x\in(-\infty,+\infty)$ defined on a straight trajectory $\hvp$ passing through a periodic array of chiral domains separated by domain walls every half period $D$, as shown in Fig. \ref{fig-Chiral_Ribbon-Mapping}.
The periodic extension of domain walls corresponds to reflections $\hp_x\rightarrow -\hp_x$, as shown in Fig. \ref{fig-Chiral_Ribbon-Trajectories}, and is represented by the order parameter,
\be\label{eq-Periodic_Chiral-OP}
\Delta(\hvp,x) = S(x)\Delta \hp_x + i \Delta \hp_y
\,,
\ee
where $S(x)=1$ if $2nD < x <(2n+1)D$, $S(x)=-1$ if $(2n-1)D < x < 2nD$. Periodicity allows us to express the Andreev amplitudes as a Bloch wave solution of the form,
\be
\begin{pmatrix}	u \cr \tilde{v}	\end{pmatrix}
= 
e^{i\,kx/\hp_x}
\begin{pmatrix}	U(x) \cr V(x) \end{pmatrix}
\,,
\label{eq-Bloch_Waves}
\ee
where $k$ is the Bloch wave number, $U(x)$ and $V(x)$ are periodic functions that satisfy
\ber
\label{eq-Andreev_U}
&&
i\partial U + (\varepsilon - \hbar v_f k) U - \,\Delta(\hvp,x) V   = 0	
\,,
\\
&&
i \partial V - (\varepsilon + \hbar v_f k) V + \Delta^*(\hvp,x) U 	= 0
\,,
\label{eq-Andreev_V}
\eer
and boundary conditions: $U(2D) = U(0)$, $V(2D) = V(0)$.

The solution of Eqs. \ref{eq-Andreev_U} and \ref{eq-Andreev_V}, with periodic boundary conditions, can be carried out (see Appendix \ref{appendix-Andreev-Bloch_Solution}) to obtain the eigenfunctions, $U(x)$ and $V(x)$, and the eigenvalues, $\varepsilon(p_{||},k)$, for both positive and negative energy branches of the spectrum, which depend on the conserved momentum along the channel, $p_{||}\in[-p_f, +p_f]$, and the Bloch wave number, $k\in[-\pi/2D,+\pi/2D]$, defined on the first Brilloun zone.

The branches of the sub-gap spectrum, $|\varepsilon| < \Delta$, are the solutions of
\be
\varepsilon^2 = \Delta^2 \hp_y^2 + \frac{\Delta^2 \hp_x^2 (1 - \cos(2kD))}{\cosh(2\lambda D) - \cos(2kD)}
\,,
\label{eq-sub-gap_energy-levels}
\ee
where $\hbar\,v_x\,\lambda(\varepsilon) = \sqrt{\Delta^2-\varepsilon^2}$ and $v_x=v_f\hat{p}_x$. For $k=0$ we obtain \emph{two} linearly dispersing branches, 
\be
\varepsilon_{\pm}(p_{||},0) = \pm \Delta\hp_y \equiv \pm c\,p_{||}
\,,
\ee
corresponding to Weyl fermions with opposite chirality on the opposing boundaries (see Fig. \ref{fig-Sub-Gap_Band-Structure}). Hybridization of the pair of Weyl branches is evident for $k\ne 0$ with band formation of the levels at fixed $p_{||}$; the band edges are defined by the solutions of Eq. \ref{eq-sub-gap_energy-levels} at $k=0$ and the Brilloun zone boundary, $k=\pi/2D$.  
The asymptote, $\epsilon_{+} = +\,c\,p_{||}$, is the $k = 0$ dispersion for states confined on the upper edge $x=D$, while $\varepsilon_{-}=- c\,p_{||}$ corresponds to states confined near $x=0$. In the limit $D\gg\hbar v_f/2\Delta$, the maximum bandwidth is exponentially small, $w\simeq\Delta\,e^{-2D\,\Delta/\hbar v_f}$, which reflects the exponential decay of edge states into the channel. The bandwidth increases with confinement and is largest for states with $p_x = p_f$, i.e. trajectories normal to the confining boundaries. The bandwidths are exponentially small for trajectories nearly grazing incidence, i.e. $p_{||}\approx p_f$, which follows from the divergence of the effective period, $2D_{\text{eff}}=2D/\hp_x$, for near grazing incidence trajectories. 

\vspace*{-5mm}
\section{Propagator and Spectral Function}\label{sec-hybridization-Eilenberger} 
\vspace*{-5mm}

By solving the Andreev equations, we obtain the sub-gap eigenstates for the confined film. The states are coherent superpositions of bound states confined at both edges with opposite group velocities, which decay into the channel region on the length scale $\sim\xi_0$. The overlap of the edge-state wave functions does not open a gap in the subgap spectrum; rather the isolated bound states with spectral function $2 \pi\Delta \delta(\eps - \eps_\pm)$ develops into a band with a bandwidth dependent on both confinement length $D$ and trajectory $\vp$.
In what follows we discuss the origin of the additional states that form the sub-gap band structure, and the spectral weight redistribution under confinement. The edge-state spectral function is obtained from the local, retarded single-particle Green's function, which in the quasiclassical limit satisfies Eilenberger's transport equation,\cite{eil68}
\be\label{equ-eilenberger-transport}
\com{\eps \tauz - \widehat \Delta}{\widehat \g (\vp, \vr; \eps)}
+ i \hbar \vv_\vp \cdot \nabla \widehat \g(\vp, \vr; \eps) = 0 
\,,
\ee
and normalization condition, $\widehat \g^2 = -\pi^2 \widehat 1$, where $\widehat\Delta$ is the pairing self-energy and $\widehat\g$ is the quasiclassical Green's function in Nambu space, and 
$\com{\widehat{A}}{\widehat{B}} = \widehat{A}\widehat{B} -\widehat{B}\widehat{A}$.
In this section we present results for the quasiclassical propagator in laterally confined \Hea\ thin films for a spatially uniform order parameter $\Delta(\vp)$. These results provide insight into the 
spectrum of the laterally confined chiral phase. Self-consistent numerical solutions of the quasiclassical transport equation are discussed in Secs. \ref{sec-Self-consistent_QC}-\ref{sec-Free_Energy} to follow.

For a laterally confined film, the solution for $\widehat{\g}$ (see Appendix \ref{appendix-quasiclassical_solution}) is
\be\label{eq-QC_propagator-channel}
\widehat{\g}(\vp, \vr; \eps) = \frac{1}{M}\left(
\widehat{\g}_0 + 
C^1(\vp) e^{-2 \lambda x / v_x}\widehat{\g}_+ + 
C^2(\vp) e^{2\lambda x / v_x} \widehat{\g}_-
\right)
\,,
\ee 
where $\widehat{\g}_0$ is the bulk propagator and $\widehat{\g}_\pm$ are the Nambu matrices representing the states generated by the edges. Coefficients $C^1$ and $C^2$ are obtained by matching boundary conditions at both edges for the trajectory $\vp$ and its specularly reflected counterpart $\ul\vp$. The factor $M$ is given by
\be
\label{equ_M_square}
M^2 = 1 + 2 C^1 C^2
= 1 - \frac{\Delta^2 \hp_x^2}{(\eps^2 - \Delta^2 \hp_y^2) \cosh(\lambda D)^2}
\,.
\ee
The physical Green's function satisfies Eilenberger's normalization condition, which is guaranteed by the commuatation relations, 
\ber
&&
(\widehat{\g}_0)^2 = -\pi^2\,\tone
\,,
\quad\quad
(\widehat{\g}_\pm)^2 = 0
\,,\\
&&
[\widehat{\g}_0, \widehat{\g}_\pm]_+ = 0
\,,
\quad\quad
[\widehat{\g}_+, \widehat{\g}_-]_+ = -2\pi^2\,\tone
\,.
\eer

The propagator $\widehat \g$ can also be expanded in Pauli matrices in Nambu space, 
$\{\taux,\tauy,\tauz\}$, 
\be
\widehat \g(\vp, \vr) = \g(\vp, \vr) \tauz + \sigma_x(\f_1(\vp, \vr) \taux + \f_2(\vp, \vr) \tauy)
\,.
\ee
The off-diagonal propagators, $\f_{1,2}(\vr,\vp)$, contain information about the Cooper pair spectrum, while the diagonal component, 
\begin{eqnarray}
\g(\vp, x; \eps)
&=& 
\frac{1}{M}
\Big(
-\frac{\pi \eps}{\Lambda} 
-\frac{\pi \Delta_1 \left(\eps \Delta_2 -\Lambda \Delta_1 \right)}{\Lambda \left(\Lambda \eps - \Delta_1 \Delta_2 \right)} 
\frac{e^{2 \lambda (D - x)} - e^{-2 \lambda x}}{e^{2 \lambda D} - e^{-2 \lambda D}} 
\nonumber
\\
&-&
\frac{\pi \Delta_1 \left(\eps \Delta_2 + \Lambda \Delta_1 \right)}{\Lambda \left(- \Lambda \eps - \Delta_1 \Delta_2 \right)} 
\frac{e^{2 \lambda x} - e^{2 \lambda(x - D)}}{e^{2 \lambda D} - e^{- 2 \lambda D}} 
\Big)
\,,
\label{equ-quasiclassical_g}
\end{eqnarray}
determines the local fermionic density of states,
\be
\cN(\vp, x; \eps) = -\frac{1}{\pi} Im\, \g(\vp, x; \eps + i 0)
\,,
\ee
where $\Lambda \equiv \sqrt{\Delta^2-\eps^2}$, $\Delta_1=\Delta\hp_x$ is the $\hp_x$ component of the order parameter for a constant gap amplitude, and $\Delta_2 = \mp\Delta\hp_y$ is the $\hp_y$ orbital component.
Here and thoughout the manuscript the notation, $\Delta_{1,2}$ refers to the $\taux$,$\tauy$ component of $\widehat\Delta$, respectively. For the undistorted chiral A phase, $\Delta_{1}=\Delta\hp_x$ and $\Delta_{2}=\mp\hp_y$ for the ground state with $L_z = \pm\hbar$. 
For laterally confined \He\ the amplitudes of the two components differ. 
Thus, for the laterally confined A-phase with right-handed chirality, and with $\hat\vx$ defining the normal to the confining edge, $\Delta_{1}=\Delta_{\perp}\hat\vp_x$ and $\Delta_{2}=-\Delta_{||}\hat\vp_y$, where $\Delta_{\perp}$ ($\Delta_{||}$) is the amplitude of the orbital state that is normal (parallel) to the pair-breaking edge. We address the effects of anisotropy induced by pairbreaking in Secs. \ref{sec-Self-consistent_QC}-\ref{sec-Free_Energy}.

\begin{figure}
\includegraphics[width=\columnwidth]{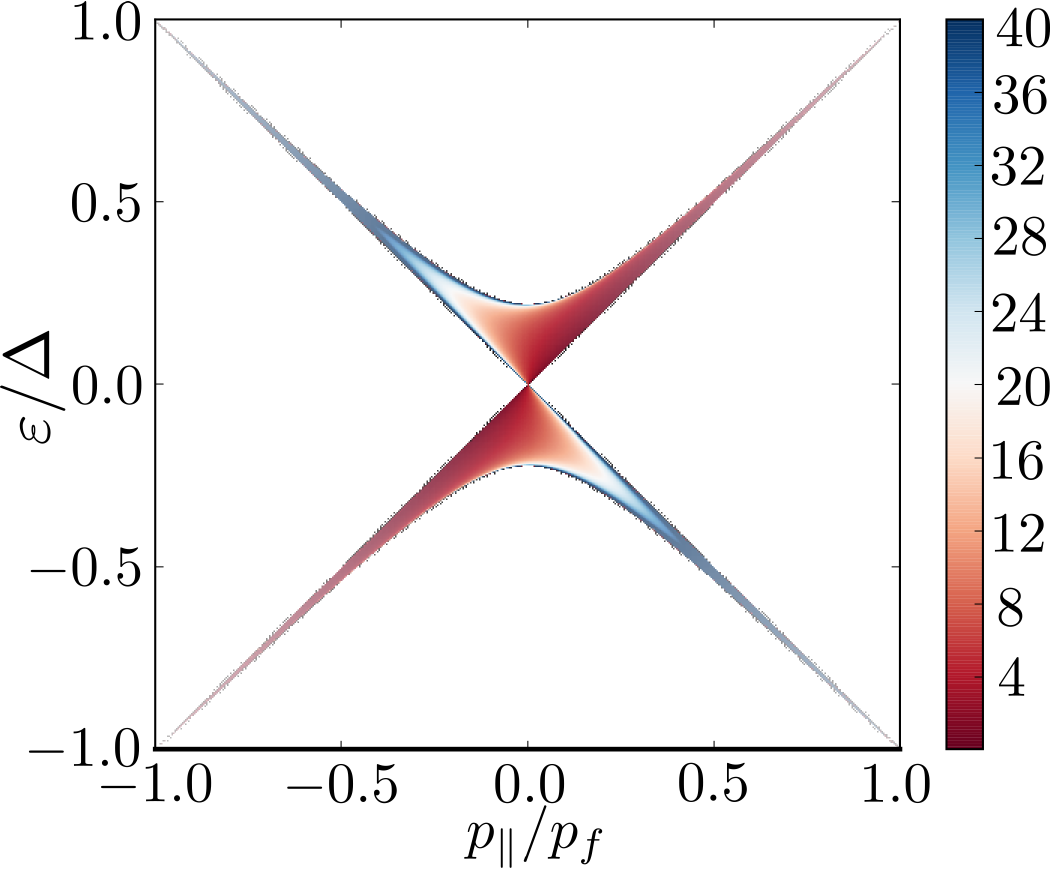}
\caption{Bound state spectral density at edge $x = D$ for lateral confinement $D = 8\xi_0$ 
	as a function of parallel momentum, $p_\parallel$, and energy, $\eps$.
	The spectral density is given by the color map and key in units of $N_f$.
	The parallel momentum is in units of $p_f$ and the bound-state energy is in units 
        of $\Delta$. 
}
\label{fig-Sub-Gap_Band-Structure}
\end{figure}

In the limit, $D/\xi_0\rightarrow\infty$, the factor $M\rightarrow 1$, and the $\tauz$ component reduces to 
\ber
\g^\infty 
=
-\frac{\pi \eps}{\Lambda}
&-&
\frac{\pi \Delta_1 \left(\eps \Delta_2 -\Lambda \Delta_1 \right)}{\Lambda 
\left(\Lambda \eps - \Delta_1 \Delta_2 \right)}\,e^{-2\lambda x}
\nonumber
\\ 
&-&
\frac{\pi \Delta_1 \left(\eps \Delta_2 + \Lambda \Delta_1 \right)}{\Lambda 
\left(-\Lambda \eps - \Delta_1 \Delta_2 \right)}\,e^{-2 \lambda (D - x)}
\,.
\eer

\begin{figure}[t]
\includegraphics[width=0.5\textwidth,trim={1cm 0cm 1.5cm 0cm}]{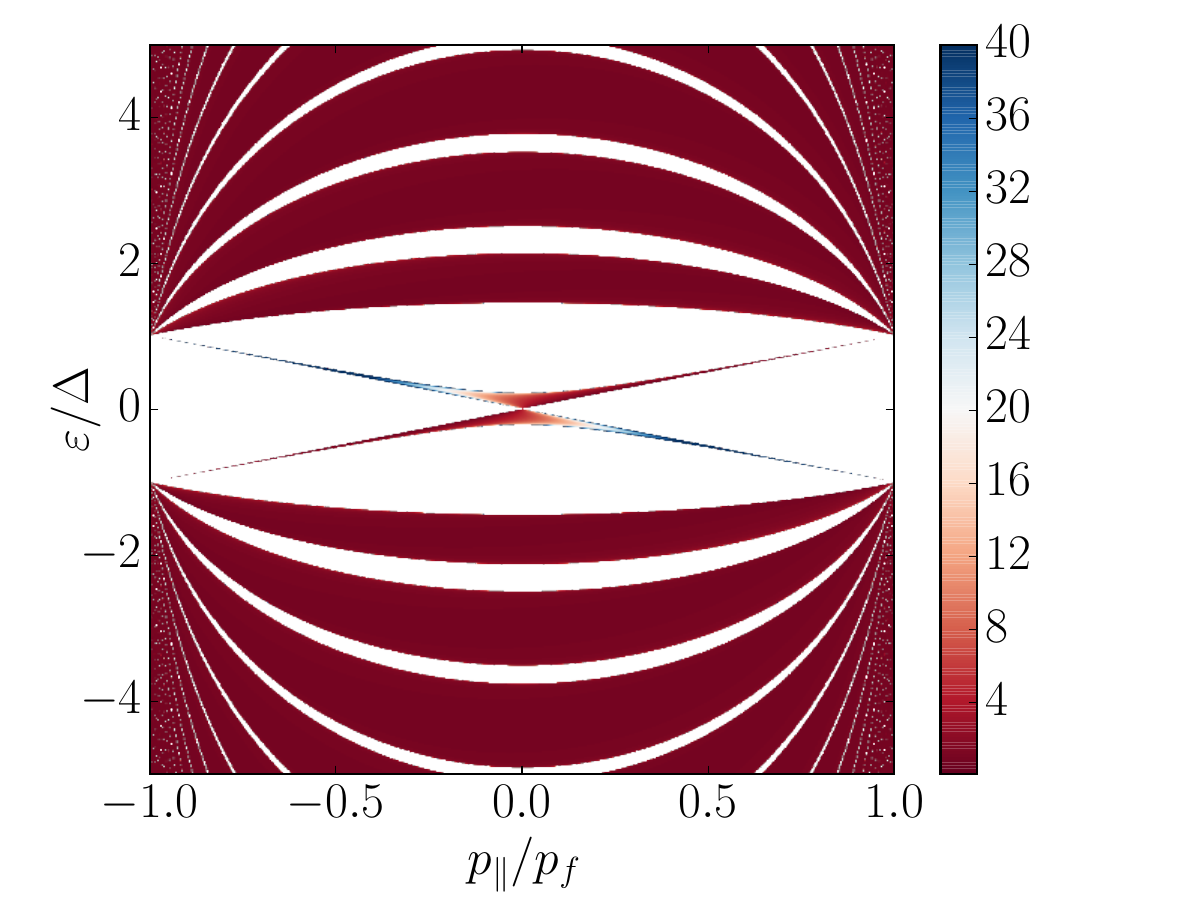}
\caption{Fermionic spectrum for a laterally confined film with $D = 8 \xi_0$ at the edge $x = D$. The spectral function is shown as a function of parallel momentum $p_\parallel$ and energy $\eps$. The energy ranges from $-5 \Delta$ to $5 \Delta$, which contains both the bound-state spectrum shown in detail in Fig. \ref{fig-Sub-Gap_Band-Structure}, and the continuum $|\eps| > \Delta$. For each $p_{\parallel}$, the continuum forms a band structure with band gaps generated by hybridization and multiple reflection.
}
\label{fig-continuum-gap}
\end{figure}

For each $\vp$, the imaginary part of $\g^\infty$ for $\eps < \Delta$ only comes from two simple poles: One at $\eps = \Delta_2$ with spectral weight proportional to $e^{-2 \Delta x / v_f}$, corresponds to the edge states confined near $x = 0$, and the other pole at $\eps = - \Delta_2$ with spectral weight $\propto e^{-2 \Delta (D - x) / v_f}$, confined near $x = D$. 
For finite $D$, the band structure of sub-gap spectrum is encoded in the normalization factor $M$, which becomes imaginary when the right-hand side of Eq. \ref{equ_M_square} is negative. For sub-gap states, $|\eps| < \Delta$, $M$ is imaginary only if $\eps$ satisfies
\be
\Delta^2 \hp_y^2 < \eps ^2 < \Delta^2 \hp_y^2 + \frac{\Delta^2 \hp_x^2} { \cosh(\lambda D)^2} 
\,.
\ee 
The range in energy over which the sub-gap spectral density is nonzero is identical to the bandwidth of bound-state energies obtained from the solution of Andreev's equations discussed in Sec. \ref{sec-hybridization-Andreev}. 
Note that $M$ not only modifies the edge contributions to the propagator $\widehat\g_\pm$, but also the bulk term proportional to $\widehat \g_0$. Multiple reflections not only reorganize the bound states by re-distributing spectral weight among themselves, but the formation of sub-gap band is also connected with changes in continuum spectrum with $|\eps| > \Delta$. 
The spectral density at $x=D$ for $D=8\xi_0$ is shown in Fig \ref{fig-Sub-Gap_Band-Structure} for the sub-gap spectrum, and in Fig \ref{fig-continuum-gap} over a range of energies into the continuum.

Figure \ref{fig-Sub-Gap_Band-Structure} shows the sub-gap spectrum $|\eps| < \Delta$ at $x=D$. For each $p_\parallel$, the chiral bound states form a band. The density of states is maximum at the upper and lower band edges, resulting from Van-Hove singularities of the one dimensional periodic pair potential. The bandwidth is largest for the normal incidence where the period is smallest. The spectral density is maximum for branch $\eps = c p_\parallel$, and exponentially small for the opposite chirality $\eps=-c p_\parallel$ corresponding to the hybridization with Weyl fermions on the opposing edge $x = 0$. 
The continuum spectrum for $|\eps| > \Delta$ shown in Fig. \ref{fig-continuum-gap}, also exhibits energy bands. Finite band gaps develop for each $p_\parallel$. The band structure in both sub-gap and continuum states results from the overlap of edges states, and the periodicity due to multiple reflection of quasiclassical trajectories.

\vspace*{2mm}
\section{Self-consistent solution}\label{sec-Self-consistent_QC}
\vspace*{-5mm}

The existence of boundaries and the formation of edge states modifies the pair propagators from their bulk values, which then leads to a modification of the off-diagonal mean-field paring energy, or order parameter. Specifically, the order parameter component associated with the p-wave orbital that is normal to the boundaries is suppressed to zero at the edge; $\f_1 = 0$ at both edges as can be seen from Eq. \ref{equ-quasiclassical_f1}. The other orbital component will either be enhanced or suppressed depending on the conditions imposed by boundary scattering. For a specularly reflecting edge, the  off-diagonal self-energies satisfy the following ``gap equations'',
\ber
\Delta_1(\vp,x)&=&\int dS_{p'}\,g(\vp,\vp')
	T\ns\sum_{\eps_n}^{|\eps_n|<\omega_c}\f_1(\vp',x;\eps_n)
\,,
\label{gap-equ-ss1}
\\
\Delta_2(\vp,x)&=&\int dS_{p'}\,g(\vp,\vp')
	T\ns\sum_{\eps_n}^{|\eps_n|<\omega_c}\f_2(\vp',x;\eps_n)
\,,
\label{gap-equ-ss2}
\eer
where the integration is over the Fermi surface, which in the 2D limit is angular integration around a circle of radius $p_f$.
The p-wave pairing interaction in the 2D limit, $g(\vp,\vp')=2g_1\vp\cdot\vp'$, and the pairing bandwidth, $\omega_c$, are eliminated in favor of the bulk transition temperature, $T_c$, using the linearized bulk gap equation for the pairing instability, $1/g_1=K(T_c)$, where
\be
\label{equ-digamma_function}
K(T) \equiv \pi T\sum_{\eps_n}^{|\eps_n| < \omega_c}\frac{1}{|\eps_n|}
\simeq\ln\left(1.13\frac{\omega_c}{T}\right)
\,,
\ee
is a digamma function. More generally, Eq. \ref{equ-digamma_function} is used to regulate the log-divergent Matsubara sums in Eqs. \ref{gap-equ-ss1}, \ref{gap-equ-ss2} in order to eliminate the cutoff and to express these functions in terms of the scaled temperature, $T/T_c$,
\be
\label{equ-gap-Tc-ss}
\hspace*{-4mm}\Delta_{i}\ln\left(\frac{T}{T_c}\right) 
\ns=\ns 
2\int\ns dS_{p'}(\vp\cdot\vp')\,
T\sum_{\eps_n}\left[\f_{i}(\vp',x;\eps_n)\ns-\ns\pi\frac{\Delta_{i}}{|\eps_n|}\right]
\,,
\ee
for $i=1,2$. The off-diagonal Matsubara Green's functions are obtained from the retarded propagators by analytic continuation ($\eps\rightarrow i\eps_n$). For constant gap amplitudes, $\Delta_{1,2}$, analytical results for $\f_{1,2}$ are given by,

\begin{widetext}
\ber
\f_1(\vp,x;\eps) 
&=&
\frac{1}{M}\,
\frac{\pi \Delta_1}{\Lambda}
\left(1 -
\frac{e^{2 \lambda (D - x)} - e^{-2\lambda x} + e^{2 \lambda x} - e^{2 \lambda(x - D)}}
     {e^{2 \lambda D} - e^{-2 \lambda D}}
\right)
\,,
\label{equ-quasiclassical_f1}
\\
\f_2(\vp, x; \eps) 
&=& 
\frac{1}{M} \left(\frac{\pi \Delta_2}{\Lambda} 
-\frac{\pi \Delta_1 \left(
\Delta_1^2 - \eps^2
\right)}{\Lambda \left(\Lambda \eps - \Delta_1 \Delta_2 \right)} 
\frac{e^{2 \lambda (D - x)} - e^{-2 \lambda x}}{e^{2 \lambda D} - e^{-2 \lambda D}} \right.
\left.\vphantom{}
-\frac{\pi \Delta_1 \left(
\Delta_1^2 - \eps^2
\right)}{\Lambda \left(- \Lambda \eps - \Delta_1 \Delta_2 \right)} 
\frac{e^{2 \lambda x} - e^{2 \lambda(x - D)}}{e^{2 \lambda D} - e^{- 2 \lambda D}} 
\right)
\,.
\label{equ-quasiclassical_f2}
\eer
\end{widetext}

Equations \ref{equ-quasiclassical_f1} and \ref{equ-quasiclassical_f2} for the pair propagators, as well as Eq. \ref{equ-quasiclassical_g} for the quasiparticle propagator, based on piece-wise constant order parameters, provide considerable insight into the sub-gap and continuum spectra, as well as semi-quantitative results for the spectral functions, order parameter and edge currents. 

\begin{figure}
\includegraphics[width=\columnwidth]{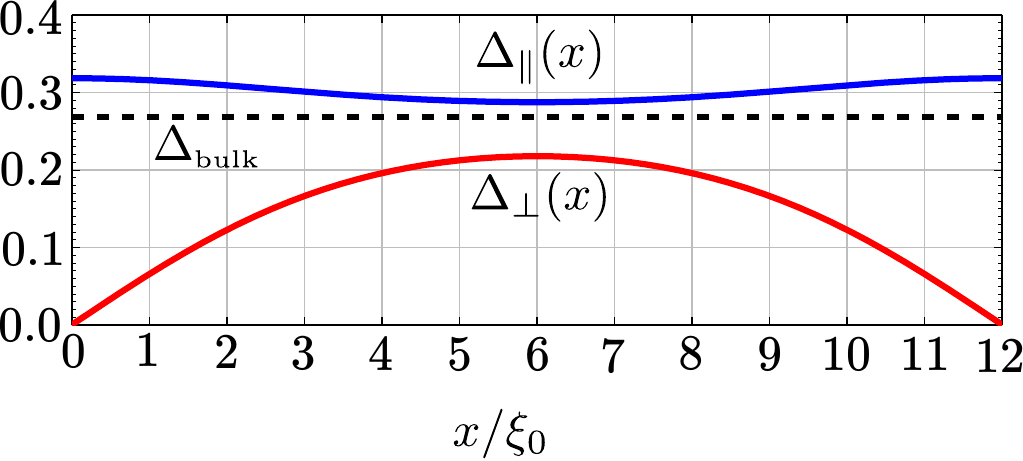}
\caption{Self-consistently determined order parameter profiles
	for lateral confinement width $D = 12\xi_0$. 
	}
\label{fig-ss_gap}
\end{figure}

\begin{figure}
\includegraphics[width=\columnwidth]{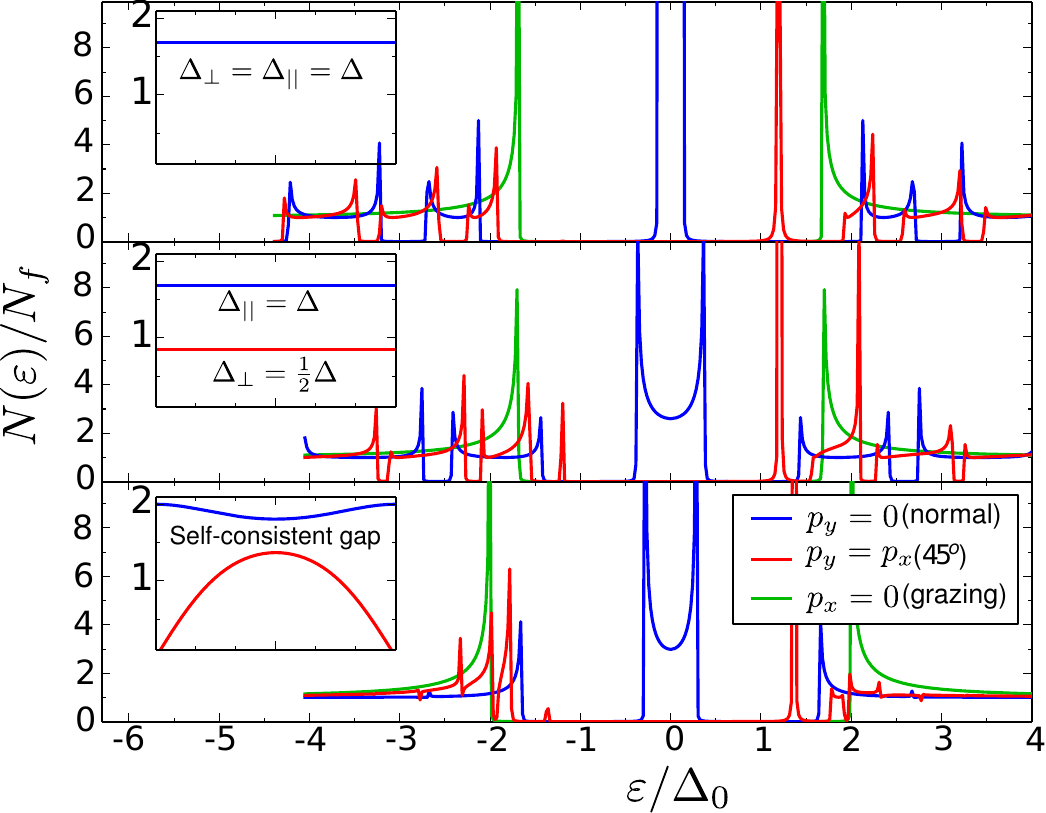}
\caption{
Spectral functions for normal incidence (blue), grazing (green) and intermediate angle (red) trajectotries. 
	Top panel: the spectrum for an isotropic chiral order parameter, 
	$\Delta(\hp_x + i\hp_y)$.
	Middle panel: the spectrum for an anisotropic chiral phase, 
	$\Delta_{\perp}=\frac{1}{2}\Delta_{||}$.
	Bottom panel: the spectrum for the self-consistently determined order parameter
	shown in Fig. \ref{fig-ss_gap} for $D = 12\xi_0$.} 
\label{fig-ss_spectrum}
\end{figure}

The gap equations (Eqs. \ref{gap-equ-ss1}, \ref{gap-equ-ss2}) are solved self-consistently with Eilenberger's transport equations.\cite{vor03,vor07} The resulting gap profiles are shown in Fig. \ref{fig-ss_gap}.
There is strong pair breaking of the amplitude for the orbital component normal to the edges; $\Delta_{\perp}(x)$ vanishes at both edges and is suppressed throughout the confined channel. Pair breaking is also reflected in the spectral density. In Fig. \ref{fig-ss_spectrum} we show the local density of states for the self-consistently determined order parameter, as well as results for a suppressed, but spatially uniform amplitude, $\Delta_{\perp}=0.5\Delta$.
As $\Delta_{\perp}(x)$ decreases the bandwidth centered around zero energy increases, and band gap to the continuum reduces. The basic structure of the spectrum, including the Van Hove singularities is already captured by the non-self-consistent order parameter.  

\vspace*{-5mm}
\section{Edge Current Suppression}\label{sec-Edge_Current_vs-TD}
\vspace*{-5mm}

The propagator, $\g(\vp, x; \eps)$, determines the mass current density, which in the Matsubara representation is given by 
\be\label{equ-mass-current}
\vj(x,T) = 2 N_f\int dS_{\vp}\,\vv_\vp\,T\,\sum_{\eps_n} \g(\vp, x; \eps_n)
\,.
\ee 
The solution for $\g$ includes the current resulting from the sub-gap edge spectrum as well as the back action of continuum to their formation. The resulting integrated sheet current for a single edge is then $J=\onefourth\,n\,\hbar$, at zero temperature.\cite{sau11}

Due to the chirality of the edge states the mass current flows in the positive $+y$ direction near the edge at $x = D$, and in $-y$ direction along the edge at $x = 0$. Thus, in the laterally confined geometry with two edges the current density vanishes at the mid point of the channel. 
Thus, we define the sheet current for a single edge as the integrated current density in half of the channel,
\be
\vJ(T, D)=\int_{D/2}^{D} dx\,\vj(x)
\,.
\ee
The reduction of the edge current at any temperature can be expressed as $|\vJ| = \onefourth\,n\,\hbar\,\times\,\cY(T, D)$. In the limit $D\rightarrow\infty$, the reduction is due to thermal excitations and is given by,
\be
\hspace*{-3mm}
\cY(T)=2\pi T\sum_{\varepsilon_n}\,\frac{\Delta^2}
		{\sqrt{\varepsilon_n^2 + \Delta^2}
		 \left(|\epsilon_n| + \sqrt{\varepsilon_n^2 + \Delta^2}\right)^2}
\,,
\label{eq-Lz_vs_T}	
\ee
where the sum runs over the Matsubara energies, $\varepsilon_n=(n+\nicefrac{1}{2})\,2\pi T$, and $\Delta$ ($\Delta_0$) is the BCS gap at temperature $T$ ($T=0$).\footnote{Equation \ref{eq-Lz_vs_T} is obtained from Eq. (56) of Ref.~\onlinecite{sau11} by carrying out the angular integration.}
Thermal excitation of the chiral edge Fermions leads to a power law reduction ($\sim T^2$) of the edge current for $T \ll \Delta_0$; in particular, $\cY(T)\simeq 1-\nicefrac{2}{3}\,\left(\pi T/\Delta_0\right)^2$.\cite{tsu12}$^{,}$\footnote{The leading order low-temperature correction is obtained by application of the Euler-Maclaurin formula to Eq. \ref{eq-Lz_vs_T}.} 

\begin{figure}[t]
\includegraphics[width=0.5\textwidth,trim={1cm 0 1cm 0}]{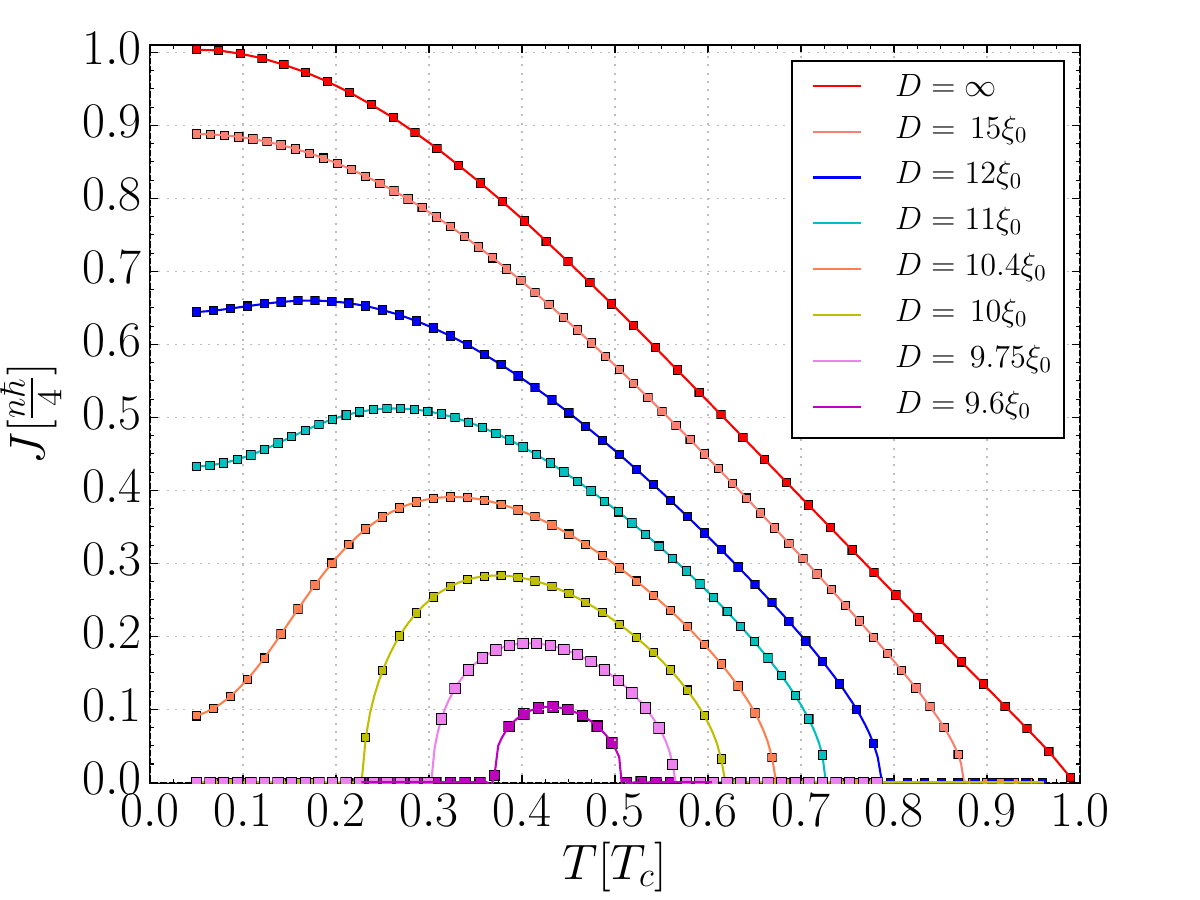}
\caption{Sheet current as a function of confinement, $D$, and temperature, $T$. Superfluidity 
	onsets at the bulk transition temperature $T_c$ for all $D$. For $D = \infty$ the 
	superfluid is the chiral A phase at all temperatures.}
\label{fig-Sheetcurrent}	
\end{figure}

The edge current calculated self-consistently as a function of $D$ and $T$ is shown in Fig. \ref{fig-Sheetcurrent}. For weak confinement, $D = 15\xi_0$, the current decreases monotonically with temperature similar to that for $D\rightarrow\infty$; however, $J(0,D)$ is suppressed relative to $\onefourth n\hbar$, and $J(T,D)$ vanishes at $T_{c_2} = 0.89 T_c$, i.e. \emph{below} the superfluid transition at $T_c$.
The vanishing of the edge current for $T_{c_2} < T \le T_c$ indicates that the high-temperature superfluid phase under confinement is a \emph{non-chiral} phase. For  $10.4\xi_0\lesssim D < 12\xi_0$, the mass current onsets at a relatively low $T_{c_2}$, increases to maximum and then decreases. This non-monotonic behavior implies that the chiral phase is also suppressed at low temperatures for sufficiently strong confinement. For $D \approx 10 \xi_0$ the low-temperature phase is also a non-chiral phase, which undergoes a transition to a chiral phase at finite temperature ($T_{c_3} = 0.23 T_c$ for $D = 10 \xi_0$), then re-enters a non-chiral phase at a higher temperature ($T_{c_2} = 0.62T_c$ for $D = 10 \xi_0$).
The temperature window, $T_{c_3} < T < T_{c_2}$, for a stable chiral phase decreases with confinement. Below a critical thickness, the mass current is zero and the superfluid phase is non-chiral at all temperatures.

\vspace*{-5mm}
\section{Phase transitions under confinement}\label{sec-Phase_Diagram}
\vspace*{-5mm}

Based on the supression of the edge currents, we conclude that for the confined channel, the chiral A phase is suppressed near $T_c$ in favor of a non-chiral phase, but for weak to moderate confinement undergoes a second order transition at $T_{c_2}$ to a chiral phase. The only possible non-chiral superfluid phase that is not suppressed by lateral confinement is the \emph{polar} phase with only one p-wave orbital component, $i\Delta_{\parallel} p_y$, aligned along the direction of the channel. The component $\propto p_x$ is completely suppressed.
To investigate the polar to chiral phase transition, we carry out linear stability analysis starting from the polar state, $i\Delta_\parallel p_y$. In particular, we allow for nucleation of additional order parameter components at a second-order transition,  
\be
\label{equ-general-instabilities}
\Delta(\vp, \vr) = A_{1} (\vr) p_x + A_{2}(\vr) p_y + i\Delta_\parallel p_y
\,,
\ee
where $|A_{1}|, |A_{2}| \ll \Delta_{\parallel}$ are perturbations. For specular boundary scattering, the amplitudes satisfy the boundary conditions: $A_{1}(0, y) = A_{1}(D, y) = 0$, $\partial_x A_{2}(x, y)\vert_{x=0} = \partial_x A_{2}(x, y)\vert_{x=D} = 0$ for all $y$. 

Eilenberger's equations are solved to linear order in $A_1$ and $A_2$. In zeroth order, the polar phase order parameter and the Green's function are given by 
\be
\Delta^{(0)} = i \Delta_\parallel p_y, \qquad \whg^{(0)} = -\pi\frac{i\eps_n - \hat \Delta^{(0)}}{\sqrt{\eps^2_n + \Delta_\parallel^2 p_y^2}}
\ee
where $\Delta_\parallel$ is computed self-consistently by solving the nonlinear gap equation, 
\be
\ln\left(\frac{T}{T_c}\right)\ns=\ns\int\ns dS_{\vp} p_y^2 T\sum_{\eps_n} 
\left( 
\frac{2\pi}{\sqrt{\eps_n^2 + \Delta_\parallel^2 p_y^2}} - \frac{2\pi}{|\eps_n|}
\right) 
\,.
\ee
It is convenient to separate the first-order corrections to the order parameter into real and imaginary terms
\be
\Delta^{(1)} = A_{1} (x, y) p_x + A_{2}(x, y) p_y \equiv \Delta^{(1)}_R + i \Delta^{(1)}_I
\,.
\ee
The corresponding corrections to the propagator obey the linearized transport equation and normalization condition, 
\ber
[i\eps_n\tz - \whDelta^{(0)}, \whg^{(1)}] + i\vv_f\cdot\grad\whg^{(1)}
&=& 
[\whDelta^{(1)}, \whg^{(0)}]
\,,
\\ 
\whg^{(0)}\whg^{(1)}
+
\whg^{(1)}\whg^{(0)}
&=& 0 
\,.
\eer
The resulting Cooper pair propagators for the first-order perturbations satisfy second-order differential equations,
\ber
\frac{1}{4}\partial^2\f^{(1)}_R-\omega_n^2 \f^{(1)}_R
&=&
-\pi\omega_n\Delta^{(1)}_R
\,,
\label{equ-fR}
\\
\frac{1}{4}\partial^2 \f^{(1)}_I-\omega_n^2\f^{(1)}_I 
&=&
-\pi\omega_n \Delta^{(1)}_I+\frac{\pi\Delta_\parallel^2 p_y^2}{\omega_n}\Delta^{(1)}_I
\label{equ-fI}
\,.
\eer
where $\omega_n=\sqrt{\eps_n^2+\Delta_{\parallel}^2 p_y^2}$ and $\partial=\vv_\vp \cdot\grad$ is the directional derivative along trajectory $\vp$. The corresponding gap equation for $\Delta^{(1)}$ is given by
\be
\hspace*{-3mm}
\Delta^{(1)}\ln\left(\frac{T}{T_c}\right)
\ns=\ns
2\ns\int\ns dS_{\vp'}\vp\cdot\vp'T\sum_{\eps_n}
\left[
\f^{(1)}(\vp', \vr;\eps_n)
- 
\pi\frac{\Delta^{(1)}}{|\eps_n|}
\right]
,
\ee
where $\f^{(1)}_R$ generates $\Delta^{(1)}_R$ and $\f^{(1)}_I$ generates $\Delta^{(1)}_I$, i.e. the real and imaginary components decouple. These linear equations can be solved by Fourier transform.

\begin{figure}
\includegraphics[width=\columnwidth]{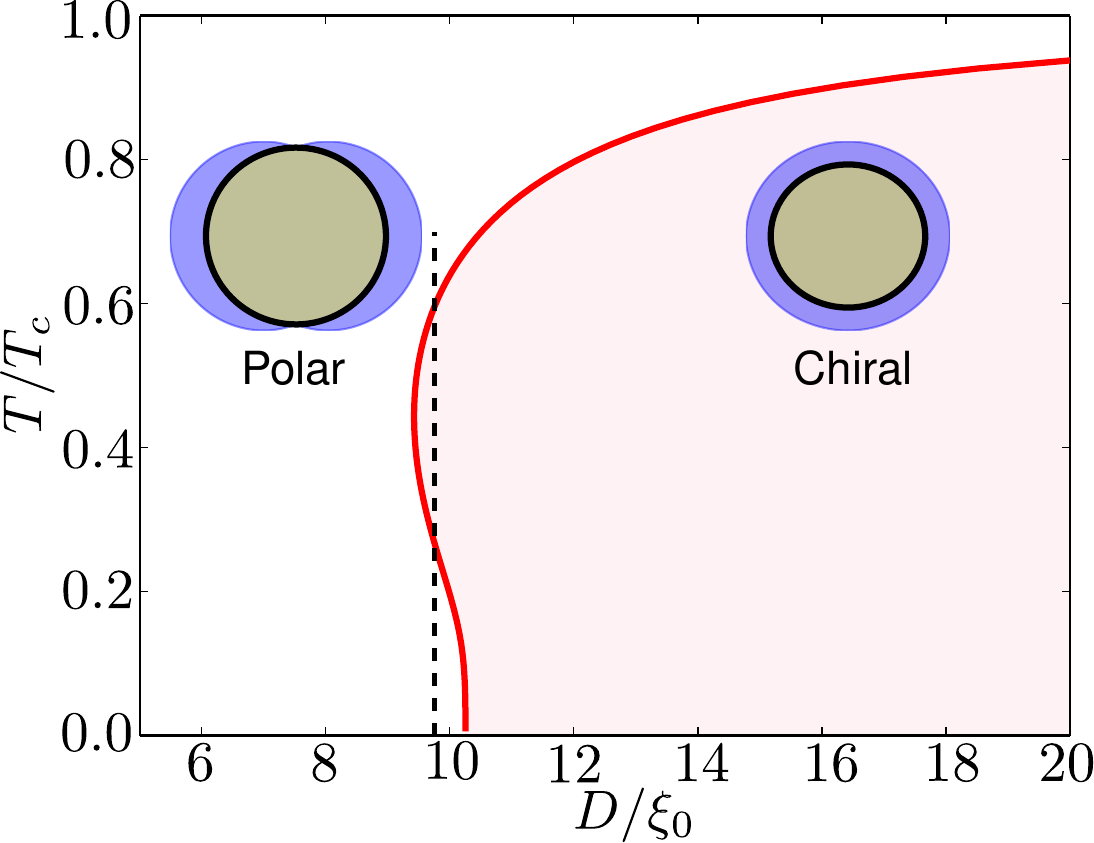}
\caption{Phase diagram showing the re-entrant Polar to A-phase transition line. The dashed line crosses the two transitions: P$\rightarrow$A at $T_{c_2} \approx 0.6 T_c$ and A$\rightarrow$P at $T_{c_3} \approx 0.3 T_c$ for $D= 9.75\xi_0$.
}
\label{fig-polar-A}
\end{figure}

\subsection{Polar to A phase transition}

The linear instability equations have a solution corresponding to a second-order transition from the polar phase to the chiral A phase. The unstable amplitude is determined by Eq. \ref{equ-fR} and $\Delta_R^{(1)}$, and is translationally invariant along the channel direction ($Q_y=0$), i.e. the only spatial dependence is in $x$ direction imposed by the edge boundary conditions. The Fourier expansion of the unstable order parameter is
\be
A_1(x) = \sum_{Q}\,a(Q)\,e^{iQ\,x}
\,,
\ee
and the solution of Eq. (\ref{equ-fR}) becomes, 
\be
\f^{(1)}_R(Q)=\frac{\pi\omega_n\,p_x}{\omega_n^2+(Q\,v_x(\vp))^2/4}\,a(Q)
\,.
\ee 
The linearized gap equation (Eq. \ref{equ-gap-Tc-ss}) can also be Fourier transformed, and reduces to the eigenvalue equation,
\ber
\label{equ-single-mode}
\left[ 
T\sum_{\eps_n}
\left\{
\int dS_{\vp}\,
\right.
\right.
&&
\hspace*{-5mm}
\frac{2\pi\omega_n p_x^2}{\omega_n^2+(Q v_f p_x)^2/4} 
\\
&&-
\left.
\left.
\frac{\pi}{|\eps_n|}
\right\}
-\ln\left(\frac{T}{T_c}\right)
\right] 
a(Q) = 0
\,.
\nonumber
\eer
The amplitude $a(Q)$ for the onset of the chiral phase has a non-vanishing solution for a $Q$ such that the coefficient defined by the terms in the brackets of Eq. (\ref{equ-single-mode}) has a solution. This eigenvalue condition, combined with the identification of $Q=n\pi/D$ to satisfy the boundary conditions $A(0) = A(D) = 0$, determines a family of critical confinement dimensions. The smallest critical dimension $D$ for a solution $Q(T)$ is $D=\pi/Q$ which defines the line $D_c(T)$ for the polar to chiral A phase transition shown in Fig \ref{fig-polar-A}.
Note that confinement results in the high temperature phase being polar. For $D\approx 10\xi_0$ there is reentrant behavior: $P\rightarrow A$ at $T_{c_2}$ followed by $A\rightarrow P$ at $T_{c_3}$, obtained earlier from the calculation of the edge current. The phase diagram agrees well with the transitions indicated in Fig \ref{fig-Sheetcurrent} for the edge current.


\begin{figure}
\includegraphics[width=\columnwidth]{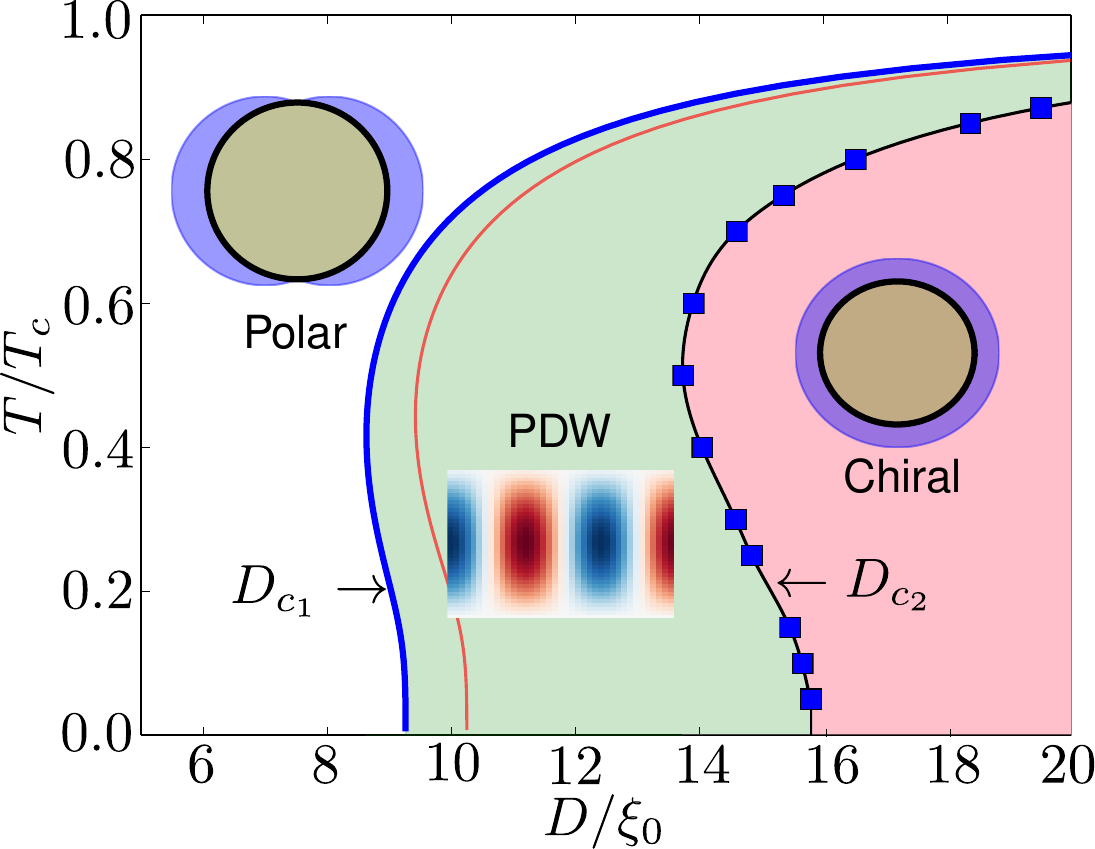}
\caption{Three stable phases exist in laterally confined superfluid \He\ films. The blue line labelled $D_{c_1}(T)$, is the critical lateral confinement curve for the polar to PDW phase transition. The red line is the critical curve for the polar to chiral A phase transition assuming translational invariance along $y$. The blue squares (black line fit) show the PDW to chiral A phase transition line, $D_{c_2}(T)$, calculated numerically from the self-consistent order parameters used as inputs to the Luttinger-Ward free energy functional.
}
\label{fig-inhomo-instability}
\end{figure}

\subsection{Polar to PDW phase transition}\label{sec-Polar-PDW_Transition}

The reentrant behavior is a hint that an inhomogeneous phase, one that accomondates both polar and chiral order but breaks translational symmetry along the channel, may be favored over a range of lateral confinement dimensions $D$.\cite{vor07} In order to investigate this possibility we consider a more general form of the order parameter for the linear instability analysis that breaks translational symmetry along the channel, and which includes both real and imaginary parts of $\Delta^{(1)}$, i.e. spatially modulated currents.
In particular, we start from Eq. \ref{equ-general-instabilities}, then represent the complex amplitudes, $A_{1,2}(\vr)$, in Fourier space as
\be
A^{R,I}_{1,2}(\vr)=\sum_{\vQ}\,a^{R,I}_{1,2}(\vQ)\,e^{i\vQ\cdot\vr}
\,,
\ee
where $A^{R,I}_{1,2}$ are the real (R) and imaginary (I) parts of $A_{1,2}$. 
Equation (\ref{equ-fR}) becomes,
\be
\f^{(1),R}(\vQ)  = \frac{\pi \omega_n (a_1^{R}(\vQ)p_x + a_2^{R}(\vQ) p_y)}
{\omega_n^2 + (\vQ \cdot \vv_\vp)^2 / 4}
\,,
\ee
and similarly for Eq. (\ref{equ-fR}),
\be
\f^{(1),I}(\vQ) = \frac{\pi \eps_n^2 (a_1^{I}(\vQ)p_x + a_2^{I}(\vQ) p_y)}
{\omega_n (\omega_n^2 + (\vQ \cdot \vv_\vp)^2 / 4)}
\,.
\ee

The resulting homogeneous equations for $a_{1,2}^{R,I}$ decouple for each mode $\vQ$, but the amplitudes for the $p_x$ and $p_y$ Cooper pairs are in general coupled,
\be
\label{equ-single-mode-pdw}
\begin{pmatrix}
 	\ln t - I_{11}^{R, I} 	& 	-I_{12}^{R, I}		\\
 	-I_{21}^{R, I}			& 		\ln t - I_{22}^{R, I}	\\
\end{pmatrix}
\begin{pmatrix}
 	a_{1}^{R, I}(\vQ)	\\
 	a_{2}^{R, I}(\vQ)
\end{pmatrix}
= 0
\,,
\ee
where $t=T/T_c$ and the coefficients, $I_{ij}^{R,I}(\vQ)$, are given by 
\ber
I^{R}_{ij} 
&=& 
2T\sum_{\eps_n}\ns\int\ns dS_{\vp}
\hat p_i \hat p_j \left(\frac{\pi \omega_n}
{\omega_n^2 + (\vQ \cdot \vp)^2 v_f^2 / 4 } - \frac{\pi}{\eps_n} \right)
\,,\quad
\label{equ-integral-R}
\\
I^{I}_{ij} 
&=& 
2T\sum_{\eps_n}\ns\int\ns dS_{\vp}
\hat p_i \hat p_j \left(\frac{\pi\eps_n^2/\omega_n}
{\omega_n^2 + (\vQ\cdot\vp)^2 v_f^2/4} - \frac{\pi}{\eps_n} \right)
\,.\quad
\label{equ-integral-I}
\eer
The first non-trivial solution to Eq. \ref{equ-single-mode-pdw} as a function of $T$ or $D$ signals the nucleation of a new superfluid phase from the translationally invariant polar phase. The spectrum of eigenvalues can be expressed in terms of a spectrum of wavevectors, $\vQ$, that satisfy the eigenvalue equation obtained from the vanishing determinants for the real and imaginary components of the linearized gap equations, 
\be
(\ln t - I^{R, I}_{11})(\ln t - I^{R, I}_{22}) - \left( I^{R, I}_{12} \right)^2 = 0
\,.
\label{equ-instability-zero} 
\ee
We find that for all temperatures, the real sector of Eqs. \ref{equ-instability-zero}, with both $Q_{x,y}\ne 0$, and $Q_y\sim\frac{\pi}{15\xi_0}$, gives the smallest critical dimension, $D_{c_1}$, for the transition from the polar phase. The resulting phase spontaneously breaks the $y$-translation symmetry along the channel. At the instability point a single mode becomes unstable leading to spatial modulation of the order parameter defined by the wavevector, $\vQ=(Q_x,Q_y)$, of the unstable mode.
In particular, the real part of the Cooper pair amplitude with orbital wave function $p_x$ develops a standing wave along the $y$ direction, $\Delta_1 \propto \sin(Q_y y)$, i.e. a pair density wave (PDW).\cite{vor07}
The critical curve for the polar to PDW transition, $D_{c_1}(T)$ (blue curve in Fig. \ref{fig-inhomo-instability}) pre-empts a polar to A-phase transition (red line), with $Q_y = 0$ enforced, at all temperatures.
Close to the transition at $D_{c_1}(T)$ the real part of the amplitude, $\Delta_1$, of the $p_x$ orbital changes sign along the channel with period $2\pi/Q_y(T)$, whereas the $p_y$ orbital has a large imaginary polar component, $i \Delta_\parallel p_y$, modulated by a small real component with the same period. 

\begin{figure}
\includegraphics[width=\columnwidth]{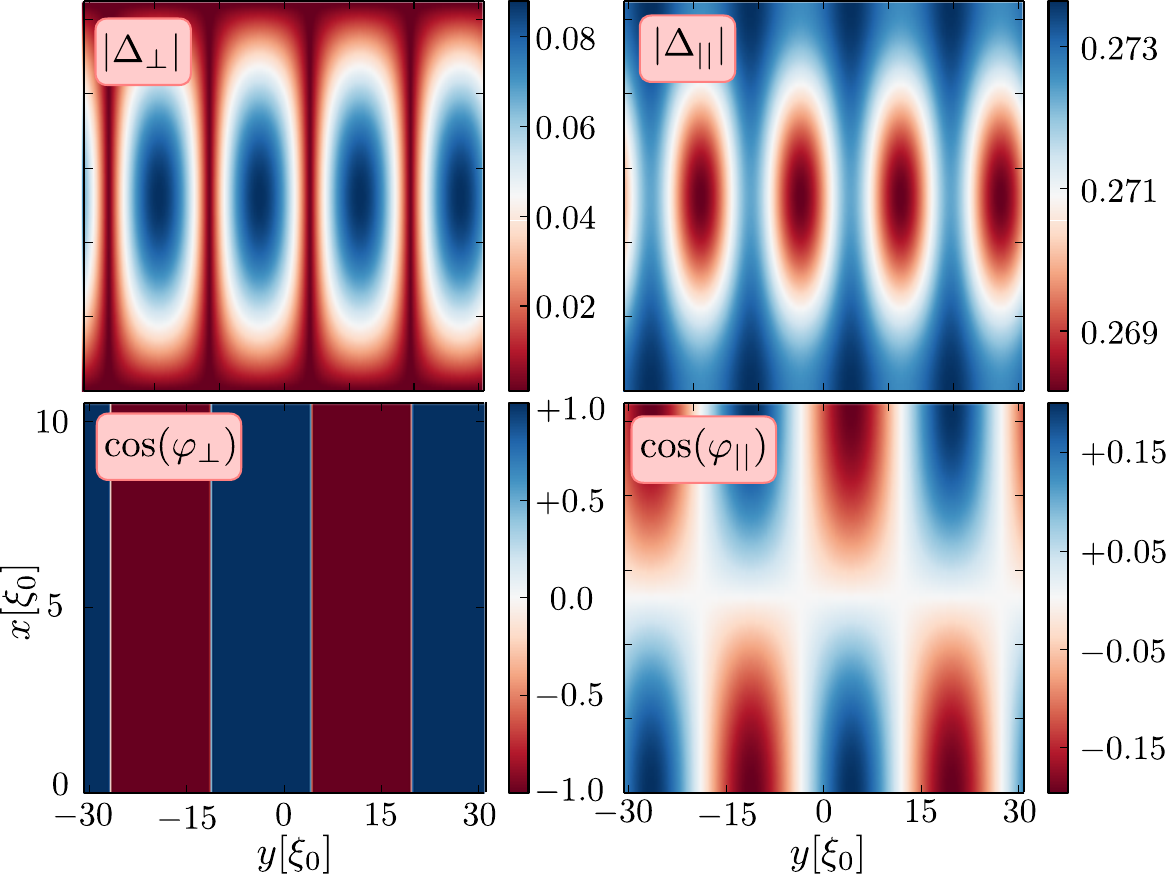}
\caption{
Self-consistent order parameters, $\Delta_{\perp,||}(\vr)=|\Delta_{\perp,||}|e^{i\varphi_{\perp,||}}$, for $D=10.5\xi_0$ and $T=0.7T_c$. The top panel shows plots of the amplitudes, and the bottom panel are plots of the cosine of the corresponding phases. Amplitudes are in the units of $2\pi T_c$. All plots are functions of real space with the horizontal axis being the direction $y$ along the channel, and the vertical axis being the distance $x$ within the laterally confined channel. The emergent period of the PDW phase is $L\approx 30\xi_0$. 
}
\label{fig-periodicPhase-amplitudePhase}
\end{figure}

\section{PDW Order Parameter}\label{sec-PDW_Phase}

For $D \gtrsim D_{c_1}(T)$, there is a proliferation of modes with $Q_y\ne 0$ that become unstable. The resulting order parameter evolves continuously into a multi-Q PDW phase. The structure of the order parameter in the PDW phase requires a self-consistent analysis that incorporates the mode-mode coupling. As we show here the PDW order parameter evolves for increasing $D > D_{c_1}(T)$ towards a periodic array of chiral domains separated by domain walls. The latter are also chiral in that the domain walls support currents flowing along the domain walls. 

The analysis is an extension of that described in Sec. \ref{sec-Self-consistent_QC} for the order parameter of the laterally confined chiral A-phase. The long period of the PDW phase, combined with the fact that there is no bulk order within the confined geometry requires a large computational cell, $L\approx 100\xi_0$, with integration along multiply-reflected classical trajectories. We start with $D\gtrsim D_{c_1}(T)$ from a ``seed'' order parameter which is the polar phase with random noise added at the level of $10^{-4}\Delta_{\parallel}$, which is unstable to the formation of the PDW phase.
The resulting self-consistent order parameter for $T=0.7T_c$ and $D=10.5\xi_0$ is shown in Fig \ref{fig-periodicPhase-amplitudePhase}. It is parametrized as $\Delta(\vr,\vp)=\Delta_{\perp}(\vr) p_x + \Delta_{||}(\vr) p_y$, where $\Delta_{\perp}(\vr)$ and $\Delta_{||}(\vr)$ are both complex amplitudes for the orbital components, $p_x$ and $p_y$, respectively.
The resulting order parameter is a periodic structure with a period $L\approx 30\xi_0$, which corresponds to the unstable wave vector, $Q_y \sim \pi / 15\xi_0$, predicted in single-mode instability analysis. 
Note that the maximum amplitude of $\Delta_{\perp}$ is at the scale of $0.08 [2 \pi T_c]$, around one-half of the bulk amplitude at that temperature. The phase changes by $\pi$ across each domain wall centered at the half period. The resulting ordered phase is both chiral and periodic along the channel with the sign of chirality alternating each half period.

\begin{figure}
\includegraphics[width=\columnwidth]{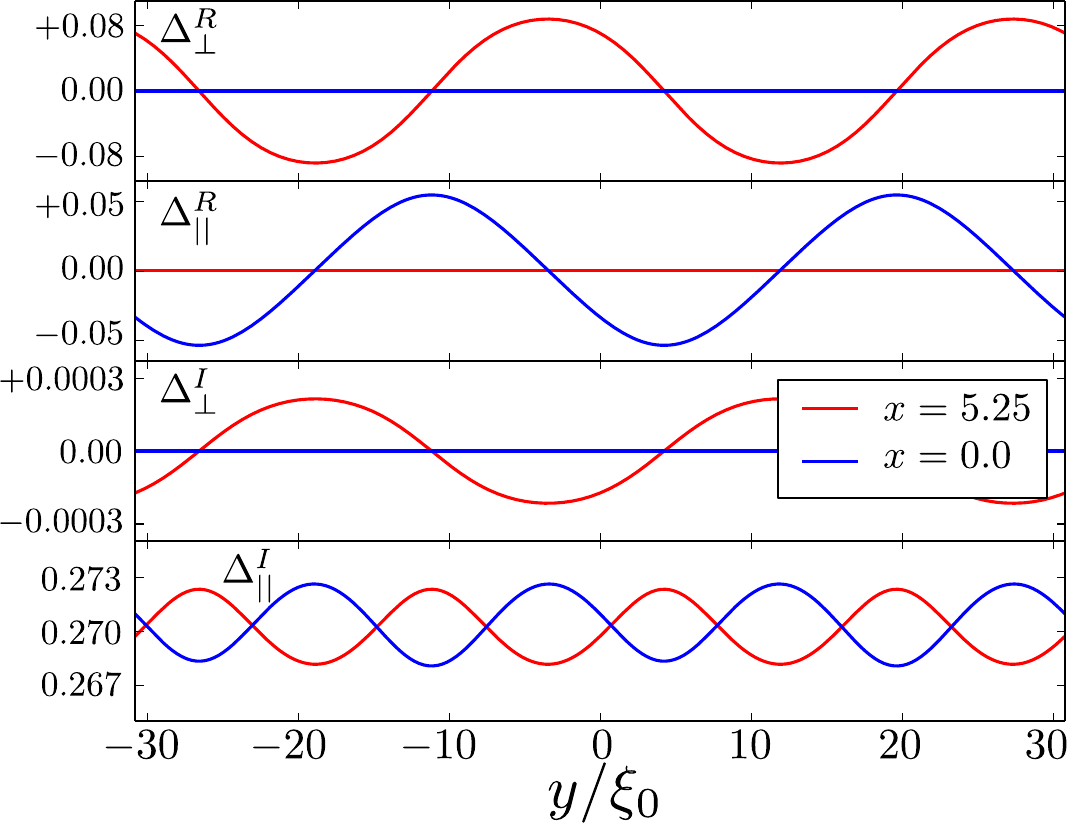}
\caption{
Self-consistent order parameter amplitudes for $T=0.7 T_c$ and $D=10.5\xi_0$. The red curves correspond to the center at $x=D/2$. Blue curves correspond the edge at $x=0$. The order parameter extends over approximately two periods $L\approx 30\xi_0$) of the PDW. The scale for vertical axes give the order parameter amplitue in units of $2\pi T_c$
}
\label{fig-Del-y}
\end{figure}

The structure and evolution of the PDW phase is most clearly shown in terms of slices of the order parameter amplitudes plotted as a function of the channel coordinate, $y$, for several lateral positions, $x$. 
Figure \ref{fig-Del-y} shows the real and imaginary parts for both obital components: $\Delta_{\perp}(\vr)=\Delta_{\perp}^R(\vr)+i\Delta^I_{\perp}(\vr)$ and $\Delta_{||}(\vr)=\Delta_{||}^R(\vr)+i\Delta_{||}^I(\vr)$. All components exhibit periodic structure, reflecting the underlying single-mode instability with $Q_y$ determining the period near the Polar to PDW phase transition.
The real part of the $p_x$ orbital, $\Delta_{\perp}^R(\vr)$, develops large amplitude modulations and changes sign within one period. By contrast the imaginary part of the $p_y$ orbital amplitude, $\Delta_{||}^I(\vr)$, is nearly constant, exhibiting small oscillations about what would otherwise be the bulk amplitude of the polar phase. The resulting structure is a periodic array of chiral domains, with the sign of the chirality switching every half period, i.e. 
an ``anti-ferromagnetic'' chiral phase 
topologically equivalent to 
$\ldots\,(+p_x + i p_y)\,|\,(-p_x + i p_y)\,|\,(+p_x + i p_y)\,\ldots$.

Both orbital components $\Delta_{\perp}(\vr)$ and $\Delta_{||}(\vr)$ have much larger spatial fluctuations in their real parts than the imaginary parts, i.e. $|\Delta_{\perp}^R|, |\Delta_{||}^R| \gg |\Delta_{\perp}^I|,|\Delta_{||}^I-\langle\Delta_{||}^I\rangle|$, which agrees with the decoupling of real and imaginary components in the linear stability analysis discussed in Sec. \ref{sec-Polar-PDW_Transition} and the fact that the first unstable mode occurs in the real sector of the order parameter fluctuations.
The much smaller flucutaions of the imaginary components are driven by mode-mode coupling once the PDW phase is established for $D\gtrsim D_{c_1}(T)$. In particular, for the amplitude of the $p_y$ orbital, $\Delta_{||}^I(\vr)$, exhibits only small oscillations about the amplitude of the parent polar phase.
The amplitude $\Delta_{||}^R$ is largely responsible for the periodic amplitude and phase modulations shown in the right panels of Fig \ref{fig-periodicPhase-amplitudePhase}.
Note also that the amplitude for the $p_x$ orbital, $\Delta_{\perp}(\vr)$, is maximum in the center of a channel, and vanishes at the edges, while in the case of the real part of the amplitude of the $p_y$ orbital, $\Delta_{||}^R(\vr)$, the converse is the case. These features are a result of specularly reflecting boundary conditions at both edges. 

\begin{figure}
\includegraphics[width=\columnwidth]{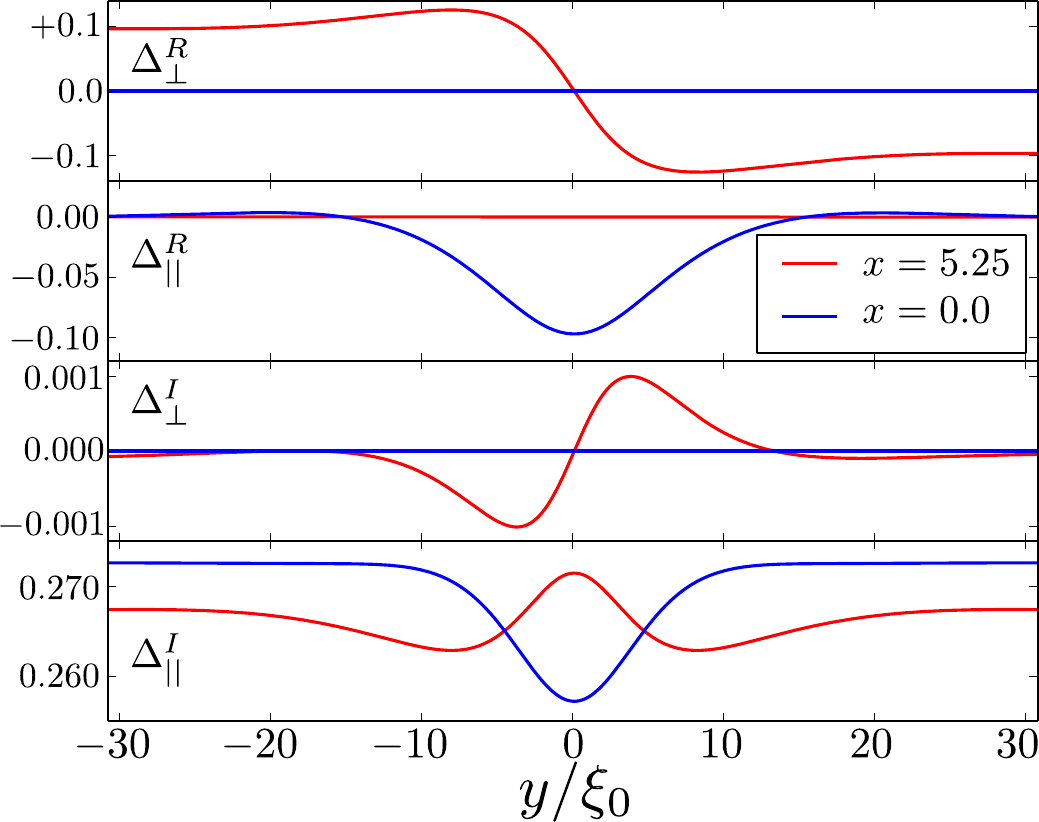}
\caption{
Self-consistent order parameter amplitudes for $T=0.7 T_c$ and $D=11.5\xi_0$. The red curves correspond to the center at $x=D/2$. Blue curves correspond the edge at $x=0$. The order parameters is shown for one-half of a period, and exhibits soliton-like domain walls as well as a ``breather mode'' confined on the domain wall.
}
\label{fig-SingleDomain-y}
\end{figure}

For weaker confinement $D=11.5\xi_0$ at the same temperature, $T=0.7T_c$, self-consistent calculations of the order parameter lead to increased periods of the PDW phase. 
As we move well away from the Polar to PDW transition at $D_{c_1}(T)$, the structure of the order parameter develops soliton-like structures, with nearly homogeneous ordered regions confined between domain walls, the latter roughly $10\xi_0$ in width.
In this limit the structure of the order parameter is dominated by the coupling of many unstable modes with different wavevectors, $\vQ$.  

The region of the channel shown in Fig. \ref{fig-SingleDomain-y} spans $60 \xi_0\approx 6\mu\mbox{m}$ and represents a half period of the PDW phase. The spatial profiles of the order parameter away from the lateral domain walls centered at $y=0$ are identical with those of laterally confined chiral A phase with the sign of the chirality switching as one crosses the domain wall. 
The period increases dramatically with weaker confinement, and as $D\rightarrow D_{c_2}(T)$ diverges towards the system size. Thus, the weak-confinement transition is the locus of points $D_{c_2}(T)$ at which it becomes energetically possible for a single domain wall to exist in the channel separating degenerate chiral ground states.

Finally, we note that there are similiarities in the energetics that lead to stabilization of the PDW phase in laterally confined very thin (quasi-2D) \He\ films considered in this report, and the ``stripe phase'' that develops from confinement in one dimension from bulk \Heb.\cite{vor07,wim16}. 
In particular, the transition from the chiral A phase to the chiral PDW phase under lateral confinement of very thin \Hea\ films, and the transition from the B-phase of \He\ to the stripe phase, both result from the competition between the cost in energy of a domain wall separating degenerate states 
- 2D \Hea\ with opposite chirality in laterally confined films and degenerate 3D B states of 
\Heb\ confined in one dimension - and the gain in condensation energy at the intersection of a domain wall and the boundaries - confining edges for the PDW phase or confining surfaces for the stripe phase. 
For both cases it is favorable for a domain wall to penetrate for sufficiently strong confinement in order to recover lost condensation energy on the boundaries. 
However, the structure of the PDW phase, and its broken symmetries, are distinctly different from that of the stripe phase of thick \He\ films. The PDW phase breaks time-reversal symmetry, while the stripe phase is time-reversal symmetric. Also, the transition from the PDW to the polar phase is second order, while the stripe to A phase is first order.
There appears to be an additional similarity if we consider the currents that flow on the domain walls and near the confining boundaries - mass currents for the PDW phase and spin currents for the stripe phase. In the section that follows we discuss the structure of the chiral domain walls that define the PDW phase near $D_{c_2}(T)$, including the mass current near the intersection of the domain wall and the edges of the laterally confined film, and how current conservation, $\dive\vj(\vr) = 0$, is maintained. Silveri et al. discuss a similar situation that arises for spin-current conservation in the case of domain walls separating degenerate B-phases that define the onset of the stripe phase.\cite{sil14}

\section{Chiral Domain Walls}\label{sec-Chiral_DWs}

\begin{figure}
\includegraphics[width=0.5\textwidth,trim={0.5cm 0 0 0}]{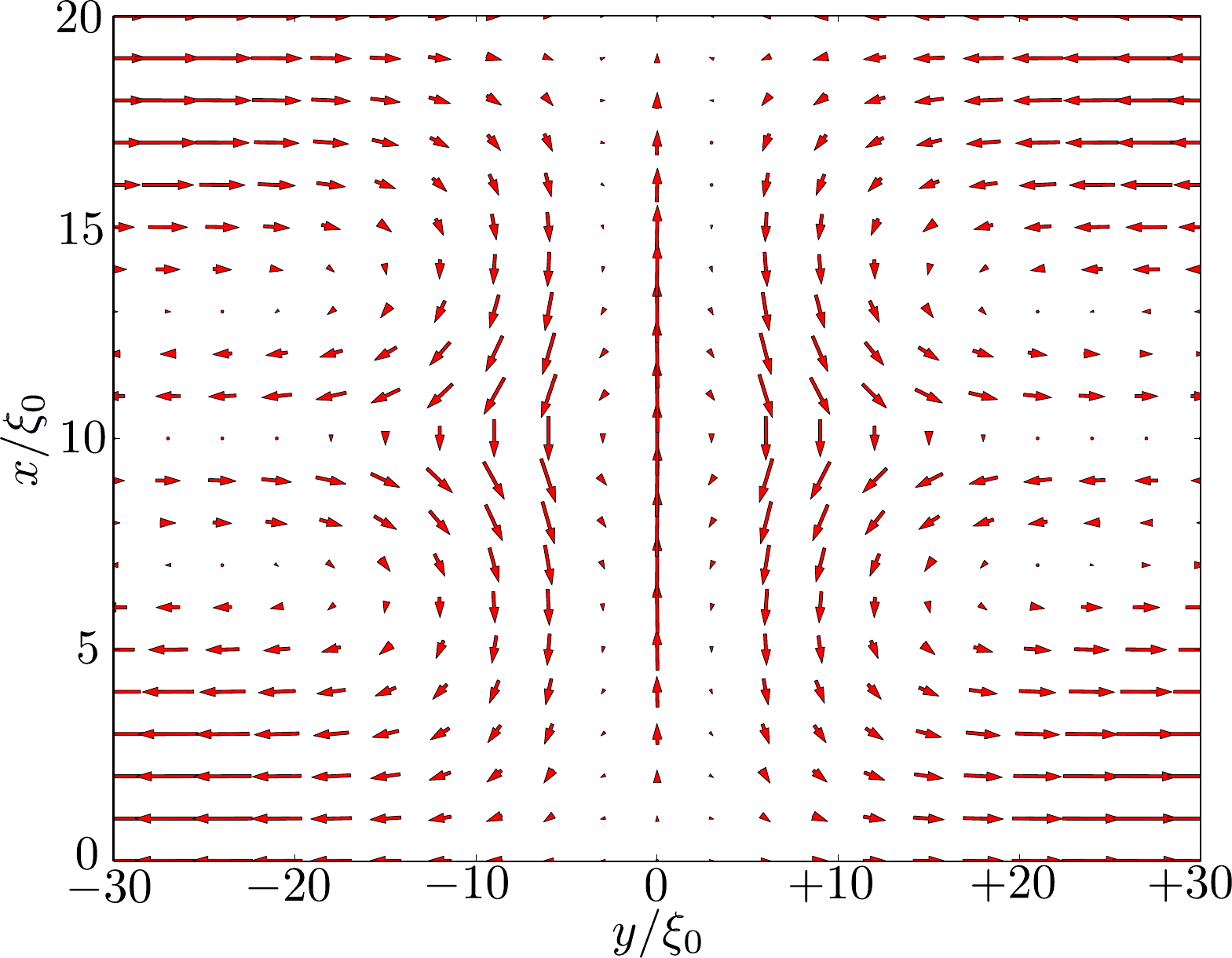}
\caption{
Current density in units of [$N_f\, 2\pi T_c\, m_3 v_f$] for a single $p_x$-type DW in an infinitely long channel of width $D=20 \xi_0$. Shown is the region $-30\xi_0\le y\le +30\xi_0$. Arrows point in the direction of flow of the local current density, and the arrow length represents the magnitude of current.
}
\label{fig-domainwall-current-selfconsistent}
\end{figure}

For laterally confined \Hea\ there are two possible chiral domain walls (DWs) separating degenerate chiral ground states: (i) the $p_x$-type DW changes sign of the amplitude of the $p_x$ orbital across the domain (located at $y=0$), i.e. $\Delta(y < 0)\propto (-p_x+ip_y)$ and $\Delta(y> 0)\propto(p_x+ip_y)$, and (ii) the $p_y$-type DW where the sign of the amplitude of the $p_y$ orbital changes sign, $\Delta(y < 0)\propto (p_x-ip_y)$ and $\Delta(y> 0)\propto(p_x+ip_y)$. Even though the left (or right) chiral domains have the same Chern numbers for both types of domain walls, the DWs are physically distinct structures.
For both DWs the Chern number changes from $N=-1$ to $N=+1$. Thus, both DWs support a branch of chiral Fermions and a ground-state current confined on the DW. However, the ground-state currents that flow along the $p_x$-type DW and the $p_y$-type DW at $y=0$ are oppositely directed; the $p_y$-type DW has current flowing along the domain wall in the direction consistent with chiralities of both chiral domains, while for the $p_x$-type DW the current flows opposite the chirality.\cite{tsu14} 

Analysis of the energy per unit length of these two DWs shows that the $p_x$-type DW has a lower free energy per unit length, and is thus the energetically favored DW, at least for weakly confined chiral domains.\cite{tsu14} Our analysis confirms that result. For a single $p_x$-type DW in laterally confined \Hea, e.g. the PDW phase near the upper critical confinement length, $D_{c_2}$, the currents along the edges are determined by the chirality of each chiral domain. Thus, the energetically favored $p_x$-type DW there appears to be a violation of current conservation at the junctions of the DW and the edges.

The apparent violation of current conservation is resolved by the self-consistent solution for the order parameter and excitation spectrum for the $p_x$-type single domain wall crossing the channel. A self-consistent calculation of the structure of a single, laterally confined $p_x$-type DW is carried out for $T=0.5 T_c$ and $D=20\xi_0$. The results include the current density at every point within the channel. We initialize the order parameter to be $-\Delta_{\perp}(x) p_x + i \Delta_{\parallel}(x) p_y$ in the region $y < 0$, and $+\Delta_{\perp}(x) p_x + i \Delta_{\parallel}(x) p_y$ in the region $y > 0$, with the domain wall centered at $y = 0$. Note that $\Delta_{\perp,||}(x)$ are real for the solutions far from the DW. The order parameter is calculated by self-consistent solution of the Eilenberger and gap equations (Eqs. \ref{equ-eilenberger-transport}, \ref{gap-equ-ss1}, \ref{gap-equ-ss2}), subject to the asymptotic boundary conditions for $y\rightarrow\pm\infty$ determined by the two possible degenerate chiral ground states in the channel. The mass current is calculated from Eq. \ref{equ-mass-current} once the self-consistent propagator and order parameter are determined.

\begin{figure}
\includegraphics[width=\columnwidth]{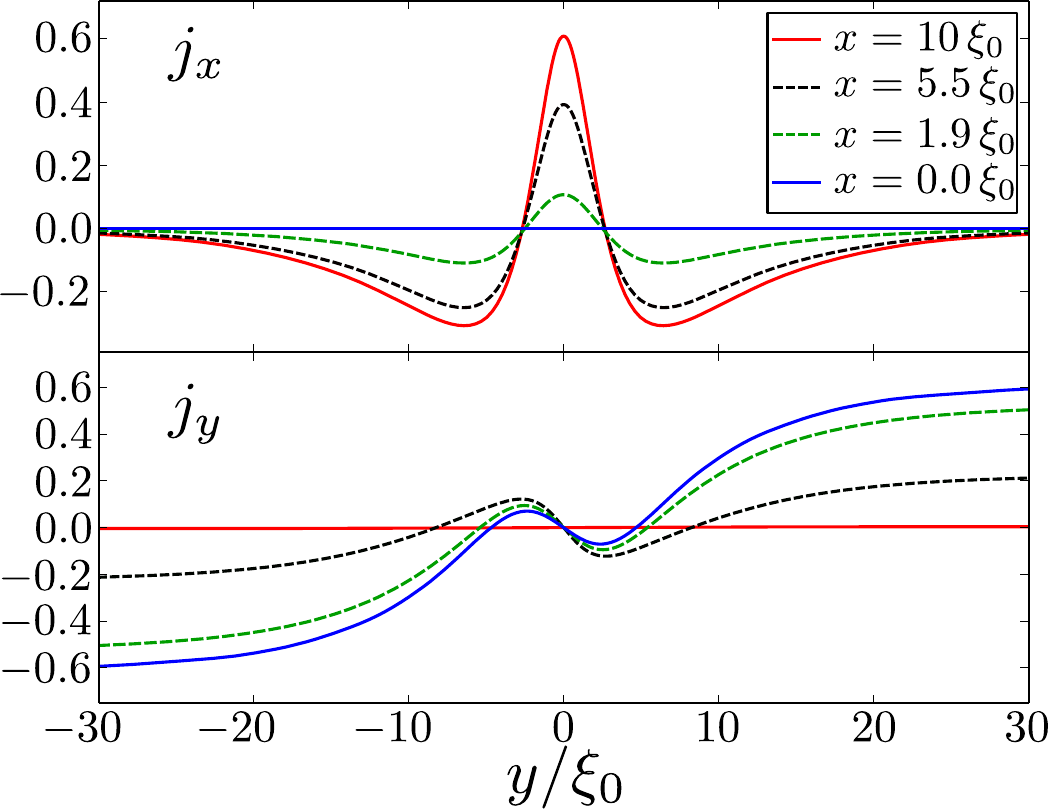}
\caption{
Slices of the components of the current density, $j_{x}(x,y)$ (top panel) and $j_y(x,y)$ (bottom panel), as a function of $y$ spanning the DW for several positions, $x$, in the channel. The vertical axes for the current density is in units of $N_f(2\pi T_c)m_3 v_f$. 
}
\label{fig-domainwall-current-cut-selfconsistent}
\end{figure}

Figure \ref{fig-domainwall-current-selfconsistent} shows the current density in a region containing a single $p_x$-type DW. As expected, the current confined near $y=0$ flows in the $+x$ direction with a magnitude that vanishes on approaching the edges at $x=0$ and $x=20\xi_0$. 
Current conservation is maintained by the appearance of two branches of current flowing in the $-x$ direction at distances of $y\approx\pm 5\xi_0$ from the DW, smoothly connecting to the currents on both the top and bottom edges as shown in Fig. \ref{fig-domainwall-current-selfconsistent}.
Finally, we note that the complex current pattern associated with the laterally confined chiral DW includes supercurrents associated with the gradient of the global phase of the PDW order parameter that are induced in the central region of the channel. The resulting current density satisfies current conservation at every point within the computational cell, i.e. $\grad\cdot\vJ(\vr)=0$ to the accuracy of our numerics which was set by an order parameter residual no greater than $10^{-4}$ at each point in the computational grid.

Finer detail of the current density of the chiral DW confined in the channel is shown in Fig. \ref{fig-domainwall-current-cut-selfconsistent}. The component $j_x$ exhibits the same sequence of current reversal on traversing the DW for all of the cuts defined by $x$. Indeed, the $p_x$-type DW appears to support \emph{three branches} of chiral Fermions, two branches with chirality $\nu =-1$ centered at $y=\pm 5\xi_0$, and a third branch with chirality $\nu=+1$ centered at $y=0$, for a net chirality of $\sum\nu=-1$. This is then consistent with the index we expect for the $p_y$-type DW with a single chiral branch bound to the DW.

\section{Stability of the PDW phase near $D_{c_2}$}\label{sec-Free_Energy}

The self-consistent solutions for the order parameter, in a region where the chiral A phase has lower energy than the polar phase, yields the PDW phase with broken translational symmetry. The PDW phase spans a wide range of channel widths, $D_{c_1} \le D \le D_{c_2}$. The period of the PDW increases with increasing $D$ (weaker confinement) and diverges as $D\rightarrow D_{c_2}(T)$; i.e. the period becomes of order the system size. The upper critical channel dimension, $D_{c2}(T)$, can be defined as the value of $D$ at which it costs zero energy to introduce a DW wall separating two otherwise degenerate domains of laterally confined \Hea. In the limit $D\lesssim D_{c_2}$, the question is ``how is it possible for a domain wall to lower the total energy of \Hea\ in a laterally confined geometry?''
The answer is that in a laterally confined geometry there is competition between the DW energy and the pair-breaking energy - the occupied branch of chiral Fermions - at edges. Each edge can be shown to be mathematically equivalent to one-half of a $p_y$-type DW. Furthermore, the energy cost of pair-breaking on the edge, equivalent to that of the $p_y$-type DW, is reduced within several coherence lengths of the points of intersection of the lateral DW ($p_x$-type) and the confining edges.
Since the energy per unit length of the $p_y$-type DW is greater than that for the $p_x$-type DW, there is a critical channel dimension at which the energy is lowered by entry of a lateral $p_x$-type DW. Once one DW can lower the energy, DWs proliferate the channel. Interactions between DWs regulate the DW density, and thereby determine the period of the PDW phase for $D\lesssim D_{c_2}$.

The stability of the PDW phase very near $D_{c_2}(T)$ is analyzed on the basis of a quasiclassical reduction of the Luttinger-Ward (LW) free-energy density for confined phases of \He.
The resulting LW free energy is then a functional of the quasiclassical Green's function, $\whg$, and mean-field pairing self energy, $\whDelta$,\cite{vor03}
\be
\Delta\Omega[\whg,\whDelta] 
= 
\frac{1}{2} \int_0^1\,d\lambda\,Sp'\left\{\whDelta(\whg_{\lambda}-\whg)\right\}
+
\Delta\Phi[\whg]
\,,
\ee 
where the functional $\Delta\Phi[\whg]$ is defined by an digrammatic expansion of the free energy functional in terms of effective interactions of low-energy quasiparticles and Cooper pairs. For our analysis we retain the leading-order weak-coupling BCS corrections to the LW free-energy for the normal Fermi-liquid, c.f. Ref. \onlinecite{ali11}. 

The operation, $Sp$, is short-hand for
\be
Sp\{\widehat{X}\}
= N_f T \sum_n\int d^3r \int \frac{d\Omega_\vp}{4 \pi} 
Tr_4
\left(\widehat{X}(\vp,\vr;\eps_n)\right)
\,.
\ee
The stationary conditions for the LW functional generate both Eilenberger's transport equation for the propagator, $\whg$, as well as the self-consistency equation for the off-diagonal self-energy, i.e. the ``gap equation''.
The auxiliary propagator, $\whg_{\lambda}$, is a function of the variable coupling constant, $\lambda$, and is defined as the solution of the Eilenberger equation with a rescaled interaction, or equivalently $\whDelta_{\lambda}\equiv \lambda\whDelta(\vp,\vr)$. For $\lambda=1$, corresponding to the full pairing interaction, we obtain the self-consistent propagator, $\whg$, and self-energy, $\whDelta$, whereas $\lambda=0$ corresponds to the normal state with $\whDelta = 0$ and $\whg_0 = -i\pi\sgn(\eps_n)\tz$.

Using the self-consistency condition from variations of the LW functional with respect to the propagator, and using the linearized gap equation to regulate the log-divergent Matsubara sums, the LW functional can be reduced to a functional of the off-diagonal pairing self-energy, $\Delta\Omega[\Delta] = \int_{\text{V}} d^3r\,\mathcal{F}[\Delta(\vr)]$, where the free-energy density is

\begin{widetext}
\be
\mathcal{F}[\Delta(\vr)] = N_f\ln\frac{T}{T_c}\int dS_{\vp}|\Delta(\vp, \vr)|^2
-
\int_0^1 d\lambda\int dS_{\vp} T\sum_n
\left(
\Delta(\vp,\vr)\f^*_\lambda(\vp,\vr;\eps_n) 
+ 
\Delta^*(\vp,\vr)\f_\lambda(\vp,\vr;\eps_n)
- 
\frac{\pi}{|\eps_n|}|\Delta(\vp,\vr)|^2
\right)
\,,
\ee
\end{widetext}
where $f_\lambda(\vp,\vr;\eps_n)$ is the solution of the Eilenberger transport equation defined with the re-scaled self-energy, $\Delta_{\lambda}$. This functional generalizes the GL free energy functional to all temperatures and to length scales $\hbar/p_f \ll D \lesssim \xi_0$.

\begin{figure}[t]
\includegraphics[width=\columnwidth]{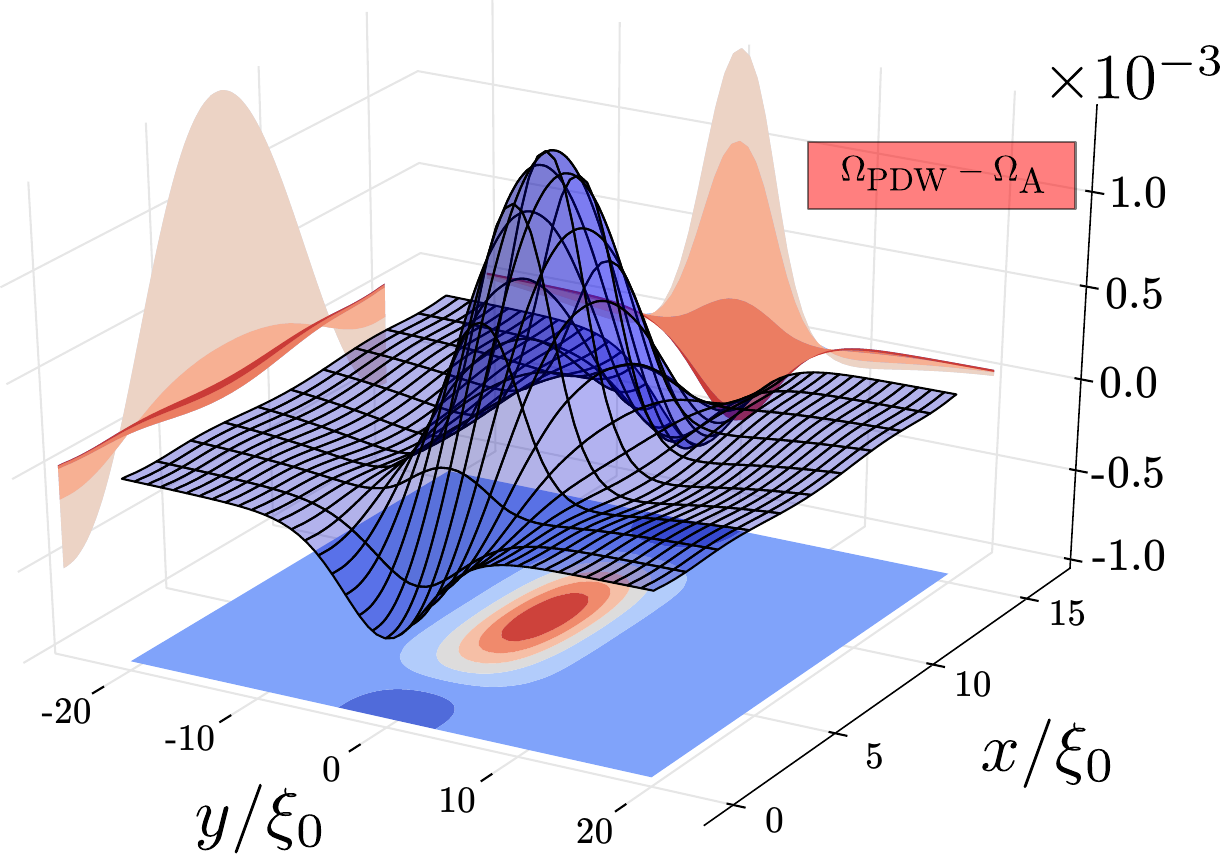}
\caption{Difference in energy density between PDW phase 
	and homogeneous A phase for confinement $D = 15\xi_0$, 
	the energy is in units of $2\pi T_c$. 
	The $x-y$ plane is the position of the channel.
}
\label{fig-energies-density}
\end{figure}

The difference in free-energy density between the PDW phase and the laterally confined A phase is shown in Fig. \ref{fig-energies-density}. The free energy density of the PDW phase is high at the center of domain wall, but the PDW phase gains in condensation energy at the junctions of the DW and the edges. 
The gain in energy at the junctions dominates as the width of the channel is reduced. Below the critical channel dimension, $D_{c2}$, the PDW phase has lower free energy, and is thus the stable ground state.
The numerical results for the upper critical channel width, $D_{c2}(T)$, are shown in Fig. \ref{fig-inhomo-instability} as the blue squares. The resulting PDW phase is the thermodynamically stable phase over a wide range of channel widths ($\Delta D \approx 7\xi_0 \approx 5.6\,\mu\mbox{m}$) and temperatures for laterally confined \He.

\section{Summary and Outlook}

The ground state of thin films of \He\ is the chiral Anderson-Morel state (\Hea), which breaks both time-reversal symmetry and mirror symmetry. The 2D A phase is a topological phase with Chern number $N=\pm 1$ for the condensate of each Fermi disk. A consequence of the topology of \Hea\ films is a spectrum of chiral edge states, their signature being a ground-state edge current. 
Under lateral confinement, edge states on opposing boundaries hybridize and develop a an intricate band structure for both the sub-gap and continuum spectrum. Hybridization of the edge spectrum leads to a reduction of the maxiumn edge current at any temperature. The non-monotomic temperature depdence of the edge current was the first signature of new ground states of superfluid \He\ films under strong confinement.  
Indeed thin films of \He\ undergo a sequence of phase transitions. At $T=0$, starting from the chiral A phase in the limit of laterally unconfined films ($D\rightarrow\infty$), superfluid \He\ undergoes a second-order transition to a PDW phase with broken translational symmetry at $D_{c2} \sim 16 \xi_0$. At this scale the PDW phase is a periodic array of chiral domains with alternating chirality, separated by domain walls, the latter also are chiral.
The PDW phase breaks time-reversal symmetry, translation invariance, but is invariant under the combination of time-reversal and translation by a one-half period of the PDW. 
Under stronger confinement a second-order phase transition occurs to a non-chiral ``polar phase'' at $D_{c1} \sim 9\xi_0$, in which a single p-wave orbital state of Cooper pairs is aligned along the channel.
This remarkable sequence of phase transitions is also mapped out in the full temperature-confinement phase diagram. 

The role of disorder on the confining boundaries has not been addressed in terms of the phase diagram for laterally confined \He, but in terms of testing the theoretical predictions of new phases of laterally confined \He\ thin films, fortunately surface disorder can be masked by pre-plating the confining surfaces with a few monolayers of superfluid \Hefour, which has been demonstrated to lead to specular surface scattering.\cite{hei21} 
In this limit thin films of superfluid \Hea\ are expected to persist well into the quasi-2D regime $\hbar/p_f\ll w\ll\xi_0$, at least until the effects of dimensional confinement modify the pairing interaction.

A key development in recent years is the marriage of ultra-low temperatures with nano-fabrication technologies, low-noise/high-precision acoustic, optical and NMR spectroscopies to study the broken symmetry ground states under strong confinement, as well as the excitations reflecting the topological nature of these novel ground states.
Finally, we note that experimental probes such as heat transport, charged impurities, nano-mechanical resonators, torsional oscillators, SQUID-NMR and sub-micron acoustic spectroscopy are some of the experimental techniques that can detect the broken symmetries of confined phases of \He, and the sub-gap spectrum that is the signature of topological superfluids.~\cite{lev10,ben10,zhe16b,lev13,lev13b,lev19,sho20,lot20,hei21,sha22,sha22a,sco23}

\section{Acknowledgement}

This research was supported by National Science Foundation Grants DMR-1106315 and DMR-1508730.


\begin{appendix}

\vspace*{-5mm}
\section{Solution to Andreev's Equation}\label{appendix-Andreev-Bloch_Solution}
\vspace*{-5mm}

We obtain an analytical solution to Andreev's Eqs. \ref{eq-Andreev_U} and \ref{eq-Andreev_V} for the piecewise-constant, periodic chiral order parameter defined in Eq. \ref{eq-Periodic_Chiral-OP}.
The solutions reveal the band-structure of the sub-gap and continuum spectrum, $\varepsilon(p_{\parallel},k)$. 
First, express the $U$ ampliutdes within each half period as
\ber\label{eq-U_V}
\hspace*{-5mm}
U_1 &=& +\frac{e^{+i\phi_{\hvp}}}{\Delta}[(\eps + \hbar v_f k)V - i \partial V] \,,\, 0 < x < D
\,,\qquad
\\
U_2 &=& -\frac{e^{-i\phi_{\hvp}}}{\Delta}[(\eps + \hbar v_f k)V - i\partial V]\,,\, D < x < 2D
\,,\qquad
\eer
where $e^{\pm i\phi_{\hvp}}\equiv\hp_x \pm i \hp_y$. The equation for $V$ becomes
\be
\partial^2 V + 2 i (\hbar v_f k) \partial V + (\eps^2 - \Delta^2 - \hbar^2 v_f^2 k^2) V = 0
\,,
\ee
with solution,
\ber\label{eq-V_solution}
V_1 &=& e^{-i k x} (C_1 e^{\lambda x} + C_2 e^{-\lambda x}) \,,\quad 0 < x < D		
\,,
\\
V_2 &=& e^{-i k x} (C'_1 e^{\lambda x} + C'_2 e^{-\lambda x})\,, \quad D < x < 2D
\,,
\eer
where $\lambda(\varepsilon)=\sqrt{\Delta^2-\varepsilon^2}/\hbar v_x$.

\eject
\noindent The boundary conditions: $V_1(x=0)=V_2(x=2D)$ and $V_1(x=D)=V_2(x=D)$, imply
\ber\label{eq-BC1}
C'_1 &=& \frac{1}{e^{2\lambda D}-1} [(e^{2ikD} - 1)C_1 + (e^{2ikD} - e^{-2\lambda D})C_2]	
\,,\qquad
\\
C'_2 &=& \frac{1}{1-e^{-2\lambda D}}[(e^{2\lambda D}-e^{2ikD})C_1 + (1-e^{2ikD}) C_2]
\,.\qquad
\label{eq-BC2}
\eer
The solution for $U(x)$ can then be expressed as
\ber\label{eq-U_solution}
\hspace*{-2mm}
U_1 &\ns=\ns&+\frac{e^{+i\phi_{\hvp}}}{\Delta} e^{-ikx}
	[(\eps-i\Lambda)C_1 e^{\lambda x}\ns+\ns(\eps+i\Lambda)C_2 e^{-\lambda x})]	
\,,\qquad
\\
U_2 &\ns=\ns&-\frac{e^{-i\phi_{\hvp}}}{\Delta} e^{-ikx}
		[(\eps-i\Lambda)C'_1 e^{\lambda x}\ns+\ns(\eps+i\Lambda)C'_2 e^{-\lambda x})]
\,,\qquad
\eer
where $\Lambda = \sqrt{\Delta^2-\varepsilon^2}$. The boundary conditions, $U_1(x=0)=U_2(x=2D)$ and $U_1(x=D)=U_2(x=D)$, then yield,
\ber\label{eq-BC3}
e^{+2i\phi_{\hvp}}[e^{-i\phi_\eps}C_1+e^{i\phi_\eps}C_2] 
=
\hspace*{3cm}
&&
\nonumber\\
-e^{-2ikD}[e^{-i\phi_\eps}C'_1 e^{2\lambda D}+e^{i\phi_\eps}C'_2 e^{-2\lambda D}]	
\,,
&&
\\
e^{+2i\phi_{\hvp}}[e^{-i\phi_\eps}C_1 e^{\lambda D}+e^{i\phi_\eps}C_2 e^{-\lambda D}] 
=
\hspace*{1.65cm}
&&
\nonumber\\
-[e^{-i\phi_\eps}C'_1 e^{\lambda D} + e^{i\phi_\eps}C'2 e^{-\lambda D}]
\,,
&&
\label{eq-BC4}
\eer
where $(\eps+i\Lambda)/\Delta\equiv e^{i\phi_\eps}$ is a phase factor for $|\eps|<\Delta$. 
Using Eqs. \ref{eq-BC1}-\ref{eq-BC2} we obtain eigenvalue equations for $C_1$ and $C_2$,
\begin{widetext}
\ber\label{eq-C1C2_1}
\left[e^{2i\phi_{\hvp}-i\phi_\eps} \ns+\ns e^{-i\phi_\eps}\frac{e^{2\lambda D}(1-e^{-2ikD})}{e^{2\lambda D}-1} 
\ns+\ns e^{i\phi_\eps} \frac{e^{2\lambda D}e^{-2ikD} - 1}{e^{2\lambda D} - 1}\right]
\ns C_1 
\ns+\ns	
\left[e^{2i\phi_{\hvp} + i\phi_\eps}+e^{- i\phi_\eps} \frac{e^{2\lambda D}-e^{-2ikD}}{e^{2\lambda D}-1} 
\ns+\ns e^{i\phi_\eps} \frac{ e^{-2ikD} - 1}{e^{2\lambda D} - 1} \right]
\ns C_2 
\ns = \ns 0 
\,,\qquad\quad
\\
\left[e^{2i\phi_{\hvp} - i\phi_\eps} \ns+\ns e^{-i\phi_\eps} \frac{e^{2ikD} - 1}{e^{2\lambda D} - 1} + 
e^{i\phi_\eps} \frac{e^{2\lambda D} - e^{2ikD}}{e^{2\lambda D} - 1}\right] C_1 
\ns+\ns
\left[e^{2i\phi_{\hvp} + i\phi_\eps} e^{-2\lambda D} \ns+\ns e^{-i\phi_\eps} \frac{e^{2ikD} - 
e^{-2\lambda D}}{e^{2\lambda D} - 1} + e^{i\phi_\eps}\frac{1 - e^{2ikD}}{e^{2\lambda D} - 1} \right] C_2 
\ns=\ns 0
\,.\qquad\quad
\label{eq-C1C2_2}
\eer
\end{widetext}
For sub-gap solutions, $|\varepsilon|<\Delta$, $\lambda(\eps)$ is real and positive. The determinant of 
Eqs. \ref{eq-C1C2_1}-\ref{eq-C1C2_2} then yields the eigenvalue equation,
\be\label{eq-E(k)}
\cos(2\phi_\eps) = \frac{(1 - \cosh(2\lambda D))\cos(2\phi_{\hvp}) - (1 - \cos(2kD))}
						{\cosh(2\lambda D) - \cos(2kD)}
\,.
\ee 

Using $\cos(2\phi_\eps)=(2\eps^2/\Delta^2)-1$ and $\cos(2\phi_{\hvp})=\hp_x^2-\hp_y^2$, Eq. \ref{eq-E(k)}
can be expressed as
\be\label{eq-E(k)_2}
\varepsilon^2 = \Delta^2 \hp_y^2 + \frac{\Delta^2 \hp_x^2 (1 - \cos(2kD))}{\cosh(2\lambda D) - \cos(2kD)}
\,,
\ee
which is Eq. \ref{eq-sub-gap_energy-levels}.
In the limit $D\rightarrow\infty$ 
the sub-gap bands collapse to the \emph{two} chiral branches, $\epsilon^{\pm} = \pm \Delta\hp_y$.
The branch with $\epsilon^{+} = +\Delta\hp_y$ is confined on the upper edge $x=D$, while the branch with dispersion $\varepsilon^{-}=-\Delta \hp_y$ corresponds to the edge state on the lower edge $x=0$.
For $D$ finite, the sub-gap bandwidth increases for stronger confinement; the exponential dependence of bandwidth originates from the exponential decay of edge states. The momentum dependence of the sub-gap band shows that the bandwidth is largest for normal incidence $p_x=p_f$, and is smallest for grazing incidence $p_x=0$. This agrees with the mapping to a periodic array of domain walls; a trajectory normal to the edges has the smallest period, $2D$, while trajectories near grazing incidence have a diverging period.

\vspace{-.5cm}
\section{Solutions for the Quasiclassical Propagator}\label{appendix-quasiclassical_solution}
\vspace{-.5cm}

Here we provide the analysis for the solution of the Eilenberger equation and normalization condition for a 
piece-wise constant chiral order parameter. 
We convert the Nambu matrix representation of the Eilenberger equation to coupled differential equations for a vector representation of propagator,\cite{sau11} 
\be
\frac{1}{2}\vv_\vp \cdot \nabla \ket{\g} = \widehat{M} \ket{\g}\,
\ee
with 
\be
\ket{\g} \equiv 
\begin{pmatrix}
	\f^{\text{R}}_1	\\
	\f^{\text{R}}_2	\\
	\g^{\text{R}}_3	\\
\end{pmatrix}\, \quad 
\widehat{M} = 
\begin{pmatrix}
	0	&	\eps 	&	\Delta_2	\\
	-\eps	&	0	&	-\Delta_1	\\
	\Delta_2	& -\Delta_1	&	0	
\end{pmatrix}
\ee
For a constant order parameter amplitude, we have $\Delta_1(\vp) = \Delta p_x$, $\Delta_2(\vp) = \Delta p_y$. 
We can express $\ket{\g}$ in terms of the eigenvectors of $\widehat{M}$, $\widehat{M} \ket{\mu} = \mu \ket{\mu}$. The eigenvector with $\mu = 0$ is
\be
\ket{0; \vp} = \frac{1}{\lambda}
\begin{pmatrix}
	-\Delta_1	\\
	-\Delta_2	\\
	\eps
\end{pmatrix}
\,.
\ee
This solution corresponds to the bulk equilibrium propagator, 
\be
\hat{\g}(\vp) = -\frac{\pi}{\lambda}(\eps \tauz - \widehat{\Delta}(\vp))
\,.
\ee
The eigenvectors corresponding to eigenvalues $\mu = \pm \lambda$ are 
\be
\ket{\pm; \vp} = \frac{1}{\sqrt{2}\lambda \lambda_1} 
\begin{pmatrix}
	\pm \lambda \eps - \Delta_1 \Delta_2	\\
	\lambda_1^2	\\
	\eps \Delta_2 \mp \lambda \Delta_1
\end{pmatrix}
\, ,
\ee
where $\lambda_1 \equiv \sqrt{\Delta_1(\vp)^2 - \eps^2}$. 
These eigenvectors correspond to Nambu matrices, 
\ber
\hat{\g}_{\pm} = \frac{-\pi}{\sqrt{2}\lambda \lambda_1}
\left[(\eps \Delta_2 \mp \lambda \Delta_1)\tauz 
\right.
\qquad\qquad\qquad
\nonumber\\
\left.
\mp 
i\hat \sigma_x (\lambda \eps \mp \Delta_1 \Delta_2) \tauy 
+ 
i\sigma_x \lambda_1^2 \taux
\right]
\,.
\eer
These are the base matrices defining the propagator in Eq. \ref{eq-QC_propagator-channel}. 
In the vector represenation, the solutions for the incident and reflected trajectories are
\ber
\ket{\g(\vp)}
&\ns=\ns&
\ket{0;\vp}\ns+\ns C^1_{in}(\vp) e^{-2\lambda x/v_x}\ket{+;\vp}\ns+\ns C^2_{in}e^{2\lambda x/v_x}\ket{-;\vp}
\,,\qquad
\\
\ket{\g(\ul\vp)} 
&\ns=\ns&
\ket{0;\ul\vp}\ns+\ns 
C^1_{out}(\ul\vp)e^{-2\lambda x/v_x}\ket{-;\ul\vp}\ns+\ns C^2_{out}e^{2\lambda x/v_x}\ket{+;\ul\vp}
\,.\qquad
\eer
The specular boundary conditions require matching the propagators $\ket{\g(\vp)}$ and $\ket{\g(\ul \vp)}$ on both edges,
\be
\ket{\g(\vp, 0)} = \ket{\g(\ul \vp, 0)}, \quad \quad
\ket{\g(\vp, D)} = \ket{\g(\ul \vp, D)}
\,,
\ee
which yields $C^1_{out} = C^1_{in}\equiv C^1$ and $C^2_{out} = C^2_{in}\equiv C^2$, with 
\ber
-\Delta_1\sqrt{2}\lambda_1
&+&
C^1(\lambda\eps-\Delta_1\Delta_2)
\nonumber\\
&-&
C^2(\lambda\eps+\Delta_1\Delta_2)
=0  
\,,
\\
-\Delta_1\sqrt{2}\lambda_1
&+&
C^1(\lambda\eps-\Delta_1\Delta_2)e^{-2\lambda D/v_x}
\nonumber\\
&-&
C^2(\lambda\eps+\Delta_1\Delta_2)e^{2\lambda D}
=0  
\,.
\eer

\eject
The coefficients defining the propagator in Eq. \ref{eq-QC_propagator-channel} are 
\ber
C^1
&=& 
\frac{\sqrt{2}\Delta_1(\vp)\lambda_1}{\lambda \eps - \Delta_1(\vp) \Delta_2(\vp)} \frac{e^{2\lambda D / v_x} - 1}{e^{2\lambda D / v_x} - e^{-2\lambda D / v_x}}	
\,,
\\
C^2
&=& 
\frac{\sqrt{2} \lambda_1 \Delta_1(\vp)}{-\lambda \eps - \Delta_1(\vp) \Delta_2(\vp)} \frac{1 - 2^{-2\lambda D / v_x}}{e^{2\lambda D / v_x} - e^{-2\lambda D / v_x}}
\,.
\eer
\end{appendix}
%

\begin{thebibliography}{75}%
\makeatletter
\providecommand \@ifxundefined [1]{%
 \@ifx{#1\undefined}
}%
\providecommand \@ifnum [1]{%
 \ifnum #1\expandafter \@firstoftwo
 \else \expandafter \@secondoftwo
 \fi
}%
\providecommand \@ifx [1]{%
 \ifx #1\expandafter \@firstoftwo
 \else \expandafter \@secondoftwo
 \fi
}%
\providecommand \natexlab [1]{#1}%
\providecommand \enquote  [1]{``#1''}%
\providecommand \bibnamefont  [1]{#1}%
\providecommand \bibfnamefont [1]{#1}%
\providecommand \citenamefont [1]{#1}%
\providecommand \href@noop [0]{\@secondoftwo}%
\providecommand \href [0]{\begingroup \@sanitize@url \@href}%
\providecommand \@href[1]{\@@startlink{#1}\@@href}%
\providecommand \@@href[1]{\endgroup#1\@@endlink}%
\providecommand \@sanitize@url [0]{\catcode `\\12\catcode `\$12\catcode
  `\&12\catcode `\#12\catcode `\^12\catcode `\_12\catcode `\%12\relax}%
\providecommand \@@startlink[1]{}%
\providecommand \@@endlink[0]{}%
\providecommand \url  [0]{\begingroup\@sanitize@url \@url }%
\providecommand \@url [1]{\endgroup\@href {#1}{\urlprefix }}%
\providecommand \urlprefix  [0]{URL }%
\providecommand \Eprint [0]{\href }%
\providecommand \doibase [0]{https://doi.org/}%
\providecommand \selectlanguage [0]{\@gobble}%
\providecommand \bibinfo  [0]{\@secondoftwo}%
\providecommand \bibfield  [0]{\@secondoftwo}%
\providecommand \translation [1]{[#1]}%
\providecommand \BibitemOpen [0]{}%
\providecommand \bibitemStop [0]{}%
\providecommand \bibitemNoStop [0]{.\EOS\space}%
\providecommand \EOS [0]{\spacefactor3000\relax}%
\providecommand \BibitemShut  [1]{\csname bibitem#1\endcsname}%
\let\auto@bib@innerbib\@empty
\bibitem [{\citenamefont {Chung}\ and\ \citenamefont {Zhang}(2009)}]{chu09}%
  \BibitemOpen
  \bibfield  {author} {\bibinfo {author} {\bibfnamefont {S.~B.}\ \bibnamefont
  {Chung}}\ and\ \bibinfo {author} {\bibfnamefont {S.-C.}\ \bibnamefont
  {Zhang}},\ }\bibfield  {title} {\bibinfo {title} {{Detecting the Majorana
  Fermion Surface State of $^3$He-B through Spin Relaxation}},\ }\href
  {https://doi.org/10.1103/PhysRevLett.103.235301} {\bibfield  {journal}
  {\bibinfo  {journal} {Phys. Rev. Lett.}\ }\textbf {\bibinfo {volume} {103}},\
  \bibinfo {pages} {235301} (\bibinfo {year} {2009})}\BibitemShut {NoStop}%
\bibitem [{\citenamefont {Volovik}(2009{\natexlab{a}})}]{vol09}%
  \BibitemOpen
  \bibfield  {author} {\bibinfo {author} {\bibfnamefont {G.~E.}\ \bibnamefont
  {Volovik}},\ }\bibfield  {title} {\bibinfo {title} {{Topological Invariant
  for Superfluid $^3$He-B and Quantum Phase Transitions}},\ }\href
  {https://doi.org/10.1134/S0021364009200089} {\bibfield  {journal} {\bibinfo
  {journal} {JETP Lett.}\ }\textbf {\bibinfo {volume} {90}},\ \bibinfo {pages}
  {587} (\bibinfo {year} {2009}{\natexlab{a}})}\BibitemShut {NoStop}%
\bibitem [{\citenamefont {Volovik}(2009{\natexlab{b}})}]{vol09a}%
  \BibitemOpen
  \bibfield  {author} {\bibinfo {author} {\bibfnamefont {G.~E.}\ \bibnamefont
  {Volovik}},\ }\bibfield  {title} {\bibinfo {title} {{Fermion zero modes at
  the boundary of superfluid $^3$He-B}},\ }\href
  {https://doi.org/10.1134/S0021364009170172} {\bibfield  {journal} {\bibinfo
  {journal} {JETP Lett}\ }\textbf {\bibinfo {volume} {90}},\ \bibinfo {pages}
  {398} (\bibinfo {year} {2009}{\natexlab{b}})}\BibitemShut {NoStop}%
\bibitem [{\citenamefont {Sauls}(2011)}]{sau11}%
  \BibitemOpen
  \bibfield  {author} {\bibinfo {author} {\bibfnamefont {J.~A.}\ \bibnamefont
  {Sauls}},\ }\bibfield  {title} {\bibinfo {title} {{Surface states, Edge
  Currents, and the Angular Momentum of Chiral $p$-wave Superfluids}},\ }\href
  {https://doi.org/10.1103/PhysRevB.84.214509} {\bibfield  {journal} {\bibinfo
  {journal} {Phys. Rev. B}\ }\textbf {\bibinfo {volume} {84}},\ \bibinfo
  {pages} {214509} (\bibinfo {year} {2011})}\BibitemShut {NoStop}%
\bibitem [{\citenamefont {Mizushima}(2012)}]{miz12}%
  \BibitemOpen
  \bibfield  {author} {\bibinfo {author} {\bibfnamefont {T.}~\bibnamefont
  {Mizushima}},\ }\bibfield  {title} {\bibinfo {title} {{Superfluid $^3$He in a
  restricted geometry with a perpendicular magnetic field}},\ }\href
  {https://doi.org/10.1103/PhysRevB.86.094518} {\bibfield  {journal} {\bibinfo
  {journal} {Phys. Rev. B}\ }\textbf {\bibinfo {volume} {86}},\ \bibinfo
  {pages} {094518} (\bibinfo {year} {2012})}\BibitemShut {NoStop}%
\bibitem [{\citenamefont {Mizushima}\ \emph {et~al.}(2012)\citenamefont
  {Mizushima}, \citenamefont {Sato},\ and\ \citenamefont {Machida}}]{miz12a}%
  \BibitemOpen
  \bibfield  {author} {\bibinfo {author} {\bibfnamefont {T.}~\bibnamefont
  {Mizushima}}, \bibinfo {author} {\bibfnamefont {M.}~\bibnamefont {Sato}},\
  and\ \bibinfo {author} {\bibfnamefont {K.}~\bibnamefont {Machida}},\
  }\bibfield  {title} {\bibinfo {title} {{Symmetry Protected Topological Order
  and Spin Susceptibility in Superfluid $^{3}He-B$}},\ }\href
  {https://doi.org/10.1103/PhysRevLett.109.165301} {\bibfield  {journal}
  {\bibinfo  {journal} {Phys. Rev. Lett.}\ }\textbf {\bibinfo {volume} {109}},\
  \bibinfo {pages} {165301} (\bibinfo {year} {2012})}\BibitemShut {NoStop}%
\bibitem [{\citenamefont {Wu}\ and\ \citenamefont {Sauls}(2013)}]{wu13}%
  \BibitemOpen
  \bibfield  {author} {\bibinfo {author} {\bibfnamefont {H.}~\bibnamefont
  {Wu}}\ and\ \bibinfo {author} {\bibfnamefont {J.~A.}\ \bibnamefont {Sauls}},\
  }\bibfield  {title} {\bibinfo {title} {{Majorana excitations, spin and mass
  currents on the surface of topological superfluid $^{3}$He-B}},\ }\href
  {https://doi.org/10.1103/PhysRevB.88.184506} {\bibfield  {journal} {\bibinfo
  {journal} {Phys. Rev. B}\ }\textbf {\bibinfo {volume} {88}},\ \bibinfo
  {pages} {184506} (\bibinfo {year} {2013})}\BibitemShut {NoStop}%
\bibitem [{\citenamefont {Vorontsov}(2018)}]{vor18}%
  \BibitemOpen
  \bibfield  {author} {\bibinfo {author} {\bibfnamefont {A.~B.}\ \bibnamefont
  {Vorontsov}},\ }\bibfield  {title} {\bibinfo {title} {{Andreev bound states
  in superconducting films and confined superfluid $^3$He}},\ }\href
  {https://doi.org/10.1098/rsta.2015.0144} {\bibfield  {journal} {\bibinfo
  {journal} {Phil. Trans. Roy. Soc. A}\ }\textbf {\bibinfo {volume} {376}},\
  \bibinfo {pages} {20150144} (\bibinfo {year} {2018})}\BibitemShut {NoStop}%
\bibitem [{\citenamefont {Mizushima}\ and\ \citenamefont
  {Machida}(2018)}]{miz18d}%
  \BibitemOpen
  \bibfield  {author} {\bibinfo {author} {\bibfnamefont {T.}~\bibnamefont
  {Mizushima}}\ and\ \bibinfo {author} {\bibfnamefont {K.}~\bibnamefont
  {Machida}},\ }\bibfield  {title} {\bibinfo {title} {{Multifaceted properties
  of Andreev bound states: interplay of symmetry and topology}},\ }\href
  {https://doi.org/10.1098/rsta.2015.0355} {\bibfield  {journal} {\bibinfo
  {journal} {Phil. Trans. Roy. Soc. A}\ }\textbf {\bibinfo {volume} {376}},\
  \bibinfo {pages} {20150355} (\bibinfo {year} {2018})}\BibitemShut {NoStop}%
\bibitem [{\citenamefont {Gonz\'alez}\ \emph {et~al.}(2011)\citenamefont
  {Gonz\'alez}, \citenamefont {Bhupathi}, \citenamefont {Moon}, \citenamefont
  {Zheng}, \citenamefont {Ling}, \citenamefont {Garcell}, \citenamefont
  {Chan},\ and\ \citenamefont {Lee}}]{gon11}%
  \BibitemOpen
  \bibfield  {author} {\bibinfo {author} {\bibfnamefont {M.}~\bibnamefont
  {Gonz\'alez}}, \bibinfo {author} {\bibfnamefont {P.}~\bibnamefont
  {Bhupathi}}, \bibinfo {author} {\bibfnamefont {B.}~\bibnamefont {Moon}},
  \bibinfo {author} {\bibfnamefont {P.}~\bibnamefont {Zheng}}, \bibinfo
  {author} {\bibfnamefont {G.}~\bibnamefont {Ling}}, \bibinfo {author}
  {\bibfnamefont {E.}~\bibnamefont {Garcell}}, \bibinfo {author} {\bibfnamefont
  {H.}~\bibnamefont {Chan}},\ and\ \bibinfo {author} {\bibfnamefont
  {Y.}~\bibnamefont {Lee}},\ }\bibfield  {title} {\bibinfo {title}
  {{Characterization of MEMS Devices for the Study of Superfluid Helium
  Films}},\ }\href {https://doi.org/10.1007/s10909-010-0247-7} {\bibfield
  {journal} {\bibinfo  {journal} {J. Low Temp. Phys.}\ }\textbf {\bibinfo
  {volume} {162}},\ \bibinfo {pages} {661} (\bibinfo {year}
  {2011})}\BibitemShut {NoStop}%
\bibitem [{\citenamefont {Levitin}\ \emph
  {et~al.}(2013{\natexlab{a}})\citenamefont {Levitin}, \citenamefont {Bennett},
  \citenamefont {Casey}, \citenamefont {Cowan}, \citenamefont {Saunders},
  \citenamefont {Drung}, \citenamefont {Schurig},\ and\ \citenamefont
  {Parpia}}]{lev13}%
  \BibitemOpen
  \bibfield  {author} {\bibinfo {author} {\bibfnamefont {L.~V.}\ \bibnamefont
  {Levitin}}, \bibinfo {author} {\bibfnamefont {R.~G.}\ \bibnamefont
  {Bennett}}, \bibinfo {author} {\bibfnamefont {A.}~\bibnamefont {Casey}},
  \bibinfo {author} {\bibfnamefont {B.}~\bibnamefont {Cowan}}, \bibinfo
  {author} {\bibfnamefont {J.}~\bibnamefont {Saunders}}, \bibinfo {author}
  {\bibfnamefont {D.}~\bibnamefont {Drung}}, \bibinfo {author} {\bibfnamefont
  {T.}~\bibnamefont {Schurig}},\ and\ \bibinfo {author} {\bibfnamefont {J.~M.}\
  \bibnamefont {Parpia}},\ }\bibfield  {title} {\bibinfo {title} {{Phase
  Diagram of the Topological Superfluid $^3$He Confined in a Nano-scale Slab
  Geometry}},\ }\href {https://doi.org/10.1126/science.1233621} {\bibfield
  {journal} {\bibinfo  {journal} {Science}\ }\textbf {\bibinfo {volume}
  {340}},\ \bibinfo {pages} {841} (\bibinfo {year}
  {2013}{\natexlab{a}})}\BibitemShut {NoStop}%
\bibitem [{\citenamefont {Levitin}\ \emph
  {et~al.}(2013{\natexlab{b}})\citenamefont {Levitin}, \citenamefont {Bennett},
  \citenamefont {Surovtsev}, \citenamefont {Parpia}, \citenamefont {Cowan},
  \citenamefont {Casey},\ and\ \citenamefont {Saunders}}]{lev13b}%
  \BibitemOpen
  \bibfield  {author} {\bibinfo {author} {\bibfnamefont {L.~V.}\ \bibnamefont
  {Levitin}}, \bibinfo {author} {\bibfnamefont {R.~G.}\ \bibnamefont
  {Bennett}}, \bibinfo {author} {\bibfnamefont {E.~V.}\ \bibnamefont
  {Surovtsev}}, \bibinfo {author} {\bibfnamefont {J.~M.}\ \bibnamefont
  {Parpia}}, \bibinfo {author} {\bibfnamefont {B.}~\bibnamefont {Cowan}},
  \bibinfo {author} {\bibfnamefont {A.~J.}\ \bibnamefont {Casey}},\ and\
  \bibinfo {author} {\bibfnamefont {J.}~\bibnamefont {Saunders}},\ }\bibfield
  {title} {\bibinfo {title} {{Surface-Induced Order Parameter Distortion in
  Superfluid $^3$He-B Measured by Nonlinear NMR}},\ }\href
  {https://doi.org/10.1103/PhysRevLett.111.235304} {\bibfield  {journal}
  {\bibinfo  {journal} {Phys. Rev. Lett.}\ }\textbf {\bibinfo {volume} {111}},\
  \bibinfo {pages} {235304} (\bibinfo {year} {2013}{\natexlab{b}})}\BibitemShut
  {NoStop}%
\bibitem [{\citenamefont {Freeman}\ \emph {et~al.}(1988)\citenamefont
  {Freeman}, \citenamefont {Germain}, \citenamefont {Thuneberg},\ and\
  \citenamefont {Richardson}}]{fre88}%
  \BibitemOpen
  \bibfield  {author} {\bibinfo {author} {\bibfnamefont {M.}~\bibnamefont
  {Freeman}}, \bibinfo {author} {\bibfnamefont {R.~S.}\ \bibnamefont
  {Germain}}, \bibinfo {author} {\bibfnamefont {E.~V.}\ \bibnamefont
  {Thuneberg}},\ and\ \bibinfo {author} {\bibfnamefont {R.~C.}\ \bibnamefont
  {Richardson}},\ }\bibfield  {title} {\bibinfo {title} {{Size effects in thin
  films of superfluid $^3$He}},\ }\href
  {https://doi.org/10.1103/PhysRevLett.60.596} {\bibfield  {journal} {\bibinfo
  {journal} {Phys. Rev. Lett.}\ }\textbf {\bibinfo {volume} {60}},\ \bibinfo
  {pages} {596} (\bibinfo {year} {1988})}\BibitemShut {NoStop}%
\bibitem [{\citenamefont {Bennett}\ \emph {et~al.}(2010)\citenamefont
  {Bennett}, \citenamefont {Levitin}, \citenamefont {Casey}, \citenamefont
  {Cowan}, \citenamefont {Parpia},\ and\ \citenamefont {Saunders}}]{ben10}%
  \BibitemOpen
  \bibfield  {author} {\bibinfo {author} {\bibfnamefont {R.~G.}\ \bibnamefont
  {Bennett}}, \bibinfo {author} {\bibfnamefont {L.}~\bibnamefont {Levitin}},
  \bibinfo {author} {\bibfnamefont {A.}~\bibnamefont {Casey}}, \bibinfo
  {author} {\bibfnamefont {B.}~\bibnamefont {Cowan}}, \bibinfo {author}
  {\bibfnamefont {J.}~\bibnamefont {Parpia}},\ and\ \bibinfo {author}
  {\bibfnamefont {J.}~\bibnamefont {Saunders}},\ }\bibfield  {title} {\bibinfo
  {title} {{Superfluid $^3$He Confined to a Single 0.6 Micron Slab Stability
  and Properties of the A-Like Phase Near the Weak Coupling Limit}},\ }\href
  {https://doi.org/10.1007/s10909-009-9941-8} {\bibfield  {journal} {\bibinfo
  {journal} {J. Low Temp. Phys.}\ }\textbf {\bibinfo {volume} {158}},\ \bibinfo
  {pages} {163} (\bibinfo {year} {2010})}\BibitemShut {NoStop}%
\bibitem [{\citenamefont {Anderson}\ and\ \citenamefont {Morel}(1961)}]{and61}%
  \BibitemOpen
  \bibfield  {author} {\bibinfo {author} {\bibfnamefont {P.~W.}\ \bibnamefont
  {Anderson}}\ and\ \bibinfo {author} {\bibfnamefont {P.}~\bibnamefont
  {Morel}},\ }\bibfield  {title} {\bibinfo {title} {{Generalized
  Bardeen-Cooper-Schrieffer States and the Proposed Low-Temperature Phase of
  $^3$He}},\ }\href {https://doi.org/10.1103/PhysRev.123.1911} {\bibfield
  {journal} {\bibinfo  {journal} {Phys. Rev.}\ }\textbf {\bibinfo {volume}
  {123}},\ \bibinfo {pages} {1911} (\bibinfo {year} {1961})}\BibitemShut
  {NoStop}%
\bibitem [{\citenamefont {Brinkman}\ \emph {et~al.}(1974)\citenamefont
  {Brinkman}, \citenamefont {Serene},\ and\ \citenamefont {Anderson}}]{bri74}%
  \BibitemOpen
  \bibfield  {author} {\bibinfo {author} {\bibfnamefont {W.~F.}\ \bibnamefont
  {Brinkman}}, \bibinfo {author} {\bibfnamefont {J.~W.}\ \bibnamefont
  {Serene}},\ and\ \bibinfo {author} {\bibfnamefont {P.~W.}\ \bibnamefont
  {Anderson}},\ }\bibfield  {title} {\bibinfo {title} {{Spin-fluctuation
  stabilization of anisotropic superfluid states}},\ }\href
  {https://doi.org/10.1103/PhysRevA.10.2386} {\bibfield  {journal} {\bibinfo
  {journal} {Phys. Rev. A}\ }\textbf {\bibinfo {volume} {10}},\ \bibinfo
  {pages} {2386} (\bibinfo {year} {1974})}\BibitemShut {NoStop}%
\bibitem [{\citenamefont {Ikegami}\ \emph
  {et~al.}(2013{\natexlab{a}})\citenamefont {Ikegami}, \citenamefont
  {Tsutsumi},\ and\ \citenamefont {Kono}}]{ike13}%
  \BibitemOpen
  \bibfield  {author} {\bibinfo {author} {\bibfnamefont {H.}~\bibnamefont
  {Ikegami}}, \bibinfo {author} {\bibfnamefont {Y.}~\bibnamefont {Tsutsumi}},\
  and\ \bibinfo {author} {\bibfnamefont {K.}~\bibnamefont {Kono}},\ }\bibfield
  {title} {\bibinfo {title} {{Chiral Symmetry in Superfluid \Hea}},\ }\href
  {https://doi.org/10.1126/science.1236509} {\bibfield  {journal} {\bibinfo
  {journal} {Science}\ }\textbf {\bibinfo {volume} {341}},\ \bibinfo {pages}
  {59} (\bibinfo {year} {2013}{\natexlab{a}})}\BibitemShut {NoStop}%
\bibitem [{\citenamefont {Shevtsov}\ and\ \citenamefont {Sauls}(2016)}]{she16}%
  \BibitemOpen
  \bibfield  {author} {\bibinfo {author} {\bibfnamefont {O.}~\bibnamefont
  {Shevtsov}}\ and\ \bibinfo {author} {\bibfnamefont {J.~A.}\ \bibnamefont
  {Sauls}},\ }\bibfield  {title} {\bibinfo {title} {{Electron Bubbles and Weyl
  Fermions in Chiral Superfluid \Hea}},\ }\href
  {https://doi.org/10.1103/PhysRevB.94.064511} {\bibfield  {journal} {\bibinfo
  {journal} {Phys. Rev. B}\ }\textbf {\bibinfo {volume} {94}},\ \bibinfo
  {pages} {064511} (\bibinfo {year} {2016})}\BibitemShut {NoStop}%
\bibitem [{\citenamefont {Rice}\ and\ \citenamefont {Sigrist}(1995)}]{ric95}%
  \BibitemOpen
  \bibfield  {author} {\bibinfo {author} {\bibfnamefont {T.~M.}\ \bibnamefont
  {Rice}}\ and\ \bibinfo {author} {\bibfnamefont {M.}~\bibnamefont {Sigrist}},\
  }\bibfield  {title} {\bibinfo {title} {{Sr$_2$RuO$_4$: an electronic analogue
  of $^3$He?}},\ }\href {https://doi.org/10.1088/0953-8984/7/47/002} {\bibfield
   {journal} {\bibinfo  {journal} {J. Phys. Cond. Mat.}\ }\textbf {\bibinfo
  {volume} {7}},\ \bibinfo {pages} {L643} (\bibinfo {year} {1995})}\BibitemShut
  {NoStop}%
\bibitem [{\citenamefont {Kallin}\ and\ \citenamefont
  {Berlinsky}(2009)}]{kal09}%
  \BibitemOpen
  \bibfield  {author} {\bibinfo {author} {\bibfnamefont {C.}~\bibnamefont
  {Kallin}}\ and\ \bibinfo {author} {\bibfnamefont {A.~J.}\ \bibnamefont
  {Berlinsky}},\ }\bibfield  {title} {\bibinfo {title} {{Is Sr$_2$RuO$_4$ a
  chiral p-wave superconductor?}},\ }\href
  {https://doi.org/10.1088/0953-8984/21/16/164210} {\bibfield  {journal}
  {\bibinfo  {journal} {J Phys-Condens Mat}\ }\textbf {\bibinfo {volume}
  {21}},\ \bibinfo {pages} {164210} (\bibinfo {year} {2009})}\BibitemShut
  {NoStop}%
\bibitem [{\citenamefont {Raghu}\ \emph {et~al.}(2010)\citenamefont {Raghu},
  \citenamefont {Kapitulnik},\ and\ \citenamefont {Kivelson}}]{rag10}%
  \BibitemOpen
  \bibfield  {author} {\bibinfo {author} {\bibfnamefont {S.}~\bibnamefont
  {Raghu}}, \bibinfo {author} {\bibfnamefont {A.}~\bibnamefont {Kapitulnik}},\
  and\ \bibinfo {author} {\bibfnamefont {S.~A.}\ \bibnamefont {Kivelson}},\
  }\bibfield  {title} {\bibinfo {title} {{Hidden Quasi-One-Dimensional
  Superconductivity in Sr$_2$RuO$_4$}},\ }\href
  {https://doi.org/10.1103/PhysRevLett.105.136401} {\bibfield  {journal}
  {\bibinfo  {journal} {Phys. Rev. Lett.}\ }\textbf {\bibinfo {volume} {105}},\
  \bibinfo {pages} {136401} (\bibinfo {year} {2010})}\BibitemShut {NoStop}%
\bibitem [{\citenamefont {Scaffidi}\ \emph {et~al.}(2014)\citenamefont
  {Scaffidi}, \citenamefont {Romers},\ and\ \citenamefont {Simon}}]{sca14}%
  \BibitemOpen
  \bibfield  {author} {\bibinfo {author} {\bibfnamefont {T.}~\bibnamefont
  {Scaffidi}}, \bibinfo {author} {\bibfnamefont {J.~C.}\ \bibnamefont
  {Romers}},\ and\ \bibinfo {author} {\bibfnamefont {S.~H.}\ \bibnamefont
  {Simon}},\ }\bibfield  {title} {\bibinfo {title} {{Pairing symmetry and
  dominant band in Sr$_2$RuO$_4$}},\ }\href
  {https://doi.org/10.1103/PhysRevB.89.220510} {\bibfield  {journal} {\bibinfo
  {journal} {Phys. Rev. B}\ }\textbf {\bibinfo {volume} {89}},\ \bibinfo
  {pages} {220510} (\bibinfo {year} {2014})}\BibitemShut {NoStop}%
\bibitem [{\citenamefont {Pustogow}\ \emph {et~al.}(2019)\citenamefont
  {Pustogow}, \citenamefont {Luo}, \citenamefont {Chronister}, \citenamefont
  {Su}, \citenamefont {Sokolov}, \citenamefont {Jerzembeck}, \citenamefont
  {Mackenzie}, \citenamefont {Hicks}, \citenamefont {Kikugawa}, \citenamefont
  {Raghu}, \citenamefont {Bauer},\ and\ \citenamefont {Brown}}]{pus19}%
  \BibitemOpen
  \bibfield  {author} {\bibinfo {author} {\bibfnamefont {A.}~\bibnamefont
  {Pustogow}}, \bibinfo {author} {\bibfnamefont {Y.}~\bibnamefont {Luo}},
  \bibinfo {author} {\bibfnamefont {A.}~\bibnamefont {Chronister}}, \bibinfo
  {author} {\bibfnamefont {Y.-S.}\ \bibnamefont {Su}}, \bibinfo {author}
  {\bibfnamefont {D.~A.}\ \bibnamefont {Sokolov}}, \bibinfo {author}
  {\bibfnamefont {F.}~\bibnamefont {Jerzembeck}}, \bibinfo {author}
  {\bibfnamefont {A.~P.}\ \bibnamefont {Mackenzie}}, \bibinfo {author}
  {\bibfnamefont {C.~W.}\ \bibnamefont {Hicks}}, \bibinfo {author}
  {\bibfnamefont {N.}~\bibnamefont {Kikugawa}}, \bibinfo {author}
  {\bibfnamefont {S.}~\bibnamefont {Raghu}}, \bibinfo {author} {\bibfnamefont
  {E.~D.}\ \bibnamefont {Bauer}},\ and\ \bibinfo {author} {\bibfnamefont
  {S.~E.}\ \bibnamefont {Brown}},\ }\bibfield  {title} {\bibinfo {title}
  {{Constraints on the superconducting order parameter in Sr$_2$RuO$_4$ from
  Oxygen-17 nuclear magnetic resonance}},\ }\href
  {https://doi.org/10.1038/s41586-019-1596-2} {\bibfield  {journal} {\bibinfo
  {journal} {Nature}\ }\textbf {\bibinfo {volume} {574}},\ \bibinfo {pages}
  {72} (\bibinfo {year} {2019})}\BibitemShut {NoStop}%
\bibitem [{\citenamefont {Ngampruetikorn}\ and\ \citenamefont
  {Sauls}(2020)}]{nga20}%
  \BibitemOpen
  \bibfield  {author} {\bibinfo {author} {\bibfnamefont {V.}~\bibnamefont
  {Ngampruetikorn}}\ and\ \bibinfo {author} {\bibfnamefont {J.~A.}\
  \bibnamefont {Sauls}},\ }\bibfield  {title} {\bibinfo {title}
  {{Impurity-induced Anomalous Thermal Hall Effect in Chiral
  Superconductors}},\ }\href {https://doi.org/10.1103/PhysRevLett.124.157002}
  {\bibfield  {journal} {\bibinfo  {journal} {Phys. Rev. Lett.}\ }\textbf
  {\bibinfo {volume} {124}},\ \bibinfo {pages} {157002} (\bibinfo {year}
  {2020})},\ \Eprint {https://arxiv.org/abs/2004.06551} {2004.06551}
  \BibitemShut {NoStop}%
\bibitem [{\citenamefont {Kivelson}\ \emph {et~al.}(2020)\citenamefont
  {Kivelson}, \citenamefont {Yuan}, \citenamefont {Ramshaw},\ and\
  \citenamefont {Thomale}}]{kiv20}%
  \BibitemOpen
  \bibfield  {author} {\bibinfo {author} {\bibfnamefont {S.~A.}\ \bibnamefont
  {Kivelson}}, \bibinfo {author} {\bibfnamefont {A.~C.}\ \bibnamefont {Yuan}},
  \bibinfo {author} {\bibfnamefont {B.}~\bibnamefont {Ramshaw}},\ and\ \bibinfo
  {author} {\bibfnamefont {R.}~\bibnamefont {Thomale}},\ }\bibfield  {title}
  {\bibinfo {title} {{A proposal for reconciling diverse experiments on the
  superconducting state in {\sro}}},\ }\bibfield  {journal} {\bibinfo
  {journal} {NJP Quantum Materials}\ }\textbf {\bibinfo {volume} {5}},\ \href
  {https://doi.org/10.1038/s41535-020-0245-1} {10.1038/s41535-020-0245-1}
  (\bibinfo {year} {2020})\BibitemShut {NoStop}%
\bibitem [{\citenamefont {Leggett}\ and\ \citenamefont {Liu}(2021)}]{leg21}%
  \BibitemOpen
  \bibfield  {author} {\bibinfo {author} {\bibfnamefont {A.~J.}\ \bibnamefont
  {Leggett}}\ and\ \bibinfo {author} {\bibfnamefont {Y.}~\bibnamefont {Liu}},\
  }\bibfield  {title} {\bibinfo {title} {{Symmetry Properties of
  Superconducting Order Parameter in {\sro}}},\ }\href
  {https://doi.org/10.1007/s10948-020-05717-6} {\bibfield  {journal} {\bibinfo
  {journal} {Journal of Superconductivity and Novel Magnetism}\ }\textbf
  {\bibinfo {volume} {34}},\ \bibinfo {pages} {1647} (\bibinfo {year}
  {2021})}\BibitemShut {NoStop}%
\bibitem [{\citenamefont {Sauls}(1994)}]{sau94}%
  \BibitemOpen
  \bibfield  {author} {\bibinfo {author} {\bibfnamefont {J.~A.}\ \bibnamefont
  {Sauls}},\ }\bibfield  {title} {\bibinfo {title} {The order parameter for the
  superconducting phases of {UPt$_3$}},\ }\href
  {https://arxiv.org/abs/1812.09984} {\bibfield  {journal} {\bibinfo  {journal}
  {Adv. Phys.}\ }\textbf {\bibinfo {volume} {43}},\ \bibinfo {pages} {113}
  (\bibinfo {year} {1994})}\BibitemShut {NoStop}%
\bibitem [{\citenamefont {Strand}\ \emph {et~al.}(2009)\citenamefont {Strand},
  \citenamefont {Van~Harlingen}, \citenamefont {Kycia},\ and\ \citenamefont
  {Halperin}}]{str09}%
  \BibitemOpen
  \bibfield  {author} {\bibinfo {author} {\bibfnamefont {J.~D.}\ \bibnamefont
  {Strand}}, \bibinfo {author} {\bibfnamefont {D.~J.}\ \bibnamefont
  {Van~Harlingen}}, \bibinfo {author} {\bibfnamefont {J.~B.}\ \bibnamefont
  {Kycia}},\ and\ \bibinfo {author} {\bibfnamefont {W.~P.}\ \bibnamefont
  {Halperin}},\ }\bibfield  {title} {\bibinfo {title} {{Evidence for Complex
  Superconducting Order Parameter Symmetry in the Low-Temperature Phase of
  {\upt} from Josephson Interferometry}},\ }\href
  {https://doi.org/10.1103/PhysRevLett.103.197002} {\bibfield  {journal}
  {\bibinfo  {journal} {Phys. Rev. Lett.}\ }\textbf {\bibinfo {volume} {103}},\
  \bibinfo {pages} {197002} (\bibinfo {year} {2009})}\BibitemShut {NoStop}%
\bibitem [{\citenamefont {Schemm}\ \emph {et~al.}(2014)\citenamefont {Schemm},
  \citenamefont {Gannon}, \citenamefont {Wishne}, \citenamefont {Halperin},\
  and\ \citenamefont {Kapitulnik}}]{sch14}%
  \BibitemOpen
  \bibfield  {author} {\bibinfo {author} {\bibfnamefont {E.~R.}\ \bibnamefont
  {Schemm}}, \bibinfo {author} {\bibfnamefont {W.~J.}\ \bibnamefont {Gannon}},
  \bibinfo {author} {\bibfnamefont {C.~M.}\ \bibnamefont {Wishne}}, \bibinfo
  {author} {\bibfnamefont {W.~P.}\ \bibnamefont {Halperin}},\ and\ \bibinfo
  {author} {\bibfnamefont {A.}~\bibnamefont {Kapitulnik}},\ }\bibfield  {title}
  {\bibinfo {title} {{Observation of Broken Time-reversal Symmetry in the
  Heavy-Fermion Superconductor \upt}},\ }\href
  {https://doi.org/10.1126/science.1248552} {\bibfield  {journal} {\bibinfo
  {journal} {Science}\ }\textbf {\bibinfo {volume} {345}},\ \bibinfo {pages}
  {190} (\bibinfo {year} {2014})}\BibitemShut {NoStop}%
\bibitem [{\citenamefont {Goswami}\ and\ \citenamefont
  {Nevidomskyy}(2015)}]{gos15}%
  \BibitemOpen
  \bibfield  {author} {\bibinfo {author} {\bibfnamefont {P.}~\bibnamefont
  {Goswami}}\ and\ \bibinfo {author} {\bibfnamefont {A.~H.}\ \bibnamefont
  {Nevidomskyy}},\ }\bibfield  {title} {\bibinfo {title} {{Topological Weyl
  superconductor to diffusive thermal Hall metal crossover in the B phase of
  UPt$_3$}},\ }\href {https://doi.org/10.1103/PhysRevB.92.214504} {\bibfield
  {journal} {\bibinfo  {journal} {Phys. Rev. B}\ }\textbf {\bibinfo {volume}
  {92}},\ \bibinfo {pages} {214504} (\bibinfo {year} {2015})}\BibitemShut
  {NoStop}%
\bibitem [{\citenamefont {Agterberg}\ and\ \citenamefont
  {Tsunetsugu}(2008)}]{agt08}%
  \BibitemOpen
  \bibfield  {author} {\bibinfo {author} {\bibfnamefont {D.~F.}\ \bibnamefont
  {Agterberg}}\ and\ \bibinfo {author} {\bibfnamefont {H.}~\bibnamefont
  {Tsunetsugu}},\ }\bibfield  {title} {\bibinfo {title} {Dislocations and
  vortices in pair-density-wave superconductors},\ }\href
  {https://doi.org/10.1038/nphys999} {\bibfield  {journal} {\bibinfo  {journal}
  {Nature Physics}\ }\textbf {\bibinfo {volume} {4}},\ \bibinfo {pages} {639}
  (\bibinfo {year} {2008})}\BibitemShut {NoStop}%
\bibitem [{\citenamefont {Holmvall}\ \emph {et~al.}(2018)\citenamefont
  {Holmvall}, \citenamefont {Vorontsov}, \citenamefont {Fogelstr\"om},\ and\
  \citenamefont {L\"ofwander}}]{hol18a}%
  \BibitemOpen
  \bibfield  {author} {\bibinfo {author} {\bibfnamefont {P.}~\bibnamefont
  {Holmvall}}, \bibinfo {author} {\bibfnamefont {A.~B.}\ \bibnamefont
  {Vorontsov}}, \bibinfo {author} {\bibfnamefont {M.}~\bibnamefont
  {Fogelstr\"om}},\ and\ \bibinfo {author} {\bibfnamefont {T.}~\bibnamefont
  {L\"ofwander}},\ }\bibfield  {title} {\bibinfo {title} {{Broken translational
  symmetry at edges of high-temperature superconductors}},\ }\href
  {https://doi.org/10.1038/s41467-018-04531-y} {\bibfield  {journal} {\bibinfo
  {journal} {Nature Communications}\ }\textbf {\bibinfo {volume} {9}},\
  \bibinfo {pages} {2190} (\bibinfo {year} {2018})}\BibitemShut {NoStop}%
\bibitem [{\citenamefont {Jiang}\ and\ \citenamefont {Barlas}(2023)}]{jia23}%
  \BibitemOpen
  \bibfield  {author} {\bibinfo {author} {\bibfnamefont {G.}~\bibnamefont
  {Jiang}}\ and\ \bibinfo {author} {\bibfnamefont {Y.}~\bibnamefont {Barlas}},\
  }\bibfield  {title} {\bibinfo {title} {Pair {Density} {Waves} from {Local}
  {Band} {Geometry}},\ }\href {https://doi.org/10.1103/PhysRevLett.131.016002}
  {\bibfield  {journal} {\bibinfo  {journal} {Phys. Rev. Lett.}\ }\textbf
  {\bibinfo {volume} {131}},\ \bibinfo {pages} {016002} (\bibinfo {year}
  {2023})}\BibitemShut {NoStop}%
\bibitem [{\citenamefont {Barkman}\ \emph {et~al.}(2019)\citenamefont
  {Barkman}, \citenamefont {Zyuzin},\ and\ \citenamefont {Babaev}}]{bar23}%
  \BibitemOpen
  \bibfield  {author} {\bibinfo {author} {\bibfnamefont {M.}~\bibnamefont
  {Barkman}}, \bibinfo {author} {\bibfnamefont {A.~A.}\ \bibnamefont
  {Zyuzin}},\ and\ \bibinfo {author} {\bibfnamefont {E.}~\bibnamefont
  {Babaev}},\ }\bibfield  {title} {\bibinfo {title} {{Antichiral and
  nematicity-wave superconductivity}},\ }\href
  {https://doi.org/10.1103/PhysRevB.99.220508} {\bibfield  {journal} {\bibinfo
  {journal} {Phys. Rev. B}\ }\textbf {\bibinfo {volume} {99}},\ \bibinfo
  {pages} {220508} (\bibinfo {year} {2019})}\BibitemShut {NoStop}%
\bibitem [{\citenamefont {Santos}\ \emph {et~al.}(2019)\citenamefont {Santos},
  \citenamefont {Wang},\ and\ \citenamefont {Fradkin}}]{san19}%
  \BibitemOpen
  \bibfield  {author} {\bibinfo {author} {\bibfnamefont {L.~H.}\ \bibnamefont
  {Santos}}, \bibinfo {author} {\bibfnamefont {Y.}~\bibnamefont {Wang}},\ and\
  \bibinfo {author} {\bibfnamefont {E.}~\bibnamefont {Fradkin}},\ }\bibfield
  {title} {\bibinfo {title} {Pair-{Density}-{Wave} {Order} and {Paired}
  {Fractional} {Quantum} {Hall} {Fluids}},\ }\href
  {https://doi.org/10.1103/PhysRevX.9.021047} {\bibfield  {journal} {\bibinfo
  {journal} {Phys. Rev. X}\ }\textbf {\bibinfo {volume} {9}},\ \bibinfo {pages}
  {021047} (\bibinfo {year} {2019})}\BibitemShut {NoStop}%
\bibitem [{\citenamefont {Volovik}(2016)}]{vol16}%
  \BibitemOpen
  \bibfield  {author} {\bibinfo {author} {\bibfnamefont {G.~E.}\ \bibnamefont
  {Volovik}},\ }\bibfield  {title} {\bibinfo {title} {{Topological
  Superfluids}},\ }\href@noop {} {\bibfield  {journal} {\bibinfo  {journal}
  {arXiv}\ }\textbf {\bibinfo {volume} {1602.02595}},\ \bibinfo {pages} {1}
  (\bibinfo {year} {2016})},\ \Eprint {https://arxiv.org/abs/arxiv:1602.02595}
  {arxiv:1602.02595} \BibitemShut {NoStop}%
\bibitem [{\citenamefont {Mizushima}\ \emph {et~al.}(2016)\citenamefont
  {Mizushima}, \citenamefont {Tsutsumi}, \citenamefont {Kawakami},
  \citenamefont {Sato}, \citenamefont {Ichioka},\ and\ \citenamefont
  {Machida}}]{miz16}%
  \BibitemOpen
  \bibfield  {author} {\bibinfo {author} {\bibfnamefont {T.}~\bibnamefont
  {Mizushima}}, \bibinfo {author} {\bibfnamefont {Y.}~\bibnamefont {Tsutsumi}},
  \bibinfo {author} {\bibfnamefont {T.}~\bibnamefont {Kawakami}}, \bibinfo
  {author} {\bibfnamefont {M.}~\bibnamefont {Sato}}, \bibinfo {author}
  {\bibfnamefont {M.}~\bibnamefont {Ichioka}},\ and\ \bibinfo {author}
  {\bibfnamefont {K.}~\bibnamefont {Machida}},\ }\bibfield  {title} {\bibinfo
  {title} {{Symmetry Protected Topological Superfluids and Superconductors -
  From the Basics to $^3$He}},\ }\href {https://doi.org/10.7566/JPSJ.85.022001}
  {\bibfield  {journal} {\bibinfo  {journal} {J. Phys. Soc. Jpn.}\ }\textbf
  {\bibinfo {volume} {85}},\ \bibinfo {pages} {022001} (\bibinfo {year}
  {2016})}\BibitemShut {NoStop}%
\bibitem [{\citenamefont {Wiman}\ and\ \citenamefont {Sauls}(2013)}]{wim13}%
  \BibitemOpen
  \bibfield  {author} {\bibinfo {author} {\bibfnamefont {J.~J.}\ \bibnamefont
  {Wiman}}\ and\ \bibinfo {author} {\bibfnamefont {J.~A.}\ \bibnamefont
  {Sauls}},\ }\bibfield  {title} {\bibinfo {title} {{Superfluid phases of
  $^3$He in a periodic confined geometry}},\ }\href
  {https://doi.org/10.1007/s10909-013-0924-4} {\bibfield  {journal} {\bibinfo
  {journal} {J. Low Temp. Phys.}\ }\textbf {\bibinfo {volume} {174}},\ \bibinfo
  {pages} {1} (\bibinfo {year} {2013})}\BibitemShut {NoStop}%
\bibitem [{\citenamefont {Wiman}\ and\ \citenamefont {Sauls}(2015)}]{wim15}%
  \BibitemOpen
  \bibfield  {author} {\bibinfo {author} {\bibfnamefont {J.~J.}\ \bibnamefont
  {Wiman}}\ and\ \bibinfo {author} {\bibfnamefont {J.~A.}\ \bibnamefont
  {Sauls}},\ }\bibfield  {title} {\bibinfo {title} {{Superfluid phases of
  $^3$He in nanoscale channels}},\ }\href
  {https://doi.org/10.1103/PhysRevB.92.144515} {\bibfield  {journal} {\bibinfo
  {journal} {Phys. Rev. B}\ }\textbf {\bibinfo {volume} {92}},\ \bibinfo
  {pages} {144515} (\bibinfo {year} {2015})}\BibitemShut {NoStop}%
\bibitem [{\citenamefont {Wiman}\ and\ \citenamefont {Sauls}(2016)}]{wim16}%
  \BibitemOpen
  \bibfield  {author} {\bibinfo {author} {\bibfnamefont {J.~J.}\ \bibnamefont
  {Wiman}}\ and\ \bibinfo {author} {\bibfnamefont {J.~A.}\ \bibnamefont
  {Sauls}},\ }\bibfield  {title} {\bibinfo {title} {{Strong-Coupling and the
  Stripe Phase of $^3$He}},\ }\href {https://doi.org/10.1007/s10909-016-1632-7}
  {\bibfield  {journal} {\bibinfo  {journal} {J. Low Temp. Phys.}\ }\textbf
  {\bibinfo {volume} {184}},\ \bibinfo {pages} {1054} (\bibinfo {year}
  {2016})}\BibitemShut {NoStop}%
\bibitem [{\citenamefont {Wiman}\ and\ \citenamefont {Sauls}(2018)}]{wim18}%
  \BibitemOpen
  \bibfield  {author} {\bibinfo {author} {\bibfnamefont {J.~J.}\ \bibnamefont
  {Wiman}}\ and\ \bibinfo {author} {\bibfnamefont {J.~A.}\ \bibnamefont
  {Sauls}},\ }\bibfield  {title} {\bibinfo {title} {{Spontaneous Helical Order
  of a Chiral $p$-Wave Superfluid Confined in Nano-Scale Channels}},\ }\href
  {https://doi.org/10.1103/PhysRevLett.121.045301} {\bibfield  {journal}
  {\bibinfo  {journal} {Phys. Rev. Lett.}\ }\textbf {\bibinfo {volume} {121}},\
  \bibinfo {pages} {045301} (\bibinfo {year} {2018})},\ \Eprint
  {https://arxiv.org/abs/1802.08719} {1802.08719} \BibitemShut {NoStop}%
\bibitem [{\citenamefont {Zheng}\ \emph {et~al.}(2017)\citenamefont {Zheng},
  \citenamefont {Jiang}, \citenamefont {Barquist}, \citenamefont {Lee},\ and\
  \citenamefont {Chan}}]{zhe17}%
  \BibitemOpen
  \bibfield  {author} {\bibinfo {author} {\bibfnamefont {P.}~\bibnamefont
  {Zheng}}, \bibinfo {author} {\bibfnamefont {W.~G.}\ \bibnamefont {Jiang}},
  \bibinfo {author} {\bibfnamefont {C.~S.}\ \bibnamefont {Barquist}}, \bibinfo
  {author} {\bibfnamefont {Y.}~\bibnamefont {Lee}},\ and\ \bibinfo {author}
  {\bibfnamefont {H.~B.}\ \bibnamefont {Chan}},\ }\bibfield  {title} {\bibinfo
  {title} {{Anomalous Resonance Frequency Shift of a Microelectromechanical
  Oscillator in Superfluid \Heb}},\ }\href
  {https://doi.org/10.1007/s10909-017-1752-8} {\bibfield  {journal} {\bibinfo
  {journal} {J. Low Temp. Phys.}\ }\textbf {\bibinfo {volume} {187}},\ \bibinfo
  {pages} {309} (\bibinfo {year} {2017})}\BibitemShut {NoStop}%
\bibitem [{\citenamefont {Levitin}\ \emph {et~al.}(2019)\citenamefont
  {Levitin}, \citenamefont {Yager}, \citenamefont {Sumner}, \citenamefont
  {Cowan}, \citenamefont {Casey}, \citenamefont {Saunders}, \citenamefont
  {Zhelev}, \citenamefont {Bennett},\ and\ \citenamefont {Parpia}}]{lev19}%
  \BibitemOpen
  \bibfield  {author} {\bibinfo {author} {\bibfnamefont {L.~V.}\ \bibnamefont
  {Levitin}}, \bibinfo {author} {\bibfnamefont {B.}~\bibnamefont {Yager}},
  \bibinfo {author} {\bibfnamefont {L.}~\bibnamefont {Sumner}}, \bibinfo
  {author} {\bibfnamefont {B.}~\bibnamefont {Cowan}}, \bibinfo {author}
  {\bibfnamefont {A.~J.}\ \bibnamefont {Casey}}, \bibinfo {author}
  {\bibfnamefont {J.}~\bibnamefont {Saunders}}, \bibinfo {author}
  {\bibfnamefont {N.}~\bibnamefont {Zhelev}}, \bibinfo {author} {\bibfnamefont
  {R.~G.}\ \bibnamefont {Bennett}},\ and\ \bibinfo {author} {\bibfnamefont
  {J.~M.}\ \bibnamefont {Parpia}},\ }\bibfield  {title} {\bibinfo {title}
  {{Evidence for a Spatially Modulated Superfluid Phase of {\He} under
  Confinement}},\ }\href {https://doi.org/10.1103/PhysRevLett.122.085301}
  {\bibfield  {journal} {\bibinfo  {journal} {Phys. Rev. Lett.}\ }\textbf
  {\bibinfo {volume} {122}},\ \bibinfo {pages} {085301} (\bibinfo {year}
  {2019})}\BibitemShut {NoStop}%
\bibitem [{\citenamefont {Heikkinen}\ \emph {et~al.}(2021)\citenamefont
  {Heikkinen}, \citenamefont {Casey}, \citenamefont {Levitin}, \citenamefont
  {Rojas}, \citenamefont {Vorontsov}, \citenamefont {Sharma}, \citenamefont
  {Zhelev}, \citenamefont {Parpia},\ and\ \citenamefont {Saunders}}]{hei21}%
  \BibitemOpen
  \bibfield  {author} {\bibinfo {author} {\bibfnamefont {P.~J.}\ \bibnamefont
  {Heikkinen}}, \bibinfo {author} {\bibfnamefont {A.}~\bibnamefont {Casey}},
  \bibinfo {author} {\bibfnamefont {L.~V.}\ \bibnamefont {Levitin}}, \bibinfo
  {author} {\bibfnamefont {X.}~\bibnamefont {Rojas}}, \bibinfo {author}
  {\bibfnamefont {A.}~\bibnamefont {Vorontsov}}, \bibinfo {author}
  {\bibfnamefont {P.}~\bibnamefont {Sharma}}, \bibinfo {author} {\bibfnamefont
  {N.}~\bibnamefont {Zhelev}}, \bibinfo {author} {\bibfnamefont {J.~M.}\
  \bibnamefont {Parpia}},\ and\ \bibinfo {author} {\bibfnamefont
  {J.}~\bibnamefont {Saunders}},\ }\bibfield  {title} {\bibinfo {title}
  {{Fragility of surface states in topological superfluid $^3$He}},\ }\href
  {https://doi.org/10.1038/s41467-021-21831-y} {\bibfield  {journal} {\bibinfo
  {journal} {Nat. Comm.}\ }\textbf {\bibinfo {volume} {12}},\ \bibinfo {pages}
  {1574} (\bibinfo {year} {2021})}\BibitemShut {NoStop}%
\bibitem [{\citenamefont {Read}\ and\ \citenamefont {Green}(2000)}]{rea00}%
  \BibitemOpen
  \bibfield  {author} {\bibinfo {author} {\bibfnamefont {N.}~\bibnamefont
  {Read}}\ and\ \bibinfo {author} {\bibfnamefont {D.}~\bibnamefont {Green}},\
  }\bibfield  {title} {\bibinfo {title} {{Paired states of Fermions in two
  dimensions with breaking of parity and time-reversal symmetries and the
  fractional quantum Hall effect}},\ }\href
  {https://doi.org/10.1103/PhysRevB.61.10267} {\bibfield  {journal} {\bibinfo
  {journal} {Phys. Rev. B}\ }\textbf {\bibinfo {volume} {61}},\ \bibinfo
  {pages} {10267} (\bibinfo {year} {2000})}\BibitemShut {NoStop}%
\bibitem [{\citenamefont {Volovik}(1988)}]{vol88}%
  \BibitemOpen
  \bibfield  {author} {\bibinfo {author} {\bibfnamefont {G.~E.}\ \bibnamefont
  {Volovik}},\ }\bibfield  {title} {\bibinfo {title} {{An analog of the quantum
  Hall effect in a superfluid $^3$He film}},\ }\href@noop {} {\bibfield
  {journal} {\bibinfo  {journal} {Sov. Phys. JETP}\ }\textbf {\bibinfo {volume}
  {67}},\ \bibinfo {pages} {1804} (\bibinfo {year} {1988})}\BibitemShut
  {NoStop}%
\bibitem [{\citenamefont {Volovik}(1992{\natexlab{a}})}]{vol92}%
  \BibitemOpen
  \bibfield  {author} {\bibinfo {author} {\bibfnamefont {G.~E.}\ \bibnamefont
  {Volovik}},\ }\bibfield  {title} {\bibinfo {title} {{Quantum Hall state and
  chiral edge state in thin $^3$He-A film}},\ }\href@noop {} {\bibfield
  {journal} {\bibinfo  {journal} {JETP Lett.}\ }\textbf {\bibinfo {volume}
  {55}},\ \bibinfo {pages} {368} (\bibinfo {year} {1992}{\natexlab{a}})},\
  \bibinfo {note} {[Pis'ma ZhETF, 55, 363 (1992)]}\BibitemShut {NoStop}%
\bibitem [{\citenamefont {Volovik}(2003)}]{volovik03}%
  \BibitemOpen
  \bibfield  {author} {\bibinfo {author} {\bibfnamefont {G.~E.}\ \bibnamefont
  {Volovik}},\ }\href@noop {} {\emph {\bibinfo {title} {{The Universe in a
  Helium Droplet}}}}\ (\bibinfo  {publisher} {Clarendon Press},\ \bibinfo
  {address} {Clarendon, UK},\ \bibinfo {year} {2003})\BibitemShut {NoStop}%
\bibitem [{\citenamefont {McClure}\ and\ \citenamefont {Takagi}(1979)}]{mcc79}%
  \BibitemOpen
  \bibfield  {author} {\bibinfo {author} {\bibfnamefont {M.~G.}\ \bibnamefont
  {McClure}}\ and\ \bibinfo {author} {\bibfnamefont {S.}~\bibnamefont
  {Takagi}},\ }\bibfield  {title} {\bibinfo {title} {{Angular Momentum of
  Anisotropic Superfluids}},\ }\href
  {https://doi.org/10.1103/PhysRevLett.43.596} {\bibfield  {journal} {\bibinfo
  {journal} {Phys. Rev. Lett.}\ }\textbf {\bibinfo {volume} {43}},\ \bibinfo
  {pages} {596} (\bibinfo {year} {1979})}\BibitemShut {NoStop}%
\bibitem [{\citenamefont {Tsutsumi}(2014)}]{tsu14}%
  \BibitemOpen
  \bibfield  {author} {\bibinfo {author} {\bibfnamefont {Y.}~\bibnamefont
  {Tsutsumi}},\ }\bibfield  {title} {\bibinfo {title} {{Mass Current at a
  Domain Wall in Superfluid $^3$He A-Phase}},\ }\href
  {https://doi.org/10.1007/s10909-013-0968-5} {\bibfield  {journal} {\bibinfo
  {journal} {Journal of Low Temperature Physics}\ }\textbf {\bibinfo {volume}
  {175}},\ \bibinfo {pages} {51} (\bibinfo {year} {2014})}\BibitemShut
  {NoStop}%
\bibitem [{\citenamefont {Ikegami}\ \emph
  {et~al.}(2013{\natexlab{b}})\citenamefont {Ikegami}, \citenamefont {Chung},\
  and\ \citenamefont {Kono}}]{ike13b}%
  \BibitemOpen
  \bibfield  {author} {\bibinfo {author} {\bibfnamefont {H.}~\bibnamefont
  {Ikegami}}, \bibinfo {author} {\bibfnamefont {S.~B.}\ \bibnamefont {Chung}},\
  and\ \bibinfo {author} {\bibfnamefont {K.}~\bibnamefont {Kono}},\ }\bibfield
  {title} {\bibinfo {title} {{Mobility of Ions Trapped Below a Free Surface of
  Superfluid $^3${H}e}},\ }\href {https://doi.org/10.7566/JPSJ.82.124607}
  {\bibfield  {journal} {\bibinfo  {journal} {J. Phys. Soc. Jpn.}\ }\textbf
  {\bibinfo {volume} {82}},\ \bibinfo {pages} {124607} (\bibinfo {year}
  {2013}{\natexlab{b}})}\BibitemShut {NoStop}%
\bibitem [{\citenamefont {Ikegami}\ \emph {et~al.}(2015)\citenamefont
  {Ikegami}, \citenamefont {Tsutsumi},\ and\ \citenamefont {Kono}}]{ike15}%
  \BibitemOpen
  \bibfield  {author} {\bibinfo {author} {\bibfnamefont {H.}~\bibnamefont
  {Ikegami}}, \bibinfo {author} {\bibfnamefont {Y.}~\bibnamefont {Tsutsumi}},\
  and\ \bibinfo {author} {\bibfnamefont {K.}~\bibnamefont {Kono}},\ }\bibfield
  {title} {\bibinfo {title} {{Observation of Intrinsic Magnus Force and Direct
  Detection of Chirality in Superfluid $^3$He-A}},\ }\href
  {https://doi.org/10.7566/JPSJ.84.044602} {\bibfield  {journal} {\bibinfo
  {journal} {J. Phys. Soc. Jpn.}\ }\textbf {\bibinfo {volume} {84}},\ \bibinfo
  {pages} {044602} (\bibinfo {year} {2015})}\BibitemShut {NoStop}%
\bibitem [{\citenamefont {Kurkij\"arvi}\ and\ \citenamefont
  {Rainer}(1990)}]{kur90}%
  \BibitemOpen
  \bibfield  {author} {\bibinfo {author} {\bibfnamefont {J.}~\bibnamefont
  {Kurkij\"arvi}}\ and\ \bibinfo {author} {\bibfnamefont {D.}~\bibnamefont
  {Rainer}},\ }\bibfield  {title} {\bibinfo {title} {{Andreev Scattering in
  Superfluid $^3He$}},\ }in\ \href@noop {} {\emph {\bibinfo {booktitle} {Helium
  Three}}},\ \bibinfo {editor} {edited by\ \bibinfo {editor} {\bibnamefont
  {edited~by W.~P.~Halperin}}\ and\ \bibinfo {editor} {\bibfnamefont {L.~P.}\
  \bibnamefont {Pitaevskii}}}\ (\bibinfo  {publisher} {Elsevier Science
  Publishers, Amsterdam},\ \bibinfo {year} {1990})\ pp.\ \bibinfo {pages}
  {313--352}\BibitemShut {NoStop}%
\bibitem [{\citenamefont {Volovik}(1992{\natexlab{b}})}]{volovik92}%
  \BibitemOpen
  \bibfield  {author} {\bibinfo {author} {\bibfnamefont {G.~E.}\ \bibnamefont
  {Volovik}},\ }\href@noop {} {\emph {\bibinfo {title} {{Exotic Properties of
  Superfluid $^3$He}}}}\ (\bibinfo  {publisher} {World Scientific},\ \bibinfo
  {address} {Singapore},\ \bibinfo {year} {1992})\BibitemShut {NoStop}%
\bibitem [{\citenamefont {Andreev}(1964)}]{and64}%
  \BibitemOpen
  \bibfield  {author} {\bibinfo {author} {\bibfnamefont {A.~F.}\ \bibnamefont
  {Andreev}},\ }\bibfield  {title} {\bibinfo {title} {{Thermal Conductivity of
  the Intermediate State in Superconductors}},\ }\href@noop {} {\bibfield
  {journal} {\bibinfo  {journal} {Sov. Phys. JETP}\ }\textbf {\bibinfo {volume}
  {19}},\ \bibinfo {pages} {1228} (\bibinfo {year} {1964})}\BibitemShut
  {NoStop}%
\bibitem [{\citenamefont {Sauls}(2018)}]{sau18}%
  \BibitemOpen
  \bibfield  {author} {\bibinfo {author} {\bibfnamefont {J.~A.}\ \bibnamefont
  {Sauls}},\ }\bibfield  {title} {\bibinfo {title} {{Andreev Bound States and
  Their Signatures}},\ }\bibfield  {journal} {\bibinfo  {journal} {Phil. Trans.
  Roy. Soc. A}\ }\textbf {\bibinfo {volume} {376}},\ \href
  {https://doi.org/10.1098/rsta.2018.0140} {10.1098/rsta.2018.0140} (\bibinfo
  {year} {2018}),\ \Eprint {https://arxiv.org/abs/1805.11069} {1805.11069}
  \BibitemShut {NoStop}%
\bibitem [{\citenamefont {Stone}\ and\ \citenamefont {Roy}(2004)}]{sto04}%
  \BibitemOpen
  \bibfield  {author} {\bibinfo {author} {\bibfnamefont {M.}~\bibnamefont
  {Stone}}\ and\ \bibinfo {author} {\bibfnamefont {R.}~\bibnamefont {Roy}},\
  }\bibfield  {title} {\bibinfo {title} {{Edge modes, edge currents, and gauge
  invariance in {$p_x+ip_y$} superfluids and superconductors}},\ }\href
  {https://doi.org/10.1103/PhysRevB.69.184511} {\bibfield  {journal} {\bibinfo
  {journal} {Phys. Rev. B}\ }\textbf {\bibinfo {volume} {69}},\ \bibinfo
  {pages} {184511} (\bibinfo {year} {2004})}\BibitemShut {NoStop}%
\bibitem [{\citenamefont {Volovik}\ and\ \citenamefont
  {Mineev}(1981)}]{vol81b}%
  \BibitemOpen
  \bibfield  {author} {\bibinfo {author} {\bibfnamefont {G.~E.}\ \bibnamefont
  {Volovik}}\ and\ \bibinfo {author} {\bibfnamefont {V.~P.}\ \bibnamefont
  {Mineev}},\ }\bibfield  {title} {\bibinfo {title} {{Orbital Angular Momentum
  and Orbital Dynamics: $^3$He-A and the Bose Liquid}},\ }\href@noop {}
  {\bibfield  {journal} {\bibinfo  {journal} {{Sov. Phys. JETP}}\ }\textbf
  {\bibinfo {volume} {54}},\ \bibinfo {pages} {524} (\bibinfo {year}
  {1981})}\BibitemShut {NoStop}%
\bibitem [{\citenamefont {Balatsky}\ \emph {et~al.}(1986)\citenamefont
  {Balatsky}, \citenamefont {Volovik},\ and\ \citenamefont {Konyshev}}]{bal86}%
  \BibitemOpen
  \bibfield  {author} {\bibinfo {author} {\bibfnamefont {A.~V.}\ \bibnamefont
  {Balatsky}}, \bibinfo {author} {\bibfnamefont {G.}~\bibnamefont {Volovik}},\
  and\ \bibinfo {author} {\bibfnamefont {V.}~\bibnamefont {Konyshev}},\
  }\bibfield  {title} {\bibinfo {title} {{On the chiral anomaly in superfluid
  \Hea}},\ }\href@noop {} {\bibfield  {journal} {\bibinfo  {journal} {Sov.
  Phys. JETP}\ }\textbf {\bibinfo {volume} {63}},\ \bibinfo {pages} {1194}
  (\bibinfo {year} {1986})}\BibitemShut {NoStop}%
\bibitem [{\citenamefont {Balatsky}\ and\ \citenamefont
  {Konyshev}(1987)}]{bal87}%
  \BibitemOpen
  \bibfield  {author} {\bibinfo {author} {\bibfnamefont {A.~V.}\ \bibnamefont
  {Balatsky}}\ and\ \bibinfo {author} {\bibfnamefont {V.}~\bibnamefont
  {Konyshev}},\ }\bibfield  {title} {\bibinfo {title} {{The anomalous
  superfluid current in $^3$He-A and the index theorem}},\ }\href@noop {}
  {\bibfield  {journal} {\bibinfo  {journal} {Sov. Phys. JETP}\ }\textbf
  {\bibinfo {volume} {65}},\ \bibinfo {pages} {474} (\bibinfo {year}
  {1987})}\BibitemShut {NoStop}%
\bibitem [{\citenamefont {Kurkij\"arvi}\ \emph {et~al.}(1987)\citenamefont
  {Kurkij\"arvi}, \citenamefont {Rainer},\ and\ \citenamefont {Sauls}}]{kur87}%
  \BibitemOpen
  \bibfield  {author} {\bibinfo {author} {\bibfnamefont {J.}~\bibnamefont
  {Kurkij\"arvi}}, \bibinfo {author} {\bibfnamefont {D.}~\bibnamefont
  {Rainer}},\ and\ \bibinfo {author} {\bibfnamefont {J.~A.}\ \bibnamefont
  {Sauls}},\ }\bibfield  {title} {\bibinfo {title} {{Superfluid {\He} and Heavy
  Fermion Superconductors Near Surfaces and Interfaces}},\ }\href
  {https://doi.org/10.1139/p87-227} {\bibfield  {journal} {\bibinfo  {journal}
  {Can. J. Phys.}\ }\textbf {\bibinfo {volume} {65}},\ \bibinfo {pages} {1440}
  (\bibinfo {year} {1987})}\BibitemShut {NoStop}%
\bibitem [{\citenamefont {Millis}\ \emph {et~al.}(1988)\citenamefont {Millis},
  \citenamefont {Rainer},\ and\ \citenamefont {Sauls}}]{mil88}%
  \BibitemOpen
  \bibfield  {author} {\bibinfo {author} {\bibfnamefont {A.}~\bibnamefont
  {Millis}}, \bibinfo {author} {\bibfnamefont {D.}~\bibnamefont {Rainer}},\
  and\ \bibinfo {author} {\bibfnamefont {J.~A.}\ \bibnamefont {Sauls}},\
  }\bibfield  {title} {\bibinfo {title} {{Quasiclassical Theory of
  Superconductivity Near Magnetically Active Interfaces}},\ }\href
  {https://doi.org/10.1103/PhysRevB.38.4504} {\bibfield  {journal} {\bibinfo
  {journal} {Phys. Rev. B}\ }\textbf {\bibinfo {volume} {38}},\ \bibinfo
  {pages} {4504} (\bibinfo {year} {1988})}\BibitemShut {NoStop}%
\bibitem [{\citenamefont {Eilenberger}(1968)}]{eil68}%
  \BibitemOpen
  \bibfield  {author} {\bibinfo {author} {\bibfnamefont {G.}~\bibnamefont
  {Eilenberger}},\ }\bibfield  {title} {\bibinfo {title} {{Transformation of
  Gorkov's Equation for Type II Superconductors into Transport-Like
  Equations}},\ }\href {https://doi.org/10.1007/BF01379803} {\bibfield
  {journal} {\bibinfo  {journal} {Zeit. f. Physik}\ }\textbf {\bibinfo {volume}
  {214}},\ \bibinfo {pages} {195} (\bibinfo {year} {1968})}\BibitemShut
  {NoStop}%
\bibitem [{\citenamefont {Vorontsov}\ and\ \citenamefont
  {Sauls}(2003)}]{vor03}%
  \BibitemOpen
  \bibfield  {author} {\bibinfo {author} {\bibfnamefont {A.}~\bibnamefont
  {Vorontsov}}\ and\ \bibinfo {author} {\bibfnamefont {J.~A.}\ \bibnamefont
  {Sauls}},\ }\bibfield  {title} {\bibinfo {title} {{Thermodynamic Properties
  of Thin Films of Superfluid $^3$He-A}},\ }\href
  {https://doi.org/10.1103/PhysRevB.68.064508} {\bibfield  {journal} {\bibinfo
  {journal} {Phys. Rev. B}\ }\textbf {\bibinfo {volume} {68}},\ \bibinfo
  {pages} {064508} (\bibinfo {year} {2003})}\BibitemShut {NoStop}%
\bibitem [{\citenamefont {Vorontsov}\ and\ \citenamefont
  {Sauls}(2007)}]{vor07}%
  \BibitemOpen
  \bibfield  {author} {\bibinfo {author} {\bibfnamefont {A.~B.}\ \bibnamefont
  {Vorontsov}}\ and\ \bibinfo {author} {\bibfnamefont {J.~A.}\ \bibnamefont
  {Sauls}},\ }\bibfield  {title} {\bibinfo {title} {{Crystalline Order in
  Superfluid {$^3$He} Films}},\ }\href
  {https://doi.org/10.1103/PhysRevLett.98.045301} {\bibfield  {journal}
  {\bibinfo  {journal} {Phys. Rev. Lett.}\ }\textbf {\bibinfo {volume} {98}},\
  \bibinfo {pages} {045301} (\bibinfo {year} {2007})}\BibitemShut {NoStop}%
\bibitem [{\citenamefont {Tsutsumi}\ and\ \citenamefont
  {Machida}(2012)}]{tsu12}%
  \BibitemOpen
  \bibfield  {author} {\bibinfo {author} {\bibfnamefont {Y.}~\bibnamefont
  {Tsutsumi}}\ and\ \bibinfo {author} {\bibfnamefont {K.}~\bibnamefont
  {Machida}},\ }\bibfield  {title} {\bibinfo {title} {{Edge mass current and
  the role of Majorana fermions in A-phase superfluid $^3$He}},\ }\href
  {https://doi.org/10.1103/PhysRevB.85.100506} {\bibfield  {journal} {\bibinfo
  {journal} {Phys. Rev. B}\ }\textbf {\bibinfo {volume} {85}},\ \bibinfo
  {pages} {100506} (\bibinfo {year} {2012})}\BibitemShut {NoStop}%
\bibitem [{\citenamefont {Silveri}\ \emph {et~al.}(2014)\citenamefont
  {Silveri}, \citenamefont {Turunen},\ and\ \citenamefont {Thuneberg}}]{sil14}%
  \BibitemOpen
  \bibfield  {author} {\bibinfo {author} {\bibfnamefont {M.}~\bibnamefont
  {Silveri}}, \bibinfo {author} {\bibfnamefont {T.}~\bibnamefont {Turunen}},\
  and\ \bibinfo {author} {\bibfnamefont {E.}~\bibnamefont {Thuneberg}},\
  }\bibfield  {title} {\bibinfo {title} {{Hard domain walls in superfluid
  $^3$He-B}},\ }\href {https://doi.org/10.1103/PhysRevB.90.184513} {\bibfield
  {journal} {\bibinfo  {journal} {Phys. Rev. B}\ }\textbf {\bibinfo {volume}
  {90}},\ \bibinfo {pages} {184513} (\bibinfo {year} {2014})}\BibitemShut
  {NoStop}%
\bibitem [{\citenamefont {Ali}\ \emph {et~al.}(2011)\citenamefont {Ali},
  \citenamefont {Zhang},\ and\ \citenamefont {Sauls}}]{ali11}%
  \BibitemOpen
  \bibfield  {author} {\bibinfo {author} {\bibfnamefont {S.}~\bibnamefont
  {Ali}}, \bibinfo {author} {\bibfnamefont {L.}~\bibnamefont {Zhang}},\ and\
  \bibinfo {author} {\bibfnamefont {J.~A.}\ \bibnamefont {Sauls}},\ }\bibfield
  {title} {\bibinfo {title} {{Thermodynamic Potential for Superfluid $^3$He in
  Silica Aerogel}},\ }\href {https://doi.org/10.1007/s10909-010-0310-4}
  {\bibfield  {journal} {\bibinfo  {journal} {J. Low Temp. Phys.}\ }\textbf
  {\bibinfo {volume} {162}},\ \bibinfo {pages} {233} (\bibinfo {year}
  {2011})}\BibitemShut {NoStop}%
\bibitem [{\citenamefont {Levitin}\ \emph {et~al.}(2010)\citenamefont
  {Levitin}, \citenamefont {Bennett}, \citenamefont {Casey}, \citenamefont
  {Cowan}, \citenamefont {Parpia},\ and\ \citenamefont {Saunders}}]{lev10}%
  \BibitemOpen
  \bibfield  {author} {\bibinfo {author} {\bibfnamefont {L.~V.}\ \bibnamefont
  {Levitin}}, \bibinfo {author} {\bibfnamefont {R.~G.}\ \bibnamefont
  {Bennett}}, \bibinfo {author} {\bibfnamefont {A.~J.}\ \bibnamefont {Casey}},
  \bibinfo {author} {\bibfnamefont {B.}~\bibnamefont {Cowan}}, \bibinfo
  {author} {\bibfnamefont {J.}~\bibnamefont {Parpia}},\ and\ \bibinfo {author}
  {\bibfnamefont {J.}~\bibnamefont {Saunders}},\ }\bibfield  {title} {\bibinfo
  {title} {{Superfluid $^3$He Confined in a Single 0.6 Micron Slab: Phase
  Transition Between Superfluid Phases with Hysteresis}},\ }\href
  {https://doi.org/10.1007/s10909-009-9941-8} {\bibfield  {journal} {\bibinfo
  {journal} {J. Low Temp. Phys.}\ }\textbf {\bibinfo {volume} {158}},\ \bibinfo
  {pages} {159} (\bibinfo {year} {2010})}\BibitemShut {NoStop}%
\bibitem [{\citenamefont {Zhelev}\ \emph {et~al.}(2017)\citenamefont {Zhelev},
  \citenamefont {Abhilash}, \citenamefont {Smith}, \citenamefont {Bennett},
  \citenamefont {Rojas}, \citenamefont {Levitin}, \citenamefont {Saunders},\
  and\ \citenamefont {Parpia}}]{zhe16b}%
  \BibitemOpen
  \bibfield  {author} {\bibinfo {author} {\bibfnamefont {N.}~\bibnamefont
  {Zhelev}}, \bibinfo {author} {\bibfnamefont {T.~S.}\ \bibnamefont
  {Abhilash}}, \bibinfo {author} {\bibfnamefont {E.~N.}\ \bibnamefont {Smith}},
  \bibinfo {author} {\bibfnamefont {R.~G.}\ \bibnamefont {Bennett}}, \bibinfo
  {author} {\bibfnamefont {X.}~\bibnamefont {Rojas}}, \bibinfo {author}
  {\bibfnamefont {L.}~\bibnamefont {Levitin}}, \bibinfo {author} {\bibfnamefont
  {J.}~\bibnamefont {Saunders}},\ and\ \bibinfo {author} {\bibfnamefont
  {J.~M.}\ \bibnamefont {Parpia}},\ }\bibfield  {title} {\bibinfo {title} {{The
  A-B transition in superfluid Helium-3 under confinement in a thin slab
  geometry}},\ }\href {https://doi.org/10.1038/ncomms15963} {\bibfield
  {journal} {\bibinfo  {journal} {Nat. Comm.}\ }\textbf {\bibinfo {volume}
  {8}},\ \bibinfo {pages} {15963} (\bibinfo {year} {2017})}\BibitemShut
  {NoStop}%
\bibitem [{\citenamefont {Shook}\ \emph {et~al.}(2020)\citenamefont {Shook},
  \citenamefont {Vadakkumbatt}, \citenamefont {Senarath~Yapa}, \citenamefont
  {Doolin}, \citenamefont {Boyack}, \citenamefont {Kim}, \citenamefont
  {Popowich}, \citenamefont {Souris}, \citenamefont {Christani}, \citenamefont
  {Maciejko},\ and\ \citenamefont {Davis}}]{sho20}%
  \BibitemOpen
  \bibfield  {author} {\bibinfo {author} {\bibfnamefont {A.~J.}\ \bibnamefont
  {Shook}}, \bibinfo {author} {\bibfnamefont {V.}~\bibnamefont {Vadakkumbatt}},
  \bibinfo {author} {\bibfnamefont {P.}~\bibnamefont {Senarath~Yapa}}, \bibinfo
  {author} {\bibfnamefont {C.}~\bibnamefont {Doolin}}, \bibinfo {author}
  {\bibfnamefont {R.}~\bibnamefont {Boyack}}, \bibinfo {author} {\bibfnamefont
  {P.~H.}\ \bibnamefont {Kim}}, \bibinfo {author} {\bibfnamefont {.~G.~G.}\
  \bibnamefont {Popowich}}, \bibinfo {author} {\bibfnamefont {F.}~\bibnamefont
  {Souris}}, \bibinfo {author} {\bibfnamefont {H.}~\bibnamefont {Christani}},
  \bibinfo {author} {\bibfnamefont {J.}~\bibnamefont {Maciejko}},\ and\
  \bibinfo {author} {\bibfnamefont {J.~P.}\ \bibnamefont {Davis}},\ }\bibfield
  {title} {\bibinfo {title} {{Stabilized Pair Density Wave via Nanoscale
  Confinement of Superfluid {\He}}},\ }\href
  {https://doi.org/10.1103/PhysRevLett.124.015301} {\bibfield  {journal}
  {\bibinfo  {journal} {Phys. Rev. Lett.}\ }\textbf {\bibinfo {volume} {124}},\
  \bibinfo {pages} {015301} (\bibinfo {year} {2020})}\BibitemShut {NoStop}%
\bibitem [{\citenamefont {Lotnyk}\ \emph {et~al.}(2020)\citenamefont {Lotnyk},
  \citenamefont {Eyal}, \citenamefont {Zhelev}, \citenamefont {Abhilash},
  \citenamefont {Smith}, \citenamefont {Terilli}, \citenamefont {Wilson},
  \citenamefont {Mueller}, \citenamefont {Einzel}, \citenamefont {Saunders},\
  and\ \citenamefont {Parpia}}]{lot20}%
  \BibitemOpen
  \bibfield  {author} {\bibinfo {author} {\bibfnamefont {D.}~\bibnamefont
  {Lotnyk}}, \bibinfo {author} {\bibfnamefont {A.}~\bibnamefont {Eyal}},
  \bibinfo {author} {\bibfnamefont {N.}~\bibnamefont {Zhelev}}, \bibinfo
  {author} {\bibfnamefont {T.~S.}\ \bibnamefont {Abhilash}}, \bibinfo {author}
  {\bibfnamefont {E.~N.}\ \bibnamefont {Smith}}, \bibinfo {author}
  {\bibfnamefont {M.}~\bibnamefont {Terilli}}, \bibinfo {author} {\bibfnamefont
  {J.}~\bibnamefont {Wilson}}, \bibinfo {author} {\bibfnamefont
  {E.}~\bibnamefont {Mueller}}, \bibinfo {author} {\bibfnamefont
  {D.}~\bibnamefont {Einzel}}, \bibinfo {author} {\bibfnamefont
  {J.}~\bibnamefont {Saunders}},\ and\ \bibinfo {author} {\bibfnamefont
  {J.~M.}\ \bibnamefont {Parpia}},\ }\bibfield  {title} {\bibinfo {title}
  {{Thermal transport of helium-3 in a strongly confining channel}},\ }\href
  {https://doi.org/10.1038/s41467-020-18662-8} {\bibfield  {journal} {\bibinfo
  {journal} {Nat. Comm.}\ }\textbf {\bibinfo {volume} {11}},\ \bibinfo {pages}
  {4843} (\bibinfo {year} {2020})}\BibitemShut {NoStop}%
\bibitem [{\citenamefont {Sharma}\ and\ \citenamefont {Sauls}(2022)}]{sha22}%
  \BibitemOpen
  \bibfield  {author} {\bibinfo {author} {\bibfnamefont {P.}~\bibnamefont
  {Sharma}}\ and\ \bibinfo {author} {\bibfnamefont {J.~A.}\ \bibnamefont
  {Sauls}},\ }\bibfield  {title} {\bibinfo {title} {{Anomalous Thermal Hall
  Effect in Chiral Phases of {\He}-Aerogel}},\ }\href
  {https://doi.org/10.1007/s10909-021-02657-w} {\bibfield  {journal} {\bibinfo
  {journal} {J. Low Temp. Phys.}\ }\textbf {\bibinfo {volume} {208}},\ \bibinfo
  {pages} {341} (\bibinfo {year} {2022})},\ \Eprint
  {https://arxiv.org/abs/2111.02032} {2111.02032} \BibitemShut {NoStop}%
\bibitem [{\citenamefont {Sharma}\ \emph {et~al.}(2023)\citenamefont {Sharma},
  \citenamefont {Vorontsov},\ and\ \citenamefont {Sauls}}]{sha22a}%
  \BibitemOpen
  \bibfield  {author} {\bibinfo {author} {\bibfnamefont {P.}~\bibnamefont
  {Sharma}}, \bibinfo {author} {\bibfnamefont {A.}~\bibnamefont {Vorontsov}},\
  and\ \bibinfo {author} {\bibfnamefont {J.~A.}\ \bibnamefont {Sauls}},\
  }\bibfield  {title} {\bibinfo {title} {{Disorder Induced Anomalous Thermal
  Hall Effect in Chiral Phases of Superfluid {\He}}},\ }\href
  {https://doi.org/10.7566/JPSCP.38.011002} {\bibfield  {journal} {\bibinfo
  {journal} {JPS Conf. Proc.}\ }\textbf {\bibinfo {volume} {38}},\ \bibinfo
  {pages} {011002} (\bibinfo {year} {2023})},\ \bibinfo {note} {proc. of the
  LT29, Hokkaido, Japan},\ \Eprint {https://arxiv.org/abs/2209.04004}
  {2209.04004} \BibitemShut {NoStop}%
\bibitem [{\citenamefont {Scott}\ \emph {et~al.}(2023)\citenamefont {Scott},
  \citenamefont {Nguyen}, \citenamefont {Park},\ and\ \citenamefont
  {Halperin}}]{sco23}%
  \BibitemOpen
  \bibfield  {author} {\bibinfo {author} {\bibfnamefont {J.~W.}\ \bibnamefont
  {Scott}}, \bibinfo {author} {\bibfnamefont {M.~D.}\ \bibnamefont {Nguyen}},
  \bibinfo {author} {\bibfnamefont {D.}~\bibnamefont {Park}},\ and\ \bibinfo
  {author} {\bibfnamefont {W.~P.}\ \bibnamefont {Halperin}},\ }\bibfield
  {title} {\bibinfo {title} {{Magnetic Susceptibility of Andreev Bound States
  in Superfluid $^3$He-B}},\ }\href {https://doi.org/10.48550/arXiv.2302.01258}
  {\bibfield  {journal} {\bibinfo  {journal} {arXiv}\ }\textbf {\bibinfo
  {volume} {2302.01258}},\ \bibinfo {pages} {1} (\bibinfo {year}
  {2023})}\BibitemShut {NoStop}%
\end{thebibliography}
%
\end{document}